\documentclass[%
 reprint,
 amsmath,amssymb,
 aps,
]{revtex4-2}

\usepackage[english]{babel}
\usepackage{graphicx}% Include figure files
\usepackage{dcolumn}% Align table columns on decimal point
\usepackage{bm}% bold math
\usepackage{hyperref}% add hypertext capabilities
\usepackage[mathlines]{lineno}% Enable numbering of text and display math
\usepackage{booktabs}
\usepackage{multirow}
\usepackage{etoolbox}

\usepackage{braket}

\usepackage{physics}
\usepackage{siunitx}
\usepackage{amssymb}
\usepackage{parskip}
\usepackage[table]{xcolor} 
\usepackage{array}
\usepackage{tabularx}
\usepackage{newtxtext}
\usepackage{newtxmath}
\usepackage{algorithm}
\usepackage{algpseudocode}
\usepackage{makecell}
\usepackage{relsize}
\usepackage{siunitx}
\usepackage[version=3]{mhchem}

\algrenewcommand\algorithmicrequire{\textbf{Input:}}
\algrenewcommand\algorithmicensure{\textbf{Output:}}

\newcolumntype{E}[1]{%
  S[table-format=#1]@{\,}l%
}

\DeclareTextCompositeCommand{\r}{OT1}{A}{%
  \leavevmode\vbox{%
    \offinterlineskip
    \ialign{\hfil##\hfil\cr\char23\cr\noalign{\kern-1.15ex}A\cr}%
  }%
}

\begin{document}

\preprint{APS/123-QED}

\title{Discovery and recovery of crystalline materials with property-conditioned transformers}% Force line breaks with \\

\author{Cyprien Bone}
 \email{cyprien.bone.24@ucl.ac.uk}
 \author{Matthew Walker}
 \author{Bradley A. A. Martin}
 \affiliation{%
    Department of Chemistry, University College London,
    20 Gordon Street, London WC1H 0AJ, United Kingdom
}%
\author{Kuangdai Leng}
\affiliation{Earth Rover Program, 71-75 Shelton Street, Covent Garden, London, WC2H 9JQ, United Kingdom}
\author{Luis M. Antunes}
\affiliation{Independent Researcher, Canada}
\author{Ricardo Grau-Crespo}
\affiliation{%
    School of Engineering and Materials Science, Queen Mary University of London, London E1 4NS, United Kingdom
}%
\author{Amil Aligayev}
\author{Javier Dominguez}
\affiliation{
NOMATEN Centre of Excellence, National Centre for Nuclear Research, ul. A. Sołtana 7, Otwock, 05-400, Poland
}
\author{Keith T. Butler}%
 \email{k.t.butler@ucl.ac.uk}
\affiliation{%
    Department of Chemistry, University College London,
    20 Gordon Street, London WC1H 0AJ, United Kingdom
}%

\date{\today}

\begin{abstract}
Generative models have recently shown great promise for accelerating the design and discovery of new functional materials. Conditional generation enhances this capacity by allowing inverse design, where specific properties can be requested during the generation process.  However, the conditioning of transformer-based approaches is constrained by discrete tokenisation schemes and the risk of catastrophic forgetting during fine-tuning. This work introduces CrystaLLM-$\pi$ (property injection), a conditional autoregressive framework that integrates continuous property representations directly into the transformer's attention mechanism. Two architectures, Property-Key-Value (PKV) Prefix attention and PKV Residual attention, are presented. These methods bypass inefficient sequence-level tokenisation and preserve foundational knowledge from unsupervised pre-training on Crystallographic Information Files (CIFs) as textual input by minimising token and transformer-level disruptions. We evaluate these mechanisms through systematic robustness studies and test the framework across two distinct tasks. First, for materials discovery, the model is fine-tuned on a specialised photovoltaic dataset to generate novel, stable candidates validated by Density Functional Theory (DFT).  CrystaLLM-$\pi$ provides a lightweight, flexible and scalable framework for inverse materials design. Second, for structure recovery, the model processes high-dimensional, heterogeneous X-ray diffraction (XRD) patterns, achieving structural accuracy competitive with specialised models. We also demonstrate recovery of experimentally determined structures from simulated XRD patterns derived from deposited crystal structures, and show polymorph differentiation from lab-measured XRD patterns in a targeted \ce{TiO2} case study.

\end{abstract}


\maketitle

\section{Introduction}
\label{sec:main}

Developing new functional materials has historically been a time-consuming process. There are many barriers that make the discovery of new materials challenging, including navigating the vast compositional space of inorganic materials and effectively characterising newly synthesised candidates~\cite{antypov2025discovery, merchant2023scaling, davies2016computational, tiwariExploringWorldFunctional2023}. The combinatorial complexity of materials space, which scales exponentially with the number of elements, coupled with polymorphism and structural complexity, requires intelligent approaches to accelerating discovery~\cite{nwabaraHighThroughputComputational2025, allahyariCoevolutionarySearchOptimal2020, butlerMachineLearningMolecular2018}. Computational chemistry has helped ameliorate some of these problems by offering accurate predictions of the properties of new materials and complementing modern characterisation techniques~\cite{hautier2019finding, armstrong2020understanding}. However, accurate physics-based calculations are generally too computationally expensive to allow for wide and fast explorations of materials space.

Generative artificial intelligence has emerged as a powerful tool for materials design~\cite{park2024has, sanchez2018inverse, de2025generative}. In recent years, two main paradigms for generative materials modelling have emerged: diffusion/flow-based models and transformer-based models~\cite{xieCrystalDiffusionVariational2022, parkExplorationCrystalChemical2024, de2025generative, gruver2023llmtime, millerFlowMMGeneratingMaterials2024}. Diffusion and flow approaches have achieved state-of-the-art performance in generating materials both unconditionally and conditioned on specific functional properties~\cite{millerFlowMMGeneratingMaterials2024, pakornchoteDiffusionProbabilisticModels2023, yeConCDVAEMethodConditional2024, luoDeepLearningGenerative2024}. Despite these successes, conditioning these models on new datasets to target specific properties is difficult: it frequently requires important architectural modifications and a two-phase protocol for unconditional pre-training followed by adapter-specific training ~\cite{zeniGenerativeModelInorganic2025}. Additionally, many diffusion and flow models rely on graph-based representations, where scaling to larger systems is either computationally costly or requires sacrificing long-range dependencies because strong inductive biases, limiting applicability to large and complex chemistries~\cite{yangScalableDiffusionMaterials2024}.

Transformer-based generative models present an alternative framework for materials design, offering inherent architectural flexibility and favourable scalability by using token-based representations for structure generation~\cite{yanInvariantTokenizationCrystalline2024, caoSpaceGroupInformed2024, mohantyCrysTextGenerativeAI2025, breuckGenerativeMaterialTransformer2025, kazeev2503wyckoff, alampara2024mattext}. The core self-attention mechanism~\cite{vaswaniAttentionAllYou2023a} can contextualise diverse inputs in a unified manner, including compositional, structural and functional property information. Functional property conditioning is currently achieved in transformer-based large language models (LLMs), such as NatureLM and AtomGPT, by fine-tuning general-use models with token-based guidance~\cite{choudharyAtomGPTAtomisticGenerative2024, xiaNatureLMDecipheringLanguage2025, choudharyDiffractGPTAtomicStructure2024, jablonkaLeveragingLargeLanguage2023}. However, standard transformer architectures struggle to condition on continuous physical properties effectively. Discrete, digit-level tokenisation inherently disrupts ordinal relationships and prevents smooth mathematical interpolation~\cite{levyLanguageModelsEncode2025, liExposingNumeracyGaps2025}. Furthermore, using identical discrete tokens to encode continuous global properties and local spatial coordinates or physical descriptors forces a representational conflict within the embedding space. This gap can be resolved with enough data, but is problematic given the relative scarcity of specialised property-labelled crystallographic datasets~\cite{schmidtImprovingMachinelearningModels2024, lematerial2024, kirklinOpenQuantumMaterials2015}.

This work presents CrystaLLM-\scalebox{1.3}{$\pi$} (Property Injection), a transformer-based framework that aims to resolve these limitations. The central hypothesis is that introducing property embeddings as a dedicated attention channel enables targeted material generation while avoiding any disruptive expansion to the sequence length. Two mechanisms are introduced: Property-Key-Value (PKV) Prefix attention and PKV Residual attention. Both methods integrate continuous property information directly into the attention layers, bypassing sequence embedding layers. The mechanisms are designed to preserve the rich structural knowledge from the pre-trained base model, which is based on CrystaLLM~\cite{antunesCrystalStructureGeneration2024}, while efficiently enabling structure generation steered by functional property targets.

To isolate the impact of these architectural modifications, generative performance is systematically evaluated against standard sequence-based conditioning baselines across datasets ranging from 1K to 653K samples. The generality of the methods is demonstrated across two distinct materials design tasks: the generation of novel stable photovoltaic absorber material candidates with high potential efficiency, verified by Density Functional Theory (DFT), and the recovery of known experimental structures from X-ray diffraction (XRD) data.

The results establish inverse design capabilities of conditional autoregressive crystal structure generation. CrystaLLM-\scalebox{1.3}{$\pi$} provides an accessible, lightweight and flexible framework for the recovery and discovery of crystalline materials with a range of targeted functional properties and structural prompts, adaptable to diverse scenarios in materials design without the need for complex architectural redesign or extensive retraining.

\section{Results}
\label{sec:Results}

\subsection{CrystaLLM-\scalebox{1.3}{$\pi$} Framework}
\label{sec:base_model}

\begin{figure*}[htbp!]
    \centering
    \includegraphics[width=0.75\textwidth]{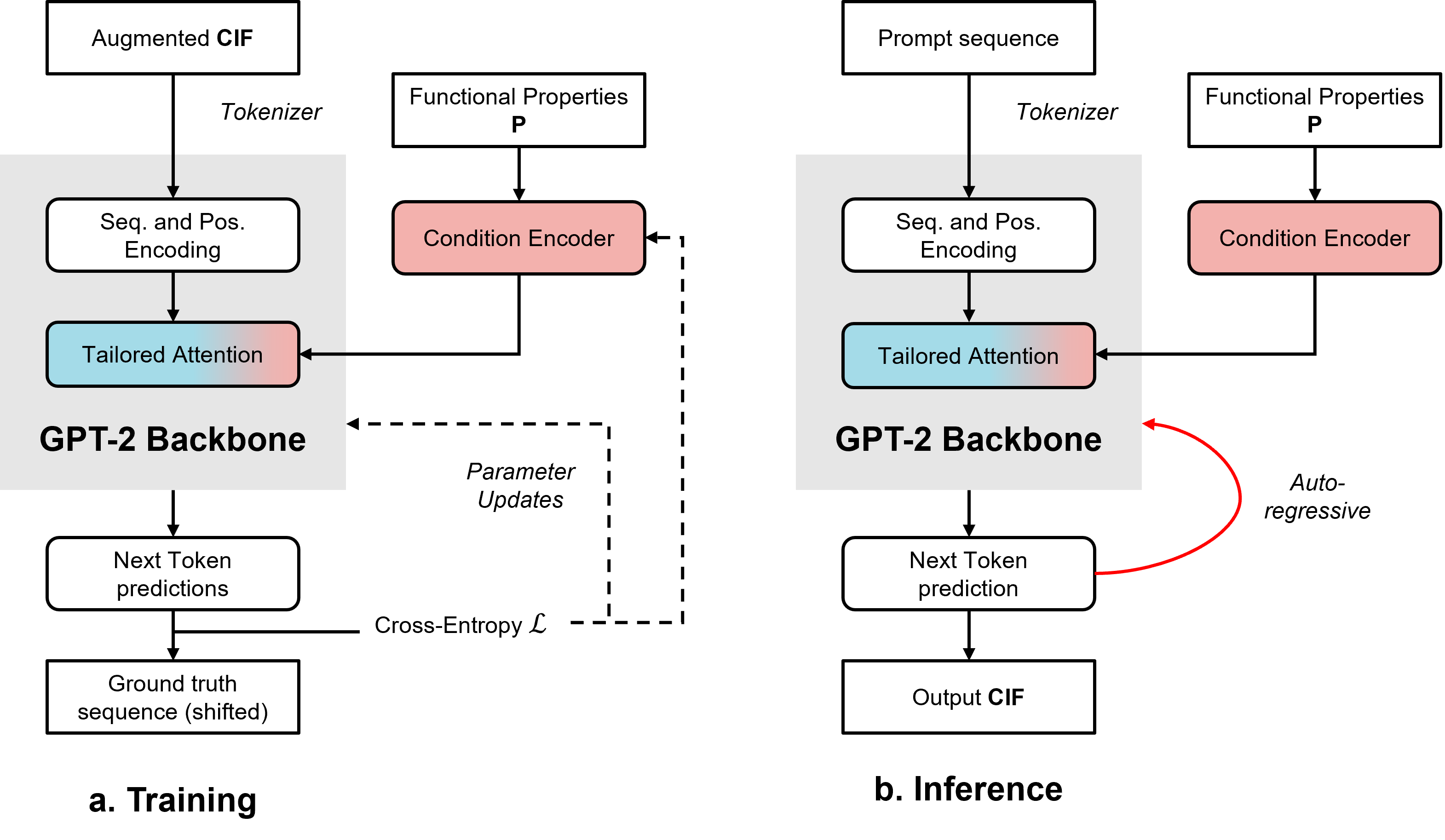}
    \caption{
    Framework for property-conditioned generation of Crystallographic Information Files (CIFs). \textbf{a.} During training, an augmented CIF is converted to token and positional embeddings, while the target property vector $P$ is mapped by a condition encoder and injected into each multi-head attention layer of the GPT-2 backbone. Next-token predictions are optimised against the shifted ground-truth sequence. \textbf{b.} During inference, the model receives an initial prompt sequence and the target property vector $P$, then autoregressively feeds each predicted token back into the model to generate the output CIF. This schematic shows how continuous property information conditions decoding during both training and generation.}
    \label{fig:LLM_training}
\end{figure*}

CrystaLLM-\scalebox{1.3}{$\pi$} is a transformer-based framework for materials discovery that sequentially generates Crystallographic Information Files (CIFs). The model builds upon the GPT-2 architecture of the original CrystaLLM model~\cite{radfordLanguageModelsAre, antunesCrystalStructureGeneration2024}. Optimisation focuses on training the model to autoregressively generate chemically viable crystal structures by internalising the syntax and physical constraints inherent to crystallographic data, the mechanisms of which are extensively discussed in the original paper~\cite{antunesCrystalStructureGeneration2024}.

Integrating continuous functional properties for target-driven inverse design requires key framework modifications compared to unconditional crystal generation. The framework expands the unlabeled pre-training corpus and introduces a condition-aligned batching mechanism (Appendix~\ref{app:pre_processing}). A new dynamic boundary tokenisation scheme and a customised format-aware loss function are implemented to stabilise early training (Methods~\ref{sec:backbone_theory}, Appendix~\ref{app:loss_landscapes}). To ensure computational scalability and reproducibility, the underlying code adheres to standard Hugging Face tokeniser, model and training protocols, with all datasets and model weights publicly accessible. 

Evaluations on the LeMat-Benchmark (Appendix~\ref{app:lemat-bench}) quantify a set of advantages and limitations of the token-based autoregressive approach. On the non-pre-relaxed MP-20 and Alex-MP-20 benchmarks, CrystaLLM-\scalebox{1.3}{$\pi$} achieves the highest structural precision with a root mean square displacement (RMSD) of~\SI{0.3619}{\angstrom} and~\SI{0.1914}{\angstrom} respectively. In the MP-20 section, the model leads competing autoregressive frameworks in primary ranking metrics with a $0.3\%$ Stable, Unique, and Novel (SUN) score and a $3.6\%$ Metastable SUN score. While diffusion architectures currently reach lower convex hull energies, the proposed framework maintains robust structural validity ($86.4\%$) and high out-of-the-box stability, recording the highest overall Metastable rate ($48.7\%$) in this section. CrystaLLM-\scalebox{1.3}{$\pi$} however shows a clear restricted novelty rate ($24.8\%$) compared to diffusion models, which often exceed $60\%$. This constraint suggests a generation distribution biased toward high-density regions of the pre-training manifold, motivating the implementation of a functional property-conditioning framework to actively steer generation toward sparsely seen chemical spaces.

\subsection{Architectures for Property-Conditioned Generation}
\label{sec:conditioning-architectures-results}

While the discrete, digit-wise tokenisation of coordinates facilitates the hierarchical generation of spatial boundaries and physical descriptors, using the same approach for global functional properties introduces substantial architectural inefficiencies. Forcing the model to interpret complex, often multi-dimensional properties through the same discrete token embeddings used for physical descriptors of the material induces a representational conflict and exposes the fundamental numeracy gaps of standard transformers~\cite{levyLanguageModelsEncode2025, liExposingNumeracyGaps2025}. 

To resolve these limitations, CrystaLLM-\scalebox{1.3}{$\pi$} implements a continuous conditioning framework (Figure~\ref{fig:LLM_training}). During training (Figure~\ref{fig:LLM_training}a), the system tokenises an input CIF into position-encoded embeddings while a dedicated condition encoder simultaneously maps continuous numerical target properties $P$ directly into an independent, non-linear embedding space. Injecting these continuous representations into every multi-head attention (MHA) layer of the GPT-2 backbone actively biases the generation trajectory, establishing persistent property conditioning with minimal disruption to pre-trained sequence representations in embedding and Feed-Forward Neural Network (FFNN) layers. A cross-entropy loss (Methods~\ref{sec:backbone_theory}) is computed between the predicted conditional probability distribution of the next token and the ground-truth sequence. Backpropagation then jointly optimises the parameters of the transformer backbone and the condition encoder.

During inference (Figure~\ref{fig:LLM_training}b), the model receives a target property vector and an initial prompt sequence. This architecture supports \textit{ab initio} generation or generation constrained by specific cell compositions and space groups in a unified manner. The model autoregressively samples subsequent tokens to produce a complete CIF, integrating both sequence-level and property-level conditions.

To fairly investigate how best to condition transformer outputs with functional properties, we implemented and benchmarked four distinct architectures. These methods were all designed to integrate a numerical condition vector, $\mathbf{c} \in \mathbb{R}^P$, where $P$ is the number of properties, to guide the generative process. They are grouped into two classes: sequence-level methods and tailored attention mechanisms developed for this work.

Preserving the extensive structural knowledge from unsupervised pre-training is critical in conditional fine-tuning to mitigate catastrophic forgetting~\cite{mccloskeyCatastrophicInterferenceConnectionist1989}. For this goal, a dual optimisation strategy was developed and applied, as detailed in Methods~\ref{sec:conditioning-methods-methods}. A conservative learning rate was applied to the pre-trained backbone parameters to favour retention of foundational knowledge, while a higher learning rate was used for the newly initialised conditioning layers to accelerate adaptation to the new task.

\subsubsection{Sequence-Level Conditioning}
\label{sec:sequence-conditioning}
We first implemented two methods that modify the model's input sequence, serving as comparative baselines to the tailored attention methods.

The \textbf{Raw Conditioning} method involves the direct, digit-level tokenisation of the numerical values in the property vector $\mathbf{c} \in \mathbb{R}^P$. This resulting sequence of $M$ tokens (where $M \ge P$, one token per digit for each condition) is prepended to the tokenised input sequence $\mathbf{x} = (x_1, \dots, x_L)$. Formally, if we denote the tokenised property string as $\mathbf{t}_{\mathrm{props}} = (t_1, \dots, t_M)$, the new input sequence becomes $\mathbf{x'} = (\texttt{<bos>}, t_1, \dots, t_M, \verb|\n|, x_1, \dots, x_L, \texttt{<eos>})$. As this approach only manipulates the input sequence, it requires no architectural modifications and handles the property values as discrete tokens.

From the hypothesis that a continuous embedding for the properties is more naturally suited to the task, the \textbf{Prepend Conditioning} method introduces a learned, continuous representation of $\mathbf{c}$. A \textbf{Prepend Encoder}, implemented as a multi-layer perceptron (MLP) with \texttt{ReLU} activation, projects $\mathbf{c}$ into a sequence of $P$ condition embeddings, $\mathbf{E}_c \in \mathbb{R}^{P \times d_{\mathrm{emb}}}$, where $d_{\mathrm{emb}}$ is the model's embedding dimension. The input to the transformer is the concatenation of these condition embeddings with the standard token embeddings, $\mathbf{E}_x \in \mathbb{R}^{(L+P) \times d_{\mathrm{emb}}}$. The regular attention mechanism ($A_{\mathrm{out}}$) operates on this concatenated sequence:
\begin{equation}
    A_{\mathrm{out}}(Q_{\mathrm{seq}}, K_{\mathrm{seq}}, V_{\mathrm{seq}}) = \text{softmax}\left(\frac{Q_{\mathrm{seq}} K_{\mathrm{seq}}^\top}{\sqrt{d_k}}\right) V_{\mathrm{seq}} ,
\end{equation}
where $Q_{\mathrm{seq}}$, $K_{\mathrm{seq}}$ and $V_{\mathrm{seq}}$ are the Query, Key and Value matrices, respectively, generated from the transformer's representation of the concatenated sequence containing both token and condition embeddings. The full architecture diagram can be found in Appendix~\ref{app:prepend_gpt}.

\subsubsection{Tailored Attention Conditioning}
\label{sec:tailored_attention}
Direct manipulation of the attention mechanism via continuous property representations is hypothesised to enhance targeted generation and mitigate transfer learning disruption more effectively than sequence-level interventions~\cite{liPrefixTuningOptimizingContinuous2021}. To evaluate this, the Prefix and Residual architectures were developed to embed property information directly into the transformer attention layers. Intuitively, the Prefix architecture extends the effective context window by concatenating learned property keys and values to the sequence, imposing a ``hard'' structural bias. On the other hand, the Residual architecture operates as an additive modification to the base self-attention, injecting a parallel property-based score dynamically. This is effectively a weighted residual term that ``softly'' steers the generation without disrupting the intrinsic sequence attention operations.

\begin{figure*}[htbp!]
    \centering
    \includegraphics[trim=1mm 1mm 1mm 1mm, clip, width=\textwidth]{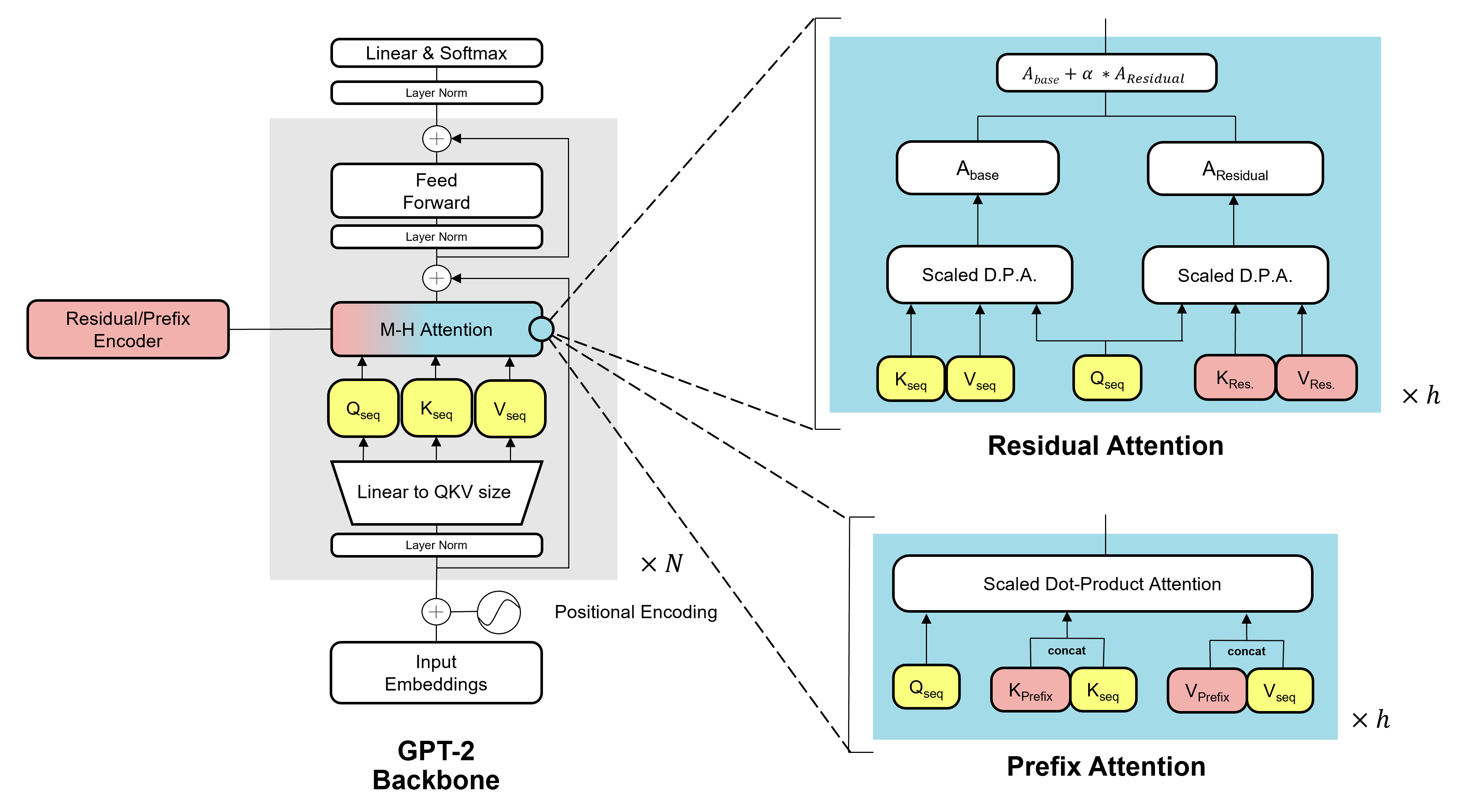}
    \caption{Schematic of the Prefix and Residual attention conditioning mechanisms. The left panel shows the GPT-2 backbone, where a conditioning encoder (Appendix~\ref{app:encoders}) injects learned property key-value representations into each of the $N$ decoder blocks. In Prefix attention (bottom right), $Q_{\mathrm{seq}}$ attends over concatenated property and sequence keys and values, $[K_{\mathrm{Prefix}}; K_{\mathrm{seq}}]$ and $[V_{\mathrm{Prefix}}; V_{\mathrm{seq}}]$. In Residual attention (top right), the base self-attention output $A_{\mathrm{base}}$ is combined with a parallel property-conditioned term $A_{\mathrm{Residual}}$ through the learned scaling parameter $\alpha$. $N$ denotes the number of decoder blocks, $h$ denotes the number of attention heads and $\alpha$ corresponds to Equation~\ref{eq:Residual_attention}. These mechanisms condition generation by either extending or augmenting the attention pathway.}
    \label{fig:Prefix_Residual_model}
\end{figure*}

\textbf{Prefix attention conditioning} (Figure~\ref{fig:Prefix_Residual_model}) directly injects property information into the self-attention mechanism of each transformer layer, bypassing sequence encoding. ``Ghost'' Key-Value (KV) pairs generated from the condition vector are concatenated with input KV pairs. This confines sequence expansion entirely to attention layers rather than FFNNs and input tokens to minimise disruption. The method is a materials-oriented adaptation of Li and Liang's Prefix Tuning natural language method~\cite{liPrefixTuningOptimizingContinuous2021}.

The \textbf{Prefix Encoder} (See Appendix~\ref{app:encoders}) transforms the input condition vector $\mathbf{c} \in \mathbb{R}^P$ into a tensor compatible with the attention mechanism. The vector $\mathbf{c}$ is first projected to a higher-dimensional latent space via a linear layer, followed by Layer Normalisation and \texttt{ReLU}. This hidden representation is projected to the KV tensor dimension for all attention heads and layers, $\mathbf{h}_{\mathrm{KV}} \in \mathbb{R}^{P \times N \times h \times 2 \times d_k}$, where $P$ is the number of conditions, $N$ being the number of decoder blocks, $h$ the number of attention heads and $2 \times d_k$ the dimension of the KV pair matrix. The resulting tensor is split into Key ($K_{\mathrm{Prefix}}$) and Value ($V_{\mathrm{Prefix}}$) tensors.

For each layer, the attention mechanism is modified. Given $Q_{\mathrm{seq}}$ from the input sequence, the attention output is computed as:
\begin{equation}
\label{eq:Prefix_attention}
A_{\mathrm{out}}(Q_{\mathrm{seq}}, K, V) = \text{softmax}\left(\frac{Q_{\mathrm{seq}} (K^\top)}{\sqrt{d_k}}\right) V,
\end{equation}
where $K = [K_{\mathrm{Prefix}}; K_{\mathrm{seq}}]$ and $V = [V_{\mathrm{Prefix}}; V_{\mathrm{seq}}]$ are the concatenations of the conditioning- and sequence-derived Key and Value matrices, respectively. The dot product is performed along the sequence length dimension with identical inner embedding dimensions.

\textbf{Residual attention conditioning} (Figure~\ref{fig:Prefix_Residual_model}) introduces a more nuanced, ``soft'' conditioning mechanism that dynamically weights the influence of the property conditions. The \textbf{Residual Encoder} (See Appendix~\ref{app:encoders}) is architecturally similar to the Prefix Encoder but employs \texttt{tanh} activations and utilises reshaping and einsum operations for more structured transformations. These are engineered so that each of the $P$ conditions initially influences a dedicated portion of the latent space before being combined. The output of the encoder is a set of KV pairs, $K_{\mathrm{Residual}}$ and $V_{\mathrm{Residual}}$, where each pair can be attributed to an input condition. This structured encoding preserves the associativity between each of the $P$ conditions and its resulting KV pair, distinct from the Prefix approach, where property representations are collectively projected.

The core of this method is the \textbf{Residual Attention} module. For each attention head, two separate attention scores are computed in parallel: a standard self-attention score over the input sequence ($A_{\mathrm{base}}$) and a ``Residual'' attention score ($A_{\mathrm{Residual}}$) between the input sequence's Queries ($Q_{\mathrm{seq}}$) and the Residual Keys ($K_{\mathrm{Residual}}$) and Values ($V_{\mathrm{Residual}}$). These are then combined via a learned, per-layer weighting parameter, $\alpha$. The modified attention output, $A_{\mathrm{out}}$, is computed as:
\begin{equation}
\label{eq:Residual_attention}
\begin{aligned}
A_{\mathrm{base}} &= \text{softmax}\left(\frac{Q_{\mathrm{seq}} K_{\mathrm{seq}}^\top}{\sqrt{d_k}}\right) V_{\mathrm{seq}}, \\
A_{\mathrm{Residual}} &= \text{softmax}\left(\frac{Q_{\mathrm{seq}} K_{\mathrm{Residual}}^\top}{\sqrt{d_k}}\right) V_{\mathrm{Residual}}, \\
A_{\mathrm{out}} &= A_{\mathrm{base}} + \alpha \cdot A_{\mathrm{Residual}}.
\end{aligned}
\end{equation}
The scalar weight, $\alpha$, is initialised at zero, similar to the ``zero-initialisation'' technique used in Low-Rank Adaptation (LoRA) fine-tuning~\cite{huLoRALowRankAdaptation2021a}. This ensures that the model initially relies only on the pre-trained self-attention mechanism, preserving its generative capabilities and mitigating catastrophic forgetting.  Optimisation of $\alpha$ during training dynamically calibrates the degree of conditioning.

This architecture handles missing or unspecified conditions, a capability enabled by the Residual Encoder's associativity and the parallel, additive structure of the attention layer. When a value of $-100$ is used in the condition vector $\mathbf{c}$, the attention scores (computed as $Q \cdot K_{\text{Residual}}^\top$) for queries attending to that specific condition variable are set to $-\infty$, effectively nullifying its contribution after $\text{softmax}$ in $A_{\text{Residual}}$, similarly to causal attention masking. The additive design bypasses an important limitation of Prefix conditioning for masking ``missing'' conditions. By concatenating keys, Prefix still changes the sequence length in the attention layers, altering the $\text{softmax}$ normalisation. The residual architecture avoids this perturbation, allowing ``missing'' condition handling while maintaining the crystal sequence's intrinsic attention distribution.

\subsection{Validation of the Conditioning Frameworks}
\label{sec:validating-cond-frameworks}
 
Having established the four conditioning architectures, we now evaluate their ability to transfer knowledge from unsupervised learning on unconditional CIF generation to the conditional property-labelled setting. The studies quantitatively compare the efficacy of sequence-level versus attention-based conditioning and how the most promising methods respond to different dataset sizes, representing realistic research settings.

\subsubsection{Advantages of pre-training and condition encoding}
\label{sec:pre-training}

\begin{figure*}[htbp!]
    \centering
    \includegraphics[width=1.0\textwidth]{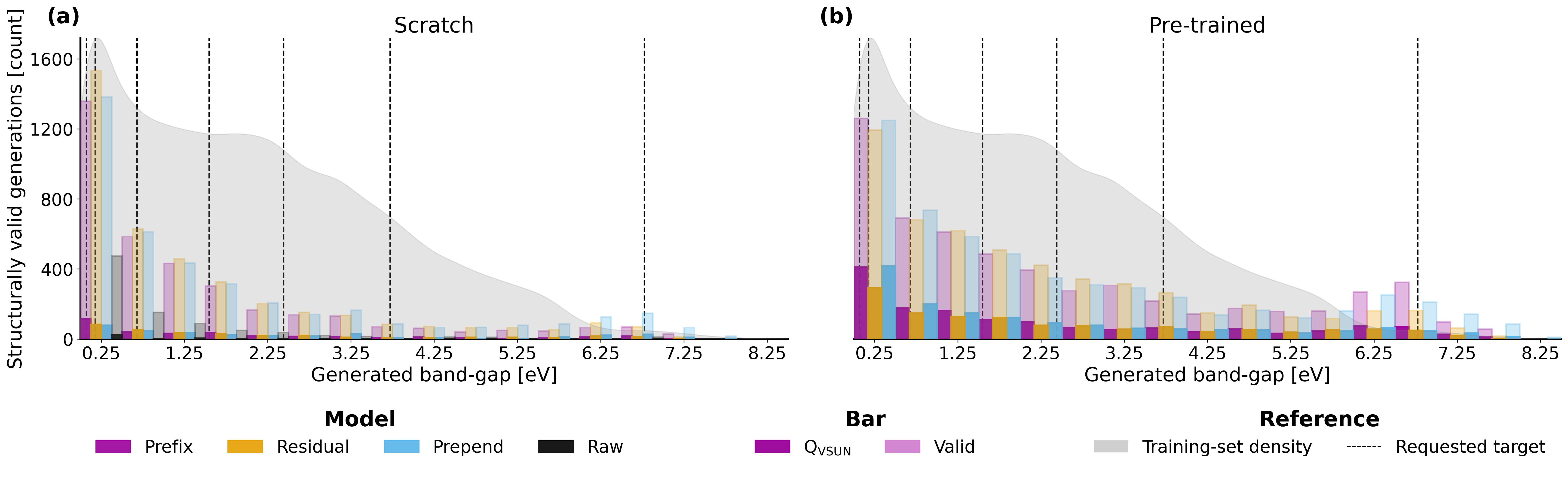}
    \caption{Output-space distributions for target band gap conditioning with and without pre-training. \textbf{a.} Models trained from scratch. \textbf{b.} Pre-trained models. For each requested band gap, coloured bars distinguish the Prefix, Residual, Prepend and Raw architectures. Lighter bars denote structurally valid generations and darker overlaid bars denote the subset satisfying the comprehensive $Q_{\mathrm{VSUN}}$ metric (Validity, Stability, Uniqueness and Structural Novelty; Methods~\ref{sec:eval_metrics}). Dashed lines mark the requested target band gaps, and the grey background shows the training-set band gap density. The~\SI{12.1}{\electronvolt} target is omitted because no qualifying generations were obtained at this value. Pre-training improves the $Q_{\mathrm{VSUN}}$ output space, but performance remains constrained by the labelled-data support.
    }
    \label{fig:pre-training_benefits_main}
\end{figure*}

Figure~\ref{fig:pre-training_benefits_main} shows how our four conditioning architectures handle specific band gap targets. We tested the models across a range from~\SIrange{0}{12}{\electronvolt}, which pushes past the high-density region (\SIrange{0}{6}{\electronvolt}) of the training set. We evaluated performance using structural Validity and the Valid-SUN quality metric ($Q_{\mathrm{VSUN}}$) under 2 training regimes. First, models trained from scratch on the MP~Bandgap dataset (53.3K structures, with $95\%$ containing 1 to 104 atoms, see Appendix~\ref{app:token_distributions}) versus models pre-trained on the broader LeMaterial dataset (4.35M structures, no band gap label) with fine-tuning on the MP~Bandgap dataset. Appendix Table~\ref{tab:pre-train-regime-metrics} summarises the quantitative impact of this pre-training.

Unsupervised pre-training improves generative capacity for generating Valid structures across all continuous architectures in this benchmark. On average, aggregate validity increases by $20.4\%$ and the number of $Q_{\mathrm{VSUN}}$ structures increases by $14.3\%$. In particular, the pre-trained models vastly outperform scratch models in terms of Valid structure generation in the tails of the target distribution. The accompanying rise in $Q_{\mathrm{VSUN}}$ MAE is consistent with a stronger structural prior that favours plausible configurations, particularly in data-sparse regions.

\begin{figure*}[htbp!]
    \centering
    \includegraphics[width=1.0\textwidth]{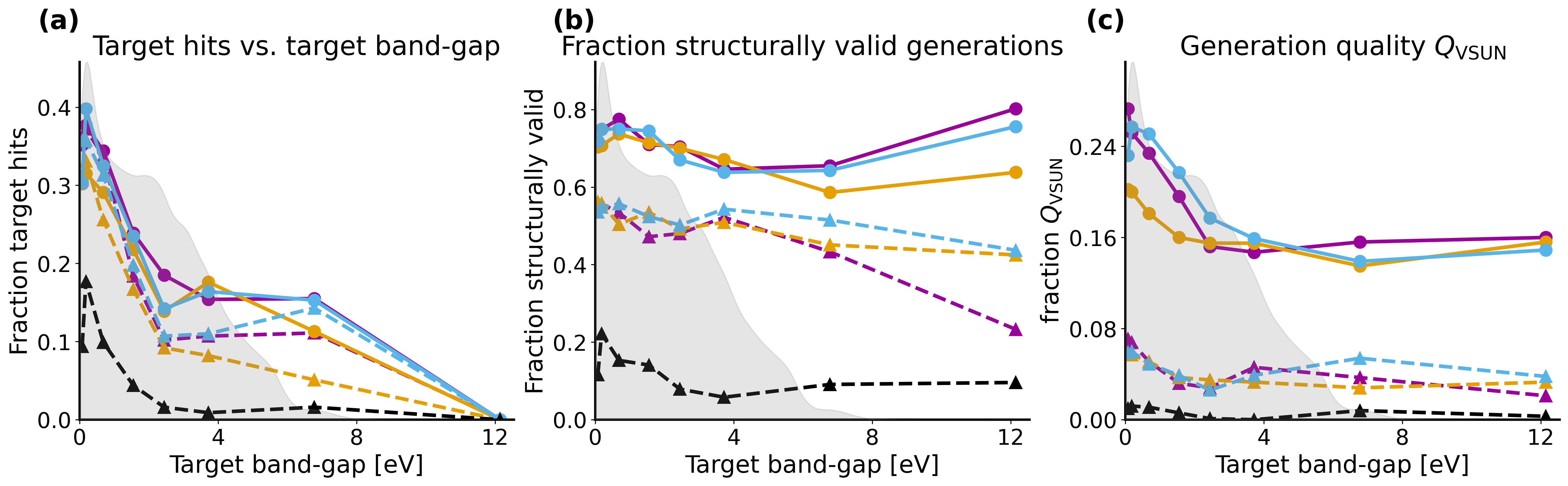}
    \newline
    \newline
    \includegraphics[width=0.7\textwidth]{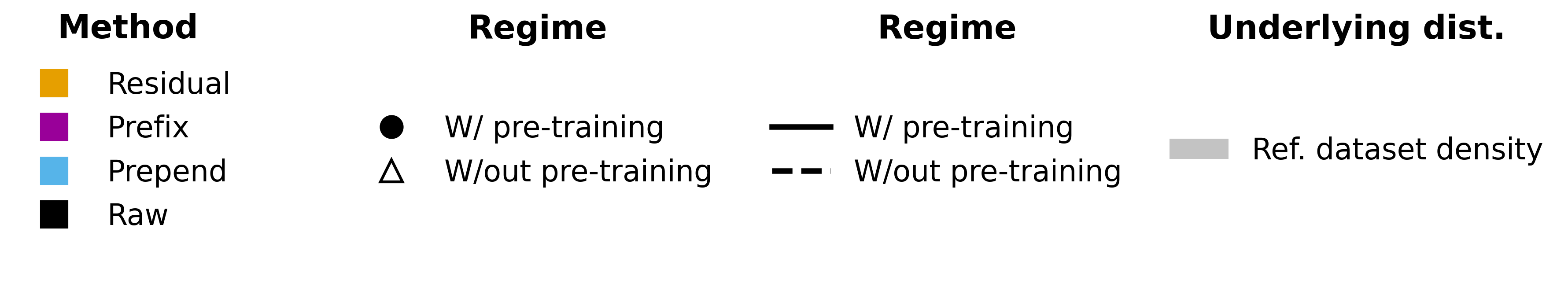}
    \caption{Target-space analysis of conditioning mechanisms with respect to pre-training initialisation. The panels show (\textbf{a}) target hit rate, (\textbf{b}) fraction of structurally valid generations and (\textbf{c}) fraction of generations satisfying the comprehensive $Q_{\mathrm{VSUN}}$ criteria (Validity, Stability, Uniqueness and Novelty; Methods~\ref{sec:eval_metrics}) across specific band gap targets, with 1K generation attempts per target (Appendix~\ref{app:experimental_protocols}). Colours denote the Residual, Prefix, Prepend and Raw models. Circles and solid lines indicate pre-trained models, whereas triangles and dashed lines indicate models trained from scratch. The grey background shows the reference training-set band gap density. As targets move farther out of distribution, hit rate declines more rapidly than structural validity or $Q_{\mathrm{VSUN}}$, indicating that the models preserve physically plausible generation more readily than exact numerical agreement at extreme targets. The Raw token-level baseline is included only as a failure reference because it does not generate Valid structures in the pre-trained regime.}
    \label{fig:pre-training_target_plot}
\end{figure*}

Target-space analysis (Appendix Figure~\ref{fig:pre-training_target_plot}) assesses the correlation of the property of interest to the conditioning value and the Valid structure generation rate to the underlying density of materials. While property hit rates decrease as targets move into sparse data regions ($r = 0.78$), $Q_{\mathrm{VSUN}}$ is more stable ($r = 0.22$). Pre-trained models still generate $Q_{\mathrm{VSUN}}$ materials near the~\SI{6.5}{\electronvolt} target despite low training density there. This indicates that the model can still use sparse boundary regions to find new Valid and $Q_{\mathrm{VSUN}}$ materials near the upper tail of the labelled distribution.

At an extreme stress test of ~\SI{12.1}{\electronvolt}, no architecture produced valid structures. This region is outside of the support of the model, meaning that there are effectively no property-structure mappings to train on. Materials with very large band gaps ($>$~\SI{12.1}{\electronvolt}) are exceedingly rare and have to be highly ionic, with the widest reported experimental band gap being LiF at ~\SI{12.1}{\electronvolt}~\cite{piacentini1976thermoreflectance}. Additionally, property labels are calculated with GGA-DFT which systematically under estimates band gaps by $30$--$40\%$ ~\cite{kimBandgapDatabaseSemiconducting2020, park2011hybrid}. Together, these make the lack of valid generations at this target unsurprising.

To benchmark our approach outside a transformer-only setting, we compared against MatterGen~\cite{zeniGenerativeModelInorganic2025}, a mixed discrete-continuous diffusion model which operates on graph data, which we will refer to as a graph diffusion model in this work. As shown in Table~\ref{tab:pre-train-regime-metrics}, there is a trade-off between volume and precision. MatterGen generates a greater total number of Valid structures ($N_{\mathrm{Valid}} = 6585$), but the conditional autoregressive models show better calibration to requested properties, achieving higher $Q_{\mathrm{VSUN}}$ $R^2$ values and lower $Q_{\mathrm{VSUN}}$ MAEs. The failure at~\SI{12.1}{\electronvolt} is also present with MatterGen~\cite{zeniGenerativeModelInorganic2025}, as both plateau near the upper edge of the labelled support. Finally, the autoregressive pipeline is substantially more efficient in this study's setting: requiring less GPU time, inference time and peak video random access memory (VRAM) than MatterGen, highlighting that text-based models are more efficient when modelling larger, more complex systems. For a detailed comparison between both conditional generation frameworks, we refer the reader to the discussion in Appendix~\ref{app:diffusion-comparison}.

The Raw baseline fails during transfer learning and is excluded from further analysis. Concatenating tokens works poorly in low-data regimes, leading to a complete loss of structural validity after pre-training. We focus the rest of our study on the three continuous encoding architectures: Prepend, Prefix and Residual.

\subsubsection{Advantages of attention-level conditioning}
\label{sec:dataset_size_study}

\begin{figure*}[htbp!]
    \centering
    \includegraphics[width=0.9\textwidth]{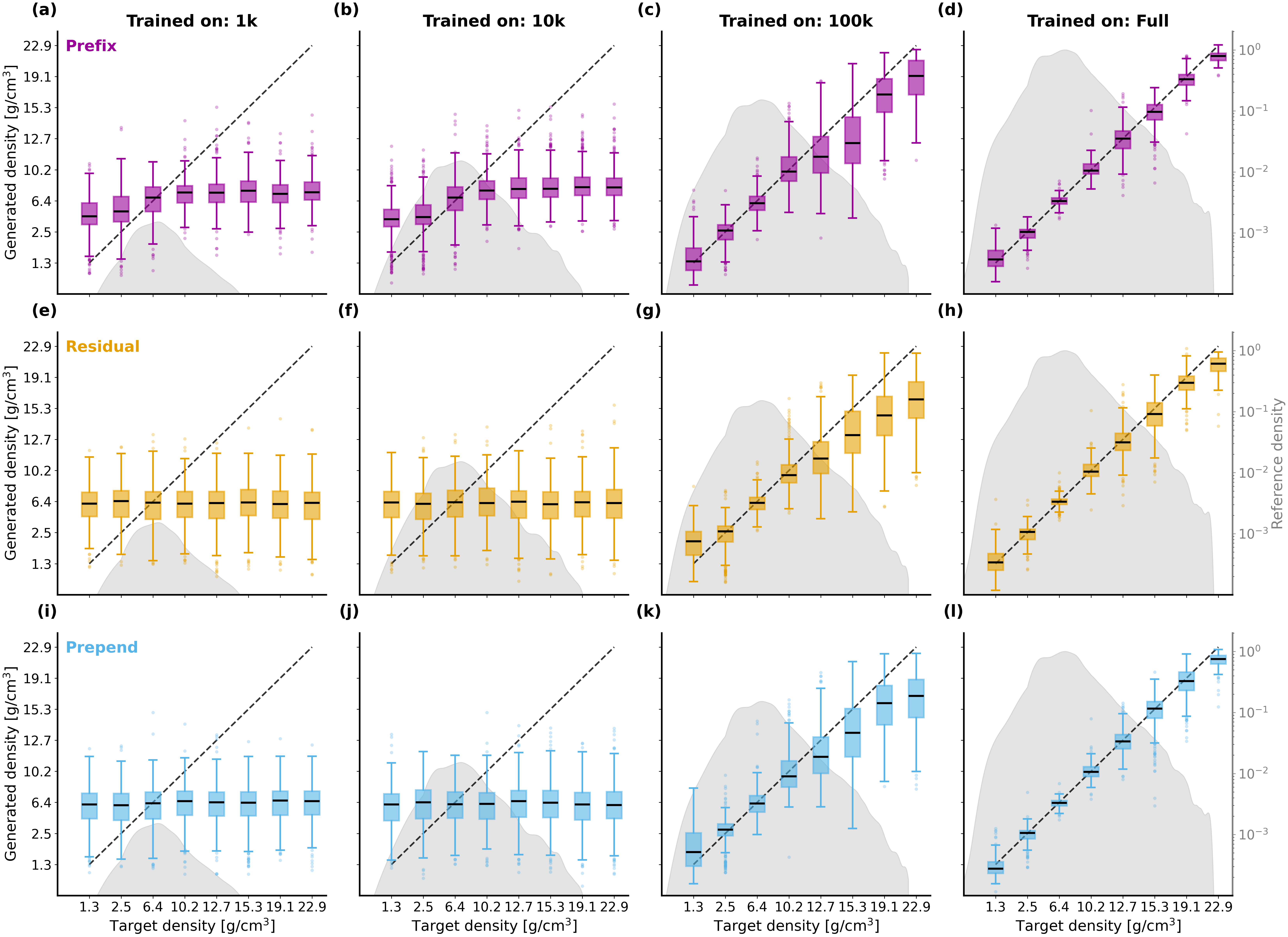}
    \caption{Density calibration across fine-tuning set sizes. Rows correspond to the Prefix, Residual and Prepend conditioning architectures, and columns correspond to 1K, 10K, 100K and Full fine-tuning sets. Box plots summarise the generated density distributions for $Q_{\mathrm{VSUN}}$ structures at each requested target density. The dashed diagonal marks perfect calibration, and the grey-shaded distribution shows the density distribution of the corresponding training subset on a log scale. Calibration is weak in the 1K and 10K regimes and becomes well aligned with the target values at 100K and Full data volumes.}
    \label{fig:dataset_size_study_main}
\end{figure*}

This section investigates continuous conditioning mechanisms via the physical density property, using the MatterGen Density fine-tuning dataset (653K labelled structures; see Appendix~\ref{app:dataset_construction}). Density provides an effective benchmark because it is available \textit{a priori} for all materials. Scaling regimes for generative performance are quantified in Figure~\ref{fig:dataset_size_study_main} and Appendix~\ref{app:dataset-size-study}. The analysis isolates an important transition: at extreme data scarcity, outputs reflect the broad statistical mean of the unconditional prior, but increasing the data volume and specifically the inclusion of long-tail distributions, allows for precise control over target properties.

Within the restricted 1K and 10K regimes, all architectures struggle to deviate from the pre-training distribution. Because these subsets are derived via random sampling, they inherently capture the dense centre of the structural manifold, limiting the variance necessary to learn extreme examples. In the 1K regime, the Residual method exhibits high structural robustness, generating the maximum number of Valid structures ($N_{\mathrm{Valid}} = 7170$). The architecture preserves foundational structural knowledge during severe data scarcity. This preservation prevents the immediate collapse of the generative distribution, an effect supported by the loss landscape analysis detailed in Appendix~\ref{app:loss_landscapes}. The Prepend method exhibits the lowest validity and targeting yields in these environments, suggesting that non-attention-level conditioning requires more data to overcome the sequence-level perturbation. Under these regimes, the Prefix model outperforms the other two methods significantly, but remains severely restricted by the lack of structure-property materials present in the tails of the complete dataset.

As the dataset scales to 100K and Full volumes, the architectures successfully overcome the unconditional bias. The Prefix mechanism exhibits the strongest calibration metrics, with the highest positive correlation ($Q_{\mathrm{VSUN}}$ $R^2 = 0.83$ and $0.97$) while maintaining the lowest $Q_{\mathrm{VSUN}}$ MAE (\SIlist{1.98;0.72}{g/cm^3}) at 100K and Full sizes respectively. On the other hand, the Prepend method required the most data to overcome sequence disruption but ultimately anneals to highly competitive performance, marginally exceeding Prefix and Residual in Q-Hit yield ($16.51\%$) at the Full data limit.  

While specific architectural advantages emerge at the extremes of data availability, the aggregate results suggest that Prefix is the strongest overall architecture in this study. It attains the best value in 15 of the 20 primary performance metrics considered in this scaling study, compared with 3 for Residual and 2 for Prepend. Taken together, these results suggest a regime-dependent pattern: Residual conditioning knowledge transfer from pre-training is more robust in the data-limited setting, whereas Prefix provides the best overall trade-off between validity, calibration and data efficiency across the broader study.

\subsection{Exploration of photovoltaic efficiency chemical space}
\label{sec:slme}

\begin{figure*}[htbp!]
    \centering
    \includegraphics[width=0.9\textwidth]{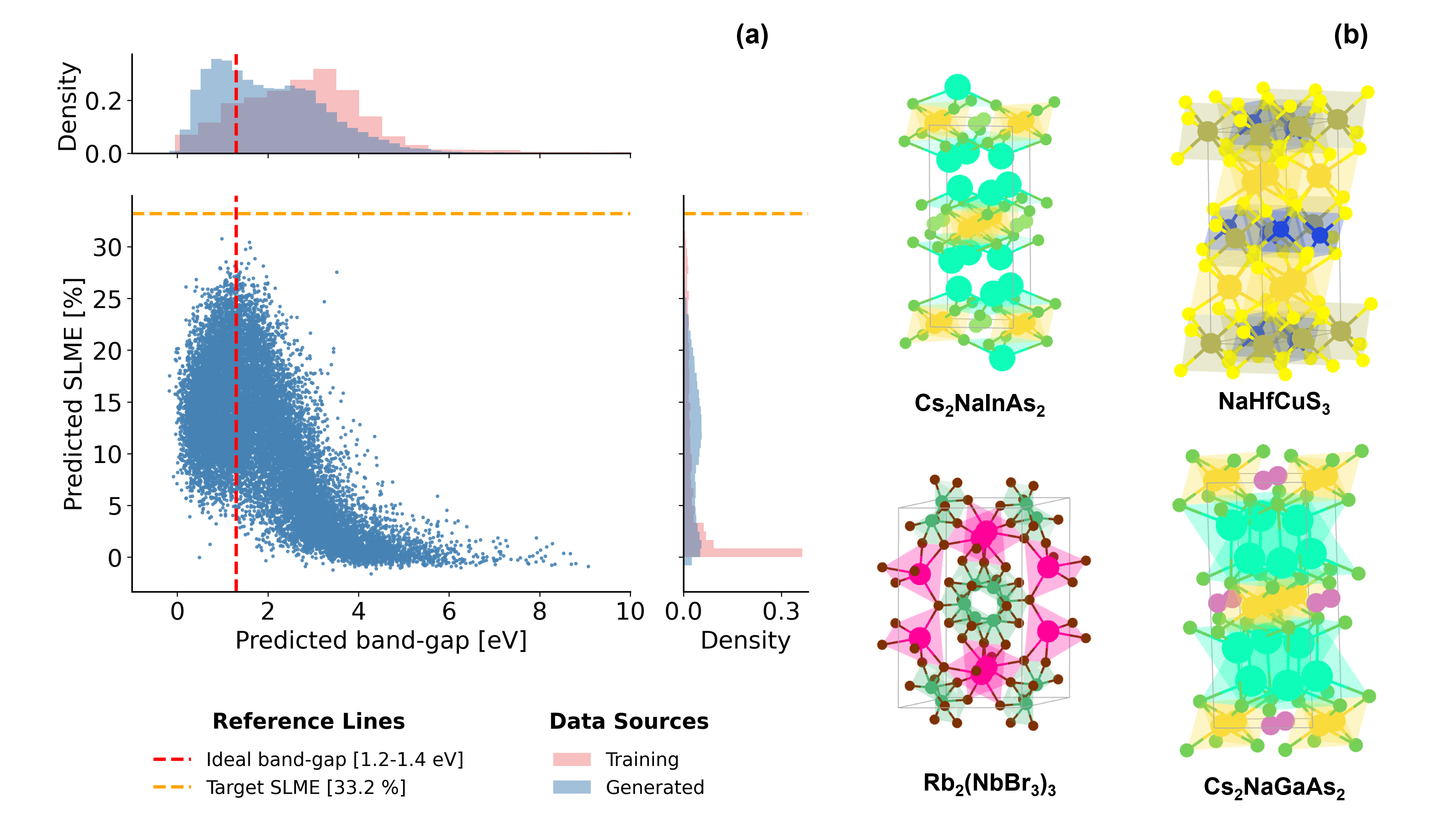}
    \caption{Photovoltaic candidate screening and representative validated structures. \textbf{a.} Predicted surrogate Spectroscopic Limited Maximum Efficiency (SLME) is plotted against predicted HSE06 band gap for generated structures, with marginal distributions comparing the generated and training datasets. Dashed reference lines mark the ideal band gap region near~\SIrange{1.2}{1.4}{\electronvolt} and the target SLME of $33.2\%$. The generated distribution concentrates near the optimal band gap range while spanning a broad range of predicted efficiencies. \textbf{b.} Four Density Functional Theory (DFT) validated candidate structures are shown: \ce{Cs2NaInAs2} (SLME $26.4\%$, $E_{\text{hull}}^{\text{PBE}} = \SI{-0.605}{\electronvolt/atom}$), \ce{Cs2NaGaAs2} (SLME $24.4\%$, $E_{\text{hull}}^{\text{PBE}} = \SI{-0.589}{\electronvolt/atom}$), \ce{NaHfCuS3} (SLME $23.3\%$, $E_{\text{hull}}^{\text{PBE}} = \SI{-0.013}{\electronvolt/atom}$) and \ce{Rb2(NbBr3)3} (SLME $13.3\%$, $E_{\text{hull}}^{\text{PBE}} = \SI{0.001}{\electronvolt/atom}$). \ce{Rb2(NbCl3)3} structures (SLME: $11.6\%$ and $11.3\%$, both $E_{\text{hull}}^{\text{PBE}}$: \SI{0.003}{\electronvolt/atom}, 2 polymorphs not shown) were also identified as stable, promising candidates. The model effectively steered generation to a promising area of chemical space to generate materials of interest.}
    \label{fig:gen_slme}
\end{figure*}

The Spectroscopic Limited Maximum Efficiency (SLME) serves as a theoretical measure of photovoltaic potential, integrating absorption profiles, band gap characteristics and non-radiative recombination losses to estimate the upper limit of power conversion efficiency~\cite{shockleyDetailedBalanceLimit1961, yuIdentificationPotentialPhotovoltaic2012}. Identifying materials with high SLME requires navigating a multi-dimensional optimisation landscape where both electronic structure and optical properties must be simultaneously considered. This section demonstrates how the CrystaLLM-\scalebox{1.3}{$\pi$} framework leverages large-scale unsupervised pre-training to map complex structure-property relationships within a data-scarce regime. The model was fine-tuned on the MP~SLME dataset~\cite{walkerCarbonCostMaterials2025}, with 5.35K SLME labelled inorganic structures. To steer the generation toward high-performance regions of chemical space, the model was conditioned on the maximum observed SLME in the training set ($33.2\%$). Because the target property set consists of a single, scalar condition with no missing values and fine-tuning data was scarce, prefix conditioning was employed.

\begin{table}[htbp!]
    \centering
    \small
    \setlength{\tabcolsep}{6pt}
    \renewcommand{\arraystretch}{1.0}
    \begin{tabular}{@{}lS[
        table-format=5.0,
        group-separator={,},
        group-minimum-digits=4
    ]@{}}
    \toprule
    \textbf{VSUN specs.} & {\textbf{$N$ Materials}} \\
    
    \midrule
    Fine-tuning, Structural     & 16463 \\
    Fine-tuning, Compositional  & 15462 \\
    Pre-training, Structural     & 8712  \\
    Pre-training, Compositional  & 1809  \\
    \bottomrule
    \end{tabular}
    \caption{Counts of Valid, Stable, Unique and Novel (VSUN) materials under different novelty definitions and reference datasets (Methods~\ref{sec:eval_metrics}). Counts are computed from 100K generations. The fine-tuning dataset is MP~SLME with 5.3K structures and the pre-training dataset is LeMaterial with 4.35M structures. Structural novelty denotes the absence of a matching structure under pymatgen's \texttt{StructureMatcher}, whereas compositional novelty denotes the absence of a material with the same reduced formula and structure in the reference dataset.}
    \label{tab:vsun_summary}
\end{table}

Table~\ref{tab:vsun_summary} details the generation of VSUN structures. Analysis of the compositional novelty profiles reveals distinct behaviour depending on the reference dataset. While 15,462 generated structures possess reduced formulas absent from the fine-tuning corpus, this yield decreases to 1,809 when compared against the expansive pre-training corpus. This disparity is consistent with the transfer of structure-property information from the fine-tuning task onto a broader structural prior acquired during pre-training.

To validate the physical properties of the conditionally generated chemical space, the predicted band gaps of the candidate structures are analysed (Figure~\ref{fig:gen_slme}a). Maximising single-junction photovoltaic efficiency requires a band gap between~\SIlist{1.2;1.4}{\electronvolt}~\cite{shockleyDetailedBalanceLimit1961}. The baseline training data exhibits an average band gap between~\SIlist{3;4}{\electronvolt}. If the framework failed to capture the structural determinants of SLME, generation outputs would default to this underlying prior. Instead, the distribution of generated structures shifts away from the dataset mode and clusters within the optimal energy range. This shift away from the dataset mode is consistent with the model learning structure-property correlations relevant to high SLME, rather than simply reproducing the dominant band gap distribution of the training set.

Following the criteria established by Yu and Zunger~\cite{yuIdentificationPotentialPhotovoltaic2012}, an SLME exceeding $20\%$ indicates a high-performance photovoltaic material. Within the 16,463 structurally novel materials with respect to the fine-tuning set (Table~\ref{tab:vsun_summary}), the generated set counts 1,546 candidates with ALIGNN-predicted SLMEs surpassing the threshold. To manage the computational cost of high-fidelity DFT validation, subsequent analysis focused on a subset of 30 top-ranking candidates. This selection prioritised candidates demonstrating both high predicted SLME and favourable sustainability profiles, quantified via the Herfindahl-Hirschman Index (HHI) for elemental supply chains (see Methods~\ref{sec:eval_metrics}). The properties and selection rationale for these candidates are detailed in Table~\ref{tab:slme_candidates_full}, with four representative structures highlighted in Figure~\ref{fig:gen_slme}b.

Analysis of the generated candidates reveals that the model output distribution exhibits chemically intuitive optimisation pathways, such as atomic substitutions on established high-performance motifs. For instance, the generation of \ce{Cs2NaInAs2} and \ce{Cs2NaGaAs2} exploits the optoelectronic properties of known \ce{InAs} and \ce{GaAs} motifs~\cite{songEnergyLevelTuned2018, papezOverviewCurrentState2021} to propose stable, high-SLME quaternary variants. Furthermore, the model identified a metastable family of rubidium-niobium halides, including \ce{Rb2(NbBr3)3} ($E_{\text{hull}}^{\text{PBE}}$: \SI{0.001}{\electronvolt/atom}) and two polymorphs of \ce{Rb2(NbCl3)3} ($E_{\text{hull}}^{\text{PBE}}$: \SI{0.003}{\electronvolt/atom}). These compositions are completely novel relative to the fine-tuning dataset, and the bromide variant is entirely absent from all training sets. While ALIGNN surrogate predictions initially suggested promising SLME values for these halides, subsequent hybrid DFT characterisation revealed that the conduction band minimum is split off from the main conduction band density of states. This electronic structure results in poor absorption properties and potentially low electron conductivity. This discrepancy highlights an important limitation of surrogate property-prediction models for materials discovery and the importance of \textit{ab initio} validation. Finally, the framework proposed \ce{NaHfCuS3}, a chalcogenide variant with a validated SLME of $22.6\%$. Although chalcogenides are established in thin-film photovoltaics~\cite{ChalcogenideCompoundsSolar2021, kuhar2017sulfide}, this specific material lacks prior photovoltaic investigation to the best of our knowledge. While high-fidelity validation was restricted to a curated subset, the broader pool of over 1,546 structurally novel candidates with predicted high SLME provides a substantial chemical space for future high-throughput characterisation. 

\subsection{Recovery of materials using X-Ray Diffraction (XRD) information conditioning}
\label{sec:recovery}

\subsubsection{Benchmarking against other ML models for XRD conditioned generation}
\label{sec:benchmarks}

To validate CrystaLLM-\scalebox{1.3}{$\pi$} on recovery tasks, its performance in XRD-conditioned structure determination was benchmarked against concurrently developed generative architectures. These baselines included DiffractGPT, a general-purpose large language model fine-tuned for XRD data via prompt engineering~\cite{choudharyDiffractGPTAtomicStructure2024}; Uni3Dar, an autoregressive model which uses compressed spatial representations with multi-modal conditioning tokens~\cite{luUni3DARUnified3D2025}; and PXRDGen, a diffusion pipeline employing contrastive learning to guide generation~\cite{liPowderDiffractionCrystal2025}. Comparative evaluations were conducted using benchmarks on established MP-20~\cite{jainCommentaryMaterialsProject2013} and Jarvis-DFT~\cite{choudharyJointAutomatedRepository2020} datasets.

For the MP-20 benchmark, CrystaLLM-\scalebox{1.3}{$\pi$} was trained from scratch following the PXRDGen protocol. With a perplexity-based ranking for candidate selection, the model achieves a conditional match rate of $62.73\%$ with an RMSD of~\SI{0.0444}{\angstrom} (Table~\ref{tab:mp20-benchmark}). As noted in the PXRDGen paper~\cite{liPowderDiffractionCrystal2025}, some of the test set structures fail standard validity checks applied post-generation, meaning that they could never be matched under these checks. Therefore, we also performed an analysis with a relaxed validity check for proposed structures (``Skip Valid.''), which shows an elevated match rate of $69.01\%$.

\begin{table}[htbp!]
    \centering
    \setlength{\tabcolsep}{5.5pt}
    \begin{tabular}{lcccc}
        \toprule
        \multirow{2}{*}{\textbf{Model}} & \multicolumn{2}{c}{\textbf{Conditional}} & \multicolumn{2}{c}{\textbf{Unconditional}} \\
        \cmidrule(lr){2-3} \cmidrule(lr){4-5} & \textbf{Match \%} & \textbf{RMSD} & \textbf{Match \%} & \textbf{RMSD} \\
        
        \midrule
        PXRDGen & 68.68 & 0.0707 & - & - \\
        Uni3Dar & \textbf{75.08} & \textbf{0.0276} & 65.48 & 0.0317 \\
        CrystaLLM-\scalebox{1.3}{$\pi$} & 62.73 & \textit{0.0444} & 55.85 & 0.0437 \\
        CrystaLLM-\scalebox{1.3}{$\pi$} \\ 
        (Skip Valid.) & \textit{69.01} & 0.0465 & - & - \\
        \bottomrule
    \end{tabular}
    
    \caption{Performance on the MP-20 benchmark for XRD-guided crystal structure prediction. Bold and italics denote the best and second-best metrics. Each ground-truth structure is evaluated against a single reported candidate per method. PXRDGen uses a 1-shot generation protocol, whereas Uni3Dar and CrystaLLM-\scalebox{1.3}{$\pi$} use ``1-perplexity'' ranking. Unconditional CrystaLLM-\scalebox{1.3}{$\pi$} metrics are taken from the reported ``1-shot'' MP-20 benchmark in Antunes et al.~\cite{antunesCrystalStructureGeneration2024}. The ``Skip Valid.'' row reports results when the benchmark validity screen is bypassed during matching.}
    \label{tab:mp20-benchmark}
\end{table}

The framework was evaluated on the Jarvis-DFT benchmark following the DiffractGPT protocol~\cite{choudharyDiffractGPTAtomicStructure2024}. To establish reproducible evaluation conditions, CrystaLLM-\scalebox{1.3}{$\pi$} was fine-tuned using a seeded randomised 90:10 train-test data split. Maintaining parity with the baseline Jarvis-XRD benchmark, pre-matching validity screens were excluded during evaluation to prevent arbitrary filtering of potential matches.

Table~\ref{tab:jarvis-benchmark} compares performance against Gradient Boosting (GBR), CNN and DiffractGPT baselines. CrystaLLM-\scalebox{1.3}{$\pi$} demonstrates improved structural accuracy: the top perplexity-ranked structure (Section~\ref{sec:sampling}) achieves an RMSD of~\SI{0.0347}{\angstrom}, representing a $50.4\%$ error reduction relative to DiffractGPT (\SI{0.0700}{\angstrom}). This fidelity is confirmed by the lowest Mean Absolute Error (MAE) across all lattice parameters. When evaluating the best match drawn from a pool of ``20-consistent'' structures, where a structure is deemed consistent if the output CIF is syntactically sensible (See Methods~\ref{sec:sampling}), the match rate reaches $85.47\%$ while maintaining a~\SI{0.0361}{\angstrom} RMSD.

A stratified analysis assessed generalisation by classifying the test set into materials observed during training phases (``Seen''), structurally novel polymorphs of known compositions (``Novel Polymorphs''), and materials with entirely unseen chemical compositions (``Novel Compositions''; Appendix~\ref{app:data_leakage}). Performance was broadly stable across these subsets, with novel compositions showing slightly better aggregate fidelity and match rates in this dataset. Appendix~\ref{app:data_leakage} discusses likely contributors to this pattern, including sample-composition effects.

\begin{table}[htbp!]
    \centering
    \setlength{\tabcolsep}{2pt} 
    \resizebox{\columnwidth}{!}{%
    \begin{tabular}{lccccc} 
        \toprule
        \textbf{Model} & \textbf{Match \%} & \textbf{RMSD} & \textbf{MAE a} & \textbf{MAE b} & \textbf{MAE c} \\
        
        \midrule
        GBR & - & - & 1.030 & 0.990 & 1.270 \\ 
        CNN & - & - & 0.280 & 0.270 & 0.280 \\
        DiffractGPT \\ 
        (formula) & - & 0.0700 & 0.170 & 0.180 & 0.270 \\ 
        CrystaLLM-\scalebox{1.3}{$\pi$} \\ 
        (20-consistent) & \textbf{85.47} & \textit{0.0361} & \textit{0.142} & \textit{0.135} & \textit{0.216} \\
        CrystaLLM-\scalebox{1.3}{$\pi$} \\ 
        (1-perplexity) & \textit{66.22} & \textbf{0.0347} & \textbf{0.096} & \textbf{0.098} & \textbf{0.183} \\
        \bottomrule
    \end{tabular}%
    }
    \caption{Performance on the Jarvis-DFT benchmark. Bold and italics denote the best and second-best metrics, respectively. CrystaLLM-\scalebox{1.3}{$\pi$} results are reported for the highest-ranked perplexity output (``1-perplexity'') and for the best structural match within 20 attempts (``20-consistent''). MAE a, MAE b and MAE c are absolute errors in the lattice parameters of the matched structures.}
    \label{tab:jarvis-benchmark}
\end{table}

\subsubsection{Experimental Structure Recovery from Chili-100K Data}
\label{sec:experimental_recovery}

The model was evaluated on a 1.5K-material test set drawn from the Chili-100K XRD dataset. The underlying crystal structures are experimentally determined COD entries, while the conditioning XRD vectors were simulated from those structures using the same preprocessing pipeline as the other XRD benchmarks. This curated subset spans a broad distribution of atom counts and reflects realistic experimental structure complexity (See Appendix~\ref{app:token_distributions}). To assess generative generalisation and control for data leakage, the test set was stratified into three 500-material subsets, as discussed in the previous section: ``Seen'', ``Novel Polymorph'', and ``Novel Composition''. The XRD-conditioned model was compared against a mirrored unconditioned baseline model trained with identical datasets and hyperparameters, but without XRD conditioning entirely.

\begin{figure*}[htbp!]
    \centering
    \includegraphics[width=0.95\textwidth]{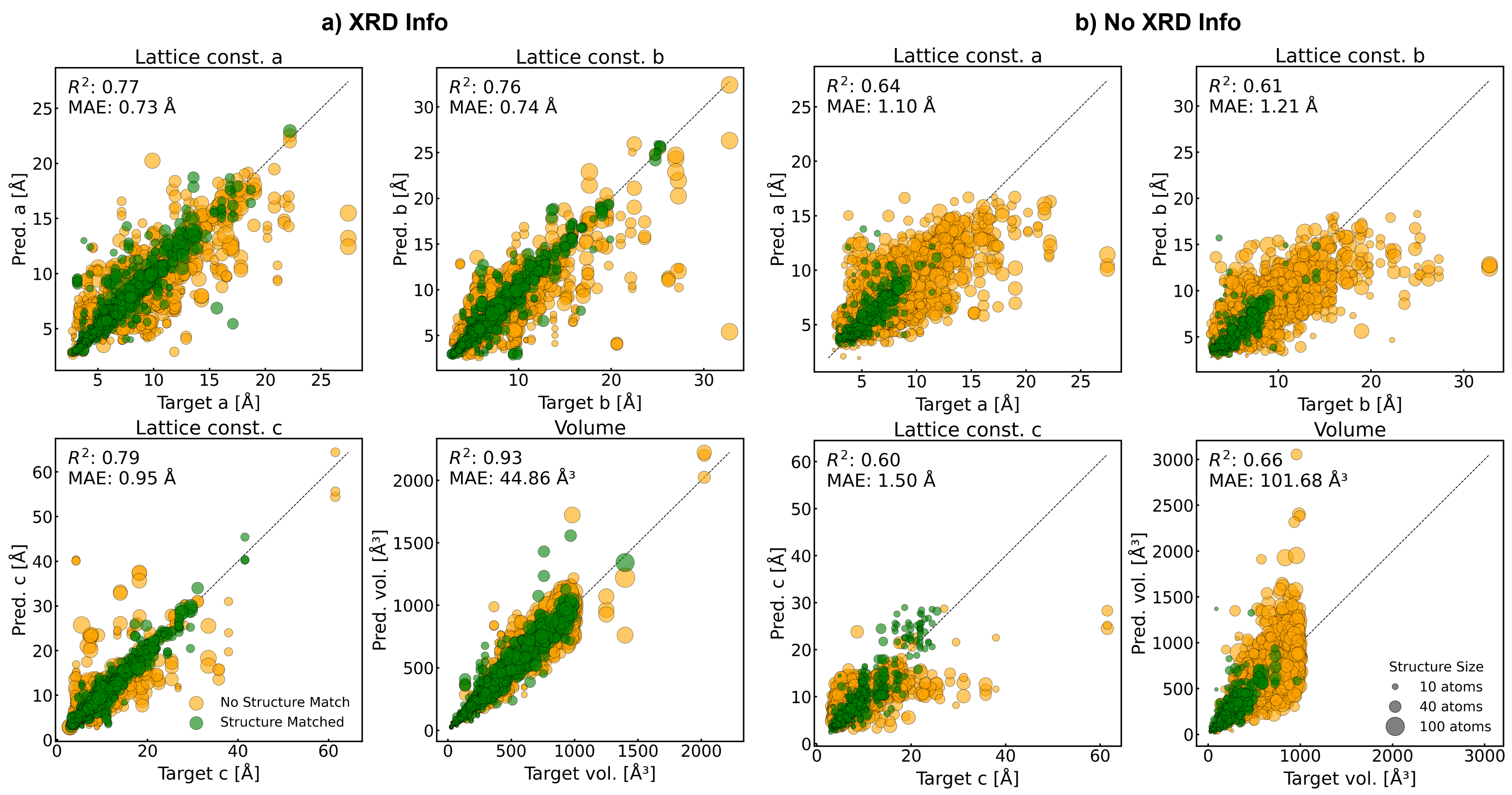}
    \vspace{5mm}
    \resizebox{\textwidth}{!}{
    \begin{tabular}{
      @{}ll
      E{2.1}
      E{1.3}
      E{3.1}
      E{2.1}
      E{2.0}
      E{1.3}
      c@{}
    }
    \toprule
    \textbf{Model Setup} &
    \textbf{Data Split} &
    \multicolumn{2}{c}{\textbf{Match \%} $(SE)$} &
    \multicolumn{2}{c}{\textbf{Mean RMSD} $(SE)$} &
    \multicolumn{2}{c}{\textbf{Vol MAE [$\AA^3$]} $(SE)$} &
    \multicolumn{2}{c}{\textbf{Avg. Atoms} $(SE)$} &
    \multicolumn{2}{c}{\textbf{Max Atoms} $(SE)$} &
    \multicolumn{2}{c}{\textbf{Avg. Perplexity} $(SE)$} &
    \textbf{Total time (h:m:s)} \\
    \midrule
    
    \multirow{5}{*}{\textbf{Conditioned}}
    & \textbf{Overall (20-consistent)}
    & 49.0 & {($\pm$0.4)}
    & 0.115 & {($\pm$0.001)}
    & 44.9 & {($\pm$0.6)}
    & 19.1 & {($\pm$0.1)}
    & 83 & {($\pm$10)}
    & 1.247 & {($\pm$0.001)}
    & \multirow{5}{*}{1:07:50 ($\pm$0:23)} \\
    
    & Seen
    & 45.5 & {($\pm$1.1)}
    & 0.119 & {($\pm$0.003)}
    & 41.4 & {($\pm$0.6)}
    & 21.9 & {($\pm$0.3)}
    & 83 & {($\pm$10)}
    & 1.268 & {($\pm$0.003)}
    & \\
    
    & Structurally Novel
    & 60.3 & {($\pm$0.3)}
    & 0.120 & {($\pm$0.002)}
    & 42.1 & {($\pm$0.9)}
    & 15.6 & {($\pm$0.0)}
    & 68 & {}
    & 1.205 & {($\pm$0.002)}
    & \\
    
    & Compositionally Novel
    & 41.4 & {($\pm$0.7)}
    & 0.103 & {($\pm$0.005)}
    & 51.1 & {($\pm$0.7)}
    & 21.2 & {($\pm$0.1)}
    & 65 & {($\pm$1)}
    & 1.268 & {($\pm$0.003)}
    & \\
    
    \cmidrule{2-14}
    
    & \textit{Overall (1-perplexity)}
    & 34.4 & {($\pm$0.2)}
    & 0.104 & {($\pm$0.000)}
    & 53.3 & {($\pm$0.4)}
    & 17.4 & {($\pm$0.2)}
    & 68 & {($\pm$2)}
    & 1.217 & {($\pm$0.001)}
    & \\
    
    \midrule
    
    \multirow{5}{*}{\textbf{Unconditioned}}
    & \textbf{Overall (20-consistent)}
    & 15.9 & {($\pm$0.1)}
    & 0.135 & {($\pm$0.001)}
    & 101.7 & {($\pm$0.2)}
    & 11.0 & {($\pm$0.2)}
    & 40 & {}
    & 1.312 & {($\pm$0.000)}
    & \multirow{5}{*}{3:07:14 ($\pm$0:37)} \\
    
    & Seen
    & 14.2 & {($\pm$0.3)}
    & 0.135 & {($\pm$0.005)}
    & 107.1 & {($\pm$2.4)}
    & 11.4 & {($\pm$0.0)}
    & 40 & {}
    & 1.329 & {($\pm$0.005)}
    & \\
    
    & Structurally Novel
    & 24.6 & {($\pm$0.0)}
    & 0.160 & {($\pm$0.003)}
    & 92.2 & {($\pm$0.3)}
    & 10.3 & {($\pm$0.3)}
    & 40 & {}
    & 1.264 & {($\pm$0.001)}
    & \\
    
    & Compositionally Novel
    & 8.9 & {($\pm$0.4)}
    & 0.064 & {($\pm$0.011)}
    & 107.2 & {($\pm$2.8)}
    & 12.4 & {($\pm$0.3)}
    & 40 & {}
    & 1.352 & {($\pm$0.007)}
    & \\
    
    \cmidrule{2-14}
    
    & \textit{Overall (1-perplexity)}
    & 8.3 & {($\pm$0.3)}
    & 0.105 & {($\pm$0.011)}
    & 165.9 & {($\pm$2.1)}
    & 11.7 & {($\pm$0.3)}
    & 40 & {}
    & 1.301 & {($\pm$0.000)}
    & \\
    
    \bottomrule
    \end{tabular}
    }
    \caption{
    Crystal structure recovery on the Chili-100K benchmark with and without XRD conditioning. \textbf{Top:} Predicted versus experimental lattice parameters ($a$, $b$, $c$) and unit-cell volumes are shown for the XRD-conditioned model (left four panels) and the text-only baseline without XRD information (right four panels). Point colour indicates whether the best candidate from ``20-consistent'' generations matches the target structure, point size scales with the target atom count and dashed lines denote perfect agreement. The text-only baseline does not produce format-compliant CIFs for targets above 40 atoms, which explains their absence from the right-hand panels. \textbf{Bottom:} Recovery metrics are reported for models with and without XRD conditioning under ``20-consistent'' and ``1-perplexity'' generation strategies, stratified into seen, structurally novel and compositionally novel test subsets. Reported quantities are match rate, mean RMSD, volume mean absolute error (MAE), average matched atoms, maximum matched atoms, sequence perplexity and total generation time per target under a 200-attempt budget on the 1.5K-material test set. Values are reported as mean $\pm$ standard error over three independent runs. XRD conditioning improves recovery accuracy and extends recovery to larger crystal systems.
    }
    \label{fig:chili_matches}
\end{figure*}

Results from three independent test-set recovery experiments (Figure~\ref{fig:chili_matches}) support the hypothesis that diffraction pattern conditioning provides improved structure generation. When conditioned on XRD profiles and evaluated over ``20-consistent'' generations, the model achieved a mean structure match rate of $49.04\%$, compared to $15.89\%$ for the unconditioned model. Within this benchmark, the stratified results are consistent with strong generalisation across seen, structurally novel and compositionally novel subsets. The structurally novel subset had the highest recovery rate; this improvement over fully seen materials is likely explained by the lower geometric complexity of the unseen polymorphs. Furthermore, for entirely novel compositions, the conditioned model maintained a $41.40\%$ match rate, while the unconditioned baseline degraded to $8.87\%$. Figure~\ref{fig:chili_matches} shows a tighter correlation between predicted and target values across all test structures for the conditioned models. The unconditional model deteriorated on larger systems, exhibiting a strict performance ceiling that prevented the recovery of any structures exceeding $40$ atoms per unit cell. In contrast, the conditioned model successfully recovered structures up to $83 \pm 10$ (SE) atoms, raising the average matched structure from $11.03$ atoms in the unconditioned setting to $19.13$ atoms per unit cell.

Because diffraction patterns explicitly encode lattice dimensions and crystallographic symmetry, this conditioning provides global structural priors. These priors guide the autoregressive generation process, mitigating the error accumulation typically observed in long sequence generation. Generating ``20-consistent'' structures across the test set was 2.75-times quicker in inference time for conditioned, compared to unconditioned models (Figure~\ref{fig:chili_matches}). This efficiency gain is supported by reduced model uncertainty during autoregression: the unconditioned baseline exhibited an average sequence perplexity of $1.312$ compared to $1.247$ with XRD signal, showing the conditioning's capability to steer generation by restricting the probability distribution at each generation step.

% Note here we also skip validity, because during dataset preprocessing, noticed structures from the dataset failed validity checks due to unreasonable bond lengths. We preserve these noting them as artifacts of experimentally observed structures which may contain some minor positional disorder

\subsubsection{Experimental Structure Recovery for \ce{TiO2} polymorphs}

Resolving structurally distinct polymorphs sharing identical reduced formulas presents a critical challenge for generative crystallography. To evaluate the capacity of the model to synthesise specific geometric configurations from diffraction data, structural recovery was tested on three experimental \ce{TiO2} polymorphs: Rutile, Anatase and Brookite. The Chili-100K training dataset contained no occurrences of these phases, with Rutile appearing only in the validation split. The unlabeled pre-training dataset contained only Anatase and Rutile, meaning Brookite was never seen at any training phase. Furthermore, the unit cell stoichiometries for Anatase (\ce{Ti4O8}) and Brookite (\ce{Ti8O16}) are not unique to these specific phases within the broader pre-training distribution. Accurate recovery, therefore, requires the model to actively interpret the XRD conditioning to differentiate the target topology from competing structures with identical atom counts.

The Chili-100K model successfully recovered all three polymorphs using only the input composition and the experimental XRD profile (Table~\ref{tab:polymorph_chili}). While match rates decline for Anatase and Brookite, producing Valid structural targets within a 20-candidate generation budget represents a tractable screening volume for experimental validation. The recovery of the entirely unseen Brookite phase supports the interpretation that the model can generate structures guided by the diffraction signal rather than relying only on memorised chemical distributions.

To test whether the model learned a direct physical mapping rather than a compositional lookup table, inference was performed targeting $2\times$ expanded supercells. The model was prompted with doubled stoichiometry and the corresponding complex XRD pattern. Successful recovery of Rutile and Anatase under doubled stoichiometry suggests that the model is responding to the composition-XRD constraint pair rather than performing a simple compositional lookup. The model was unable to match the Brookite phase, although it generated Valid outputs, likely due to the increased complexity of a lower symmetry and high atom count system.

\begin{table}[htbp!]
    \centering
    \small
    \setlength{\tabcolsep}{4pt}
    \begin{tabular}{lccc}
        \toprule
        \textbf{\ce{TiO2} Polymorph} & \textbf{Rutile} & \textbf{Anatase} & \textbf{Brookite} \\
        \textit{Atoms (\& $\times2$ SC)} & \textit{6 (12)} & \textit{12 (24)} & \textit{24 (48)} \\
        
        \midrule
        Prompts & $N$ Matches & $N$ Matches & $N$ Matches  \\
          & (RMSD) & (RMSD) & (RMSD) \\
        
        \midrule
        Comp. & 17 & 2 & 7 \\
          & ($7.70 \times 10^{-4}$) & ($3.62 \times 10^{-1}$) & ($3.96 \times 10^{-1}$) \\
        \addlinespace
        Supercell Comp. & 2 & 4 & 0 \\
          & ($2.00 \times 10^{-1}$) & ($3.39 \times 10^{-1}$) & \\
    \bottomrule
    \end{tabular}
    \caption{Recovery metrics for experimental \ce{TiO2} polymorphs using the CHILI-fine-tuned model. Rutile, Anatase and Brookite are evaluated under conventional-cell composition prompts and doubled supercell composition prompts ($\times 2$). Each entry reports the number of matched structures obtained under the ``20-consistent'' protocol, with the lowest RMSD among the matched structures given in parentheses.}
    \label{tab:polymorph_chili}
\end{table}

\section{Discussion and limitations}
\label{sec:discussion}

The central result of this study is that continuous control in autoregressive crystal generation depends strongly on where the conditioning signal enters the network. Generative materials design requires models that can map continuous physical targets onto discrete structural outputs without eroding the structural prior acquired during unsupervised pre-training. In standard autoregressive transformers, this interface is difficult because digit-level property tokenisation weakens ordinal structure and sequence-level conditioning perturbs the token representations that govern CIF generation. CrystaLLM-\scalebox{1.3}{$\pi$} addresses this limitation by injecting continuous property information directly into multi-head attention through Prefix and Residual key-value conditioning, rather than encoding targets as additional sequence tokens. Localising adaptation to the attention pathway preserves the original CIF token stream, reduces disruption to the pre-trained backbone, and maintains high structural validity during conditional fine-tuning. The contribution is therefore a conditional crystal-generation framework together with evidence that attention-level conditioning can preserve pre-trained structural priors under scarce labelled supervision. The work also addresses an evaluative gap in the field: conditional generation still lacks a comprehensive benchmark suite spanning transfer learning, data regimes and downstream task types. The controlled comparisons assembled here are intended as a step toward that more systematic standard.

A direct comparison with MatterGen (a state-of-the-art diffusion-based generative model) clarifies that the two generative frameworks favour different objectives. In the evaluated band gap setting, the MatterGen baseline produced a larger absolute number of Valid and $Q_{\mathrm{VSUN}}$ structures, consistent with the stronger geometric inductive bias of a graph-based diffusion model that operates directly on atomic environments and interatomic relationships. By contrast, CrystaLLM-\scalebox{1.3}{$\pi$} maintained tighter calibration to requested targets while requiring less memory and shorter training and inference times. These gains follow from a single-stage autoregressive pipeline in which conditioning is introduced through standard attention operations on a one-dimensional CIF representation, avoiding iterative denoising and graph construction at sampling time. The cost of this efficiency is a weaker built-in geometric prior, because the text model must infer spatial regularities from sequence statistics rather than operating directly on coordinates and neighbourhoods. Even so, the discrete CIF representation preserved crystallographic symmetry more reliably in the unrelaxed outputs examined here and remained practical for larger, chemically richer unit cells. The comparison, therefore, supports complementarity rather than a universal ranking: graph diffusion appears better suited to maximising $Q_{\mathrm{VSUN}}$ structure yield for chemistries of relatively limited complexity, whereas attention-level autoregression offers a lighter route to scalable conditional generation when target precision, symmetry retention and computational cost are the primary constraints.

The photovoltaic case study extends the conditioning framework from benchmark calibration to functional discovery in a scarce-label regime. SLME is a demanding target because it compresses absorption strength, band gap and recombination losses into a single scalar objective, so successful steering requires the model to recover indirect structure-property relationships from only 5.35K labelled materials. The results are consistent with a transfer-learning mechanism in which large-scale pre-training supplies a broad structural prior and the conditional objective redirects sampling toward motifs associated with favourable optoelectronic response. The novelty profile supports this interpretation, because many candidates are novel with respect to the fine-tuning set, whereas fewer remain compositionally novel with respect to the pre-training corpus. The DFT follow-up then sets the practical boundary of the claim. Several candidates retained promising photovoltaic characteristics after first-principles validation, whereas others exposed errors in the surrogate ranking when subtle electronic-structure features degraded absorption despite favourable surrogate evaluation. The main implication is that CrystaLLM-\scalebox{1.3}{$\pi$} can enrich the search space for plausible high-performance materials under limited labelled supervision, but functional assessment still requires high-fidelity electronic-structure validation.

The XRD experiments show that the same conditioning mechanism can absorb heterogeneous experimental constraints and convert them into useful generative bias. Unlike scalar property targets, diffraction profiles encode distributed information about lattice dimensions, symmetry and periodic ordering, so successful recovery requires the model to align a high-dimensional continuous signal with a discrete CIF sequence generation task. In this setting, conditioning improved structure recovery across theoretical and experimental benchmarks and allowed the model to recover larger and more complex materials compared to an unconditioned baseline. 

The generative range of CrystaLLM-\scalebox{1.3}{$\pi$} is limited by two factors: the coverage of the training data and the maximum sequence length of the transformer backbone. Attention-level conditioning can steer generation toward rare target regions, but it does not allow the model to move reliably far beyond the chemical space represented in the data it has learned. This limitation is clearest at extreme target values, where physically plausible structures are sparse and extrapolation remains unreliable. In practice, exploring a new region of property space still requires nearby structure-property examples in the training data. When pre-training and fine-tuning do not provide enough local support around a target region, conditional generation in that part of the space becomes less reliable. We demonstrate that this limitation also applies to graph diffusion-based conditional generation. The second constraint arises from the explicit CIF representation. Because sequence length scales with the asymmetric unit rather than the total atom count, large high-symmetry structures can remain tractable, whereas low-symmetry crystals rapidly exhaust the context window. The present framework, therefore, retains a representational bias toward compact crystallographic descriptions and against materials whose disorder or low symmetry requires long token sequences.

These limits define clear next steps. Longer-context backbones or more efficient positional schemes could expand coverage of low-symmetry and compositionally complex structures without losing the practical advantages of autoregressive decoding. Extending the training dataset to contain CIFs with fractional occupancies and substitutional disorder would also bring the model closer to experimental materials, where ideal ordered cells are often an approximation rather than the true state. Another promising direction is to combine autoregressive crystal generation with value-aware numerical representations such as xVal~\cite{golkar2024xvalcontinuousnumericaltokenization}, which could improve interpolation and extrapolation for coordinates, lattice parameters and conditioned properties. We do not evaluate such methods here because they would require changing the pre-training objective from pure next-token prediction to a mixed value-and-token learning setup, making them a distinct potential modelling direction.
Finally, mechanistic analysis of the conditioned attention layers could reveal how continuous targets alter token probabilities across composition, symmetry and coordinate fields. That level of interpretability would strengthen the scientific value of the framework by turning conditional control from an empirical capability into a tool for studying learned structure-property relations.

\section{Conclusion}
\label{sec:conclusion}

We have presented CrystaLLM-\scalebox{1.3}{$\pi$}, a conditional autoregressive model for crystal structure generation. We have demonstrated that fine-tuning a pre-trained crystal structure generator yields a model capable of generating new materials with desired characteristics. We have further shown the utility of this capacity in two important materials science settings: the discovery of new materials with target functional properties and the recovery of crystal structures from experimental characterisation data. In both settings, we have shown that CrystaLLM-\scalebox{1.3}{$\pi$} offers a lightweight, flexible and extensible framework that achieves state-of-the-art, or near state-of-the-art, performance. We have provided an easy-to-install codebase, together with a pre-built containerised API and a web-app interface, making CrystaLLM-\scalebox{1.3}{$\pi$} an accessible and powerful tool for accelerated materials development. Relevant links to these tools are provided in Section~\ref{sec:code_availability}.

\section{Methods}
\label{methods}

\subsection{Datasets}
\label{sec:datasets_curation}
Datasets were curated to study the model under different pre-training and fine-tuning conditions, as shown in Table~\ref{tab:datasets_summary}. All datasets were stored in \texttt{dataframes}, with each row containing a material's database of origin, its reduced formula, its augmented CIF string and optional numerical properties of interest along with their normalised counterparts.  All the CIF data underwent a standardised preprocessing pipeline including deduplication, data augmentation, tokenisation and specialised batching to incorporate conditioning information, with full details provided in Appendix~\ref{app:pre_processing}. For specific use cases, we applied different pre-processing; full details for exact reproducibility can be found in Appendix~\ref{app:dataset_construction}.

In the XRD task, methodological standardisation was required to ensure a fair comparison despite the variation within existing benchmarking protocols. As the original data partitions for the Jarvis-DFT benchmark were unavailable, a seeded, randomised 90:10 train-test split was established to facilitate reproducible evaluation. Furthermore, standard validity screens employed in benchmarks such as MP-20 were observed to arbitrarily discard $8.93\%$ of structure-matching test structures, a limitation discussed in the PXRDGen paper~\cite{liPowderDiffractionCrystal2025}. To isolate the intrinsic structural fidelity of the generative mechanisms from these filters, performance metrics on the MP-20 dataset are reported both with and without the filters.

\begin{table}[htbp!]
    \centering
    \small
    \setlength{\tabcolsep}{5pt}
    
    \renewcommand{\arraystretch}{1.0}
    \begin{tabular}{@{}lccl@{}}
        \toprule
        \textbf{Dataset} & \textbf{Nb Materials} & \textbf{Conditions} \\
        
        \midrule
        MP~Bandgap & 53.3K &  Band gap, $E_{\text{hull}}^{\text{PBE}}$ \\
        MatterGen Density  & 653K  & Density, $E_{\text{hull}}^{\text{PBE}}$ \\
        MP~SLME & 5.35K & SLME \\
        MP-20 XRD & 45.2K & Top-20 XRD \\
        Jarvis-DFT XRD & 76.0K &  Top-20 XRD \\
        MatterGen XRD  & 658K  & Top-20 XRD \\
        Chili-100K XRD & 14.1K & Top-20 XRD \\
        \bottomrule
    \end{tabular}
    
    \caption{Summary of property-conditioned datasets used in this work. For X-Ray Diffraction (XRD) conditioning, the condition vector stores the 20 most intense peaks and their associated $2\theta$ positions. All datasets were filtered to remove CIFs containing more than 1024 tokens, except Chili-100K, which uses a context length of 1536. All processed datasets are openly available at \href{https://huggingface.co/c-bone/datasets}{\texttt{Hugging Face}}~\cite{HuggingFacedatasets}.}
    \label{tab:datasets_summary}
\end{table}

\subsection{Transformer backbone}
\label{sec:backbone_theory}
In the pre-training phase, the model learns to generate CIFs sequentially. The model backbone is based on CrystaLLM's small model hyperparameter choices and was pre-trained on an unlabeled augmented CIF dataset~\cite{antunesCrystalStructureGeneration2024}. It consists of $\sim26$ million parameters, with an embedding size of 512, 8 decoder block layers and 8 attention heads.
For functional property conditioning, the pre-trained model was adapted for fine-tuning, where CIFs were labelled with associated properties, such as band gap or density. Architectural modifications, detailed in Section~\ref{sec:conditioning-architectures-results}, were implemented to guide CIF sequence generation using these properties. The objective of this fine-tuning stage was to enable the model to learn a mapping from a vector of material properties to chemically valid crystal structures.
Default model training protocols and hyperparameter choices can be found in Appendix~\ref{app:training_hyperparams}, and the hyperparameter search space is defined for each experiment in Appendix~\ref{param_search}. The final model configuration for every model used in this work can be found in the repository \href{https://github.com/C-Bone-UCL/CrystaLLM-pi/tree/main/_config_files}{\texttt{configuration files}}~\cite{configfiles}.

The model backbone was trained by minimising a developed variation of the cross-entropy loss, where the commonly used base loss is defined as:
\begin{equation}
\mathcal L_{CE} = \frac{1}{N}\sum_{n=1}^N\left(-\sum_{k=1}^K y_{k,n} \log(p_{k,n})\right),
\end{equation}
where $N$ is the total number of tokens in the batch for which the loss is calculated, $K$ is the size of the token vocabulary, $y_{k,n}$ is the true probability for token $k$ at position $n$ in the sequence, set to 1 for the correct target token and 0 otherwise, $p_{k,n}$ is the model's predicted probability for token $k$ at position $n$ and $\log$ is the natural logarithm.

% In practice, the predicted probabilities p_k are obtained by applying the Softmax function to the model's logits for stability apparently

The modification incorporates an additional penalty targeting \textit{fixed} tokens. These tokens are defined as the syntactically invariable elements within the CIF (e.g., ``data\_'', ``\_cell\_volume'', see Appendix~\ref{app:augmented_cif}). Conversely, \textit{variable} tokens are the parts of the CIF that can change from one structure to another (e.g.``Ge4''), which are encapsulated in brackets during CIF augmentation. The penalty on \textit{fixed} tokens aims to accelerate the learning of CIF syntax during initial training phases, diminishing gradually beyond the initial training phases as the model's syntactic understanding improves (Appendix~\ref{app:loss_landscapes}). The final loss objective is defined as:
\begin{equation}
\mathcal L = n_1\mathcal L_{\mathrm{CE}} + n_2\mathcal L_{\mathrm{Fixed}},
\end{equation}
where $\mathcal L_{\mathrm{CE}}$ is the regular cross-entropy loss and $\mathcal L_{\mathrm{Fixed}}$ is the additional cross-entropy loss applied only to $Fixed$ tokens. $n_1$ and $n_2$ are tunable hyperparameters to adjust how much each component contributes to the overall loss $\mathcal{L}$. For the large-scale pre-training, we set $\{n_1, n_2\}$ to $\{1,2\}$; for the rest of the studies, we set them to $\{1,1\}$.

\subsection{Conditioning Method Implementation}
\label{sec:conditioning-methods-methods}

To facilitate property-conditioned generation, we set up a dual optimisation strategy during fine-tuning, adapting the AdamW optimiser inspired by Howard and Rude's recommendations~\cite{loshchilovDecoupledWeightDecay2019, howardUniversalLanguageModel2018}. The parameter set of the model, $\theta$, was split into two subsets: $\theta_{backbone}$, which contains the parameters of the pre-trained backbone and $\theta_{cond}$, which contains the parameters of the newly initialised conditioning layers. This split allows for differential learning rates, with the goal of maximising the information transfer between pre-training and fine-tuning steps while enabling stronger learning of the conditioning task.

The full hyperparameter settings, including hidden dimensions, dropout, the differential learning rates applied to $\theta_{backbone}$ and $\theta_{cond}$ for fine-tuning, are provided in Appendix~\ref{app:training_hyperparams}. Experiment-specific hyperparameters are specified in the repository \href{https://github.com/C-Bone-UCL/CrystaLLM-pi/tree/main/_config_files}{\texttt{configuration files}}~\cite{configfiles}.

\subsection{CIF generation and top perplexity scoring}
\label{sec:sampling}
During model inference, sampling is used to generate tokens sequentially. This work applies sampling temperature \texttt{T}, as well as \texttt{Top-K} and nucleus \texttt{Top-p} sampling schemes to allow dynamic selection of tokens~\cite{fanHierarchicalNeuralStory2018, holtzmanCuriousCaseNeural2019}. Tuning them allows sampling from highly likely tokens (reflecting frequent training patterns) to less likely tokens, which introduces exploratory randomness during generation. The exact parameter values differ from experiment to experiment (See Appendix~\ref{app:experimental_protocols}).

For crystal structure generation, text prompts and condition pairs are established, with specified levels of verbosity in the prompts. With the Residual method, for example, we can include any combination of the following: functional property, no cell composition, cell composition, or space group without needing to retrain the model or repurpose the framework. For each prompt-condition pair, the model generates candidate CIF structures using the specified sampling parameters.

A distinct generative workflow is employed for the structure recovery tasks (Section~\ref{sec:recovery}) to maximise prediction fidelity. Outputs undergo three algorithmic consistency checks. Formula consistency verifies a coherent stoichiometry throughout the output CIF. Space-group consistency compares the stated space-group symbol against the symmetry detected by pymatgen with a precision tolerance of~\SI{0.1}{\angstrom}~\cite{ongPythonMaterialsGenomics2013}. Sensibility tests reject structures with cell lengths outside~\SIrange{0.5}{1000}{\angstrom} or cell angles outside~\SIrange{10}{170}{\degree}. To prioritise the most physically probable reconstruction, a perplexity-based scoring mechanism evaluates the generated candidates. Perplexity is computed from the token transition scores, excluding the input prompt. Lower values indicate higher model confidence based on learned structural priors. The framework generates candidates until achieving 20 self-consistent CIFs. The ``1-perplexity'' scheme then selects the single structure with the lowest perplexity for final matching. This approach balances exploration through sampling and composition prompts with condition priors, together with quality control through validation and probabilistic ranking, increasing the model's capacity to generate sensible structures for a given input (See Appendix~\ref{app:ablation_studies} for ablations).

In contrast, the discovery of novel materials requires broad exploration of the chemical space. So, perplexity ranking is not used during discovery tasks (Section~\ref{sec:slme}). Candidates are generated using pure sampling strategies with discovery-oriented parameters (Appendix~\ref{app:experimental_protocols}), and only VSUN physical filters are used to select promising candidates. This ensures the framework applies high-confidence constraints for structure solving while favouring stronger deviation from the training manifold during inverse design.

\subsection{Evaluation Metrics}
\label{sec:eval_metrics}
To evaluate the developed methods, the following performance metrics are used: CIF validity criteria for chemical consistency and structural reconstruction; structure similarity measures for novelty assessment against training data; uniqueness within generated sets; and statistics for comparative effectiveness of conditioning methods. DFT calculations then provide additional validation of thermodynamic stability and electronic properties for selected candidates.

\subsubsection{Validity Criteria}
\label{sec:validity_metric}
The validity metric follows the criteria of the original CrystaLLM paper~\cite {antunesCrystalStructureGeneration2024}. It must be possible for a structure to be reconstructed from the generated CIF, the space group must be consistent with the atomic structure, the generated bond lengths must be reasonable with respect to the atomic radii, and finally, the declared atom site multiplicity needs to be consistent with the unit cell composition.

\subsubsection{Structure Diversity}
\label{sec:diversity_metrics}
For the uniqueness metric, sets of generated structures with identical input constraints are compared. We use the $LeMaterials$ BAWLHasher (Bonding Algorithm Weisfeiler Lehman), which is a materials fingerprint method based on bonding graph structure, reduced formula and symmetry~\cite{lematerial2024}. Canonical hashes are obtained for each structure in a uniqueness pass, and structures are counted as unique if they have a unique hash. This allows the calculation to scale $\mathcal{O}(n)$, rather than $\mathcal{O}(n^{2})$ when using a structure matcher algorithm for all generated materials.

To determine novelty, we compare unique generated structures against all training-set structures with the same reduced formula. If a structure's reduced formula does not appear in the reference dataset, it is considered \textit{Compositionally} novel. If a generated structure shares a reduced formula with a reference structure but the crystal structures do not match, it is considered \textit{Structurally} novel. This matcher does not establish novelty in a broader materials-discovery sense, but serves as a practical screening tool.

Pymatgen's \texttt{StructureMatcher} is used to match two chemical structures together. Unless otherwise specified, the default matcher settings are used, namely a fractional length tolerance of $0.2$, a site tolerance of~\SI{0.3}{\angstrom} and an angle tolerance of~\SI{5}{\degree}. For normalisation purposes, structures are reduced to their primitive cell and scaled to equivalent volume before matching.

\subsubsection{Prediction models and DFT validation}
\label{sec:surrogate_models}

High-throughput property estimation for generated structures was achieved using surrogate models based on the ALIGNN architecture~\cite{choudharyAtomisticLineGraph2021}. In Section~\ref{sec:pre-training}, PBE band gaps~\cite{perdewGeneralizedGradientApproximation1996} were predicted using the standard ``mp\_gappbe\_alignn'' model trained on Materials Project data. For the specific requirements of Section~\ref{sec:slme}, we trained distinct surrogate models to predict higher-fidelity properties. HSE06 band gaps~\cite{heydHybridFunctionalsBased2003} were predicted using an \texttt{ALIGNN} model trained on $\sim 10\text{k}$ labelled structures from Kim et al.~\cite{kimBandgapDatabaseSemiconducting2020}, with hyperparameters according to Choudhary et al.~\cite{choudharyAtomisticLineGraph2021}. SLMEs were predicted using a model trained on the $\sim 5.3\text{k}$ entry MP~SLME dataset~\cite{walkerCarbonCostMaterials2025}.

The thermodynamic stability of generated materials was first evaluated using the MACE-MP calculator to determine the energy above the convex hull, denoted here as $E_{\mathrm{hull}}^{\text{MACE}}$. MACE-MP is a graph-based foundation model employing higher-order equivariant message passing, pre-trained on MP relaxation trajectories~\cite{batatiaFoundationModelAtomistic2024}. Predicted energies were processed using the \texttt{MaterialsProject2020Compatibility} scheme to ensure consistency between GGA and GGA+U calculations relative to the MP dataset.

Given the model's reported mean absolute error (MAE) of $0.057~\si{\electronvolt/atom}$ on the Matbench Discovery benchmark~\cite{riebesellFrameworkEvaluateMachine2025}, stability thresholds were adjusted to account for statistical uncertainty. For the general validation in Section~\ref{sec:pre-training}, a threshold of $E_{\mathrm{hull}}^{\text{MACE}} \leq 0.157~\si{\electronvolt/atom}$ was employed, expanding upon the typical $0.1~\si{\electronvolt/atom}$ metastability limit~\cite{jainCommentaryMaterialsProject2013}. Conversely, for the SLME Study in Section~\ref{sec:slme}, a stricter threshold of $E_{\mathrm{hull}}^{\text{MACE}} \leq 0.1~\si{\electronvolt/atom}$ was applied to aggressively funnel the discovery pipeline. The final VSUN metric integrates this stability assessment with checks for Validity, Uniqueness and Novelty~\cite{zeniGenerativeModelInorganic2025}.

To validate the stability of promising candidates identified in the SLME study, structure optimisation and energy calculations were performed at the DFT level to obtain $E_{\mathrm{hull}}^{\text{PBE}}$. We used the Vienna \textit{ab initio} Simulation Package (VASP)~\cite{kresse_ab_1993, kresse_efficiency_1996}, employing PBE exchange-correlation functionals and Projected Augmented Wave (PAW) potentials~\cite{kresse_norm-conserving_1994, PhysRevB.59.1758} matching MP v54 selections. The relaxation protocol involved sequential optimisation of ion positions and lattice parameters, followed by a final refinement step to ensure convergence within standard MP thresholds. Total energies were derived from a final static calculation with a dense \textit{k}-point mesh, ensuring energetic consistency with reference compounds for convex hull analysis. All relaxations were performed without symmetry constraints.

Detailed optical properties were subsequently computed for verified structures using a hybrid functional workflow. Following a PBE relaxation, frequency-dependent dielectric functions were calculated using the HSE06 functional~\cite{heydHybridFunctionalsBased2003} within the Independent Phase Approximation (IPA). Input files were generated using \texttt{Atomate2}~\cite{ganose_atomate2_2025}, but for the final dielectric calculations, \textit{k}-point meshes were made twice as dense as those generated by \texttt{Atomate2} in each reciprocal space direction to ensure convergence of optical properties. This produced a mesh with $\sim2000/n$ \textit{k}-points for a unit cell with $n$ atoms. Dielectric functions were converted to absorption spectra using Kramers-Kronig transformations and used to determine the SLME following the method of Yu and Zunger~\cite{yuIdentificationPotentialPhotovoltaic2012} for an absorber at 300 K and with thickness $d$ = 500 nm as is widely used~\cite{yuIdentificationPotentialPhotovoltaic2012,Fabini2019CandidateInorganicPhotovoltaics}.

Supply chain sustainability was evaluated using the Herfindahl-Hirschman Index (HHI) for elemental production ($HHI_P$) and reserves ($HHI_R$), based on data from Gaultois et al.~\cite{gaultoisDataDrivenReviewThermoelectric2013}. Material-level scores were computed as the mass-fraction weighted average of the constituent elements' HHI values extracted with \texttt{SMACT}~\cite{Davies2019}. To prioritise candidates with diverse and secure global supply chains, we employed a composite sustainability metric defined by the Euclidean distance from the origin in the HHI space: 
\begin{equation}
    HHI_{dist} = \sqrt{HHI_P^2 + HHI_R^2}
\end{equation}
A lower $HHI_{dist}$ corresponds to reduced geopolitical influence and higher supply security, serving as a selection criterion for the final candidate pool.

\subsubsection{Performance Comparison for Conditional Models}
\label{sec:performance_metrics}
When a model is evaluated on its capability to generate materials with targeted properties, different evaluation metrics are used for comparative analysis. Common statistical metrics, which include the Mean Average Error ($MAE$), standard error ($SE$), Pearson's r ($r$), correlation coefficient ($R^2$) and the dependent t-test for paired samples ($t, p$), are not reiterated here. The Hit-Rate is defined in this work as the number of materials that have a predicted property that falls within a specified range of the target value, and it is determined as:
\begin{equation}
  \mathrm{Hit-Rate}_\varepsilon=\frac{1}{N}\sum_{i=1}^{N}\left(\lvert \hat{y}_i - y_i\rvert \le \varepsilon\right),
\end{equation}
where $\varepsilon = 0.5~\si{\electronvolt}$ was used for Section~\ref{sec:pre-training} and $\varepsilon = 1.0~\si{g/cm^3}$ for Section~\ref{sec:dataset_size_study}.

\subsection{Data Availability}
Processed datasets and trained models used in this work are stored on \href{https://huggingface.co/c-bone}{\texttt{Hugging Face}}~\cite{HuggingFaceprofile}. Data preparation, training and metric calculation workflows are all detailed in the \href{https://github.com/C-Bone-UCL/CrystaLLM-pi/tree/main/notebooks}{\texttt{repository notebooks}}~\cite{repositorynotebooks}.

\subsection{Code Availability}
\label{sec:code_availability}
The code for CrystaLLM-\scalebox{1.3}{$\pi$} is available at \href{https://github.com/C-Bone-UCL/CrystaLLM-pi}{\texttt{the model repository}}~\cite{modelrepository}, which includes a detailed \texttt{README.md} on how to load and generate structures with the models mentioned in this work, as well as how to adapt the framework for other use cases. Importantly, the Prefix attention method is called ``PKV'' in the repository, and the Residual attention method is defined as ``Slider'' in the repository.  Exact training and generation hyperparameters for every experiment carried out in this work can be found in the repository \href{https://github.com/C-Bone-UCL/CrystaLLM-pi/tree/main/_config_files}{\texttt{configuration files}}~\cite{configfiles}. The pre-built \href{https://resources.psdi.ac.uk/resource-theme/37242589-7d89-462a-afad-29386bcfcda4}{\texttt{containerized API}}~\cite{containerizedAPI} and \href{https://crystallm-pi.psdi.ac.uk/}{\texttt{web-app interface}} are hosted thanks to \href{https://www.psdi.ac.uk/}{PSDI}~\cite{psdi} (Physical Sciences Data Infrastructure) resources and tools.

\subsection{Acknowledgements}
KTB acknowledges funding from EPSRC (EP/Y014405/1, EP/Y000552/1, EP/Y028775/1 and EP/Y028759/1). We acknowledge the use of University College London (UCL) Myriad High Performance Computing Facility (Myriad@UCL) and associated support services in the completion of this work. We acknowledge Young HPC access, which is partially funded by EPSRC (EP/T022213/1, EP/W032260/1 and EP/P020194/1). CB and MW are funded by a UCL start-up package. Part of this publication is based upon work from EUMINE COST Action CA21143, for the STMS supported by COST (European Cooperation in Science and Technology). KL is supported by the Bezos Earth Fund via the Earth Rover Program (G-2023-201305339). BAAM was funded by the ``AI for Science'' PSDI sub-project (“Provision of ‘AI ready’ data: prototyping data pipelines and repositories”, grant award UKRI2697). 

\subsection{Author Contributions}
CB and KTB conceptualised the project and wrote the initial manuscript, the package was developed by CB as a development from the CrystaLLM code written by LA, the Residual model was based on code provided by KL. Experimental XRD data was provided by AA and JD. DFT calculations and FGW distance evaluations were performed by MW. BM helped develop the API tools and the front-end for the web-app. RG-C, LA, BAAM and KL provided feedback and guidance on project design and were also involved in editing the final manuscript.

\subsection{Competing Interests}
The authors declare no competing interests.

\clearpage
\onecolumngrid

\section{Appendices}

% \twocolumngrid
\subsection{Property-labelled Dataset Construction Details}
\label{app:dataset_construction}

The LeMaterial dataset does not contain any functional properties; it is exclusively made up of unlabeled CIFs derived from the source material~\cite{lematerial2024}, randomly split into a train:validation ratio of 90:10. The dataset was accessed in April 2025.

For property-labelled datasets, we first consider the Alex-MP-20 data~\cite{zeniGenerativeModelInorganic2025}, accessed April 2025. Structures have at most 20 atoms per unit cell and an $E_{\text{hull}}^{\text{PBE}}$ of at most~\SI{0.1}{\electronvolt/atom}. All structures are relaxed with DFT using a PBE functional with corrections based on the Materials Project~\cite{jainCommentaryMaterialsProject2013} for energy consistency. All MatterGen datasets were randomly split into train:validation sets with a 90:10 ratio. Without any property labels, this is the \textbf{Alex-MP-20 dataset}.
The \textbf{MatterGen Density Subset} considered for the Dataset Size study was curated by calculating density using pymatgen's \texttt{Structure.density} in (\si{g/cm^3}). $E_{\text{hull}}^{\text{PBE}}$ values were taken directly from the base dataset. Density and $E_{\text{hull}}^{\text{PBE}}$ were normalised using Min-Max method.
To generate XRD spectra for the conditioning vector in the \textbf{MatterGen XRD} dataset, XRD patterns were calculated with pymatgen's XRDCalculator at the CuK$\alpha$ wavelength. The 20 most intense Bragg peaks and their 2$\theta$ positions were stored. The intensities and 2$\theta$ values were normalised with Min-Max normalisation where 0:90 were used for 2$\theta$ and 0:100 were used for min-max ranges. Patterns with under 20 Bragg peaks were padded with $[-100]$ tokens, indicating a heterogeneous ``missing value''.

The Materials Project API was queried to create the \textbf{MP~Bandgap} dataset~\cite{jainCommentaryMaterialsProject2013}, accessed in April 2025. All structures are relaxed with DFT using a PBE functional with the Materials Project corrections. Band gaps and $E_{\text{hull}}^{\text{PBE}}$ were retrieved directly from the API. Materials with null band gaps were removed. The database was randomly split into train and validation sets in a 90:10 ratio.

The \textbf{MP~SLME} dataset was curated from Walker and Butler's dataset, accessed in September 2025~\cite{walkerCarbonCostMaterials2025}, created by taking overlapping materials from the Woods-Robinson et al. and Kim et al. datasets~\cite{woods-robinsonDesigningTransparentConductors2023, kimBandgapDatabaseSemiconducting2020}.
The Spectroscopic Limited Maximum Efficiency (SLME) values were derived from high-fidelity absorption spectra, and band gaps were calculated with the Heyd-Scuseria-Ernzerhof (HSE) hybrid DFT functional.
CIFs were then obtained from Materials Project IDs associated with each property value. For hyperparameter search, the dataset was split into a 95:5 train:validation ratio, but for the final model, the full dataset was put into the training set.

The \textbf{MP-20} dataset~\cite{jainCommentaryMaterialsProject2013} was used to curate the \textbf{MP-20 XRD} dataset. The dataset was accessed in April 2025. The structures are a subset of the materials project material with at most 20 atoms per unit cell. The XRD conditioning property vectors were generated with the same methodology as the \textbf{MatterGen XRD Subset}. Both dataset splits with train:validation:test sets of 60:20:20 were taken directly from PXRDGen~\cite{liPowderDiffractionCrystal2025}.
 
The Jarvis-DFT dataset~\cite{choudharyJointAutomatedRepository2020} was used to curate the \textbf{Jarvis-DFT XRD} fine-tuning subset, accessed in January 2025. All structures are relaxed with DFT using a PBE functional. The XRD conditioning property vectors were generated with the same methodology as the \textbf{MatterGen XRD Subset}. The database was randomly split into train:validation:test sets of 80:10:10 for hyperparameter search and then 90:10 train:test for the final model to match the benchmark.

The \textbf{Chili-100K XRD} dataset contains experimentally determined crystal structures sourced from Chili-100K~\cite{Jensen2024chili}, a curated and filtered inorganic subset of the COD~\cite{grazulisCrystallographyOpenDatabase2009}. It was accessed in April 2026 and contains around 21K materials, funnelled down to 14K after our deduplication pipeline. The XRD conditioning vectors used in this benchmark were simulated from the deposited structures using the same methodology as the \textbf{MatterGen XRD Subset}. The dataset was split into train:val:test ratios of 78.6:10.7:10.7. The test set in particular was crafted to study the model's performance in a data leakage aware manner: 500 materials present in the test set were seen during a training phase (LeMaterial, MatterGen XRD or Chili-100K train:val subsets), 500 were materials whose reduced formula was seen during a training phase but the structure was never seen (Calculated using the Structure Novelty metric from Section~\ref{sec:diversity_metrics}) and 500 materials had reduced formulas and structures never seen in any phase of training.

\clearpage
\subsection{Token and Atom Distributions}
\label{app:token_distributions}

\begin{figure*}[htbp!]
    \centering
    \includegraphics[width=0.44\textwidth]{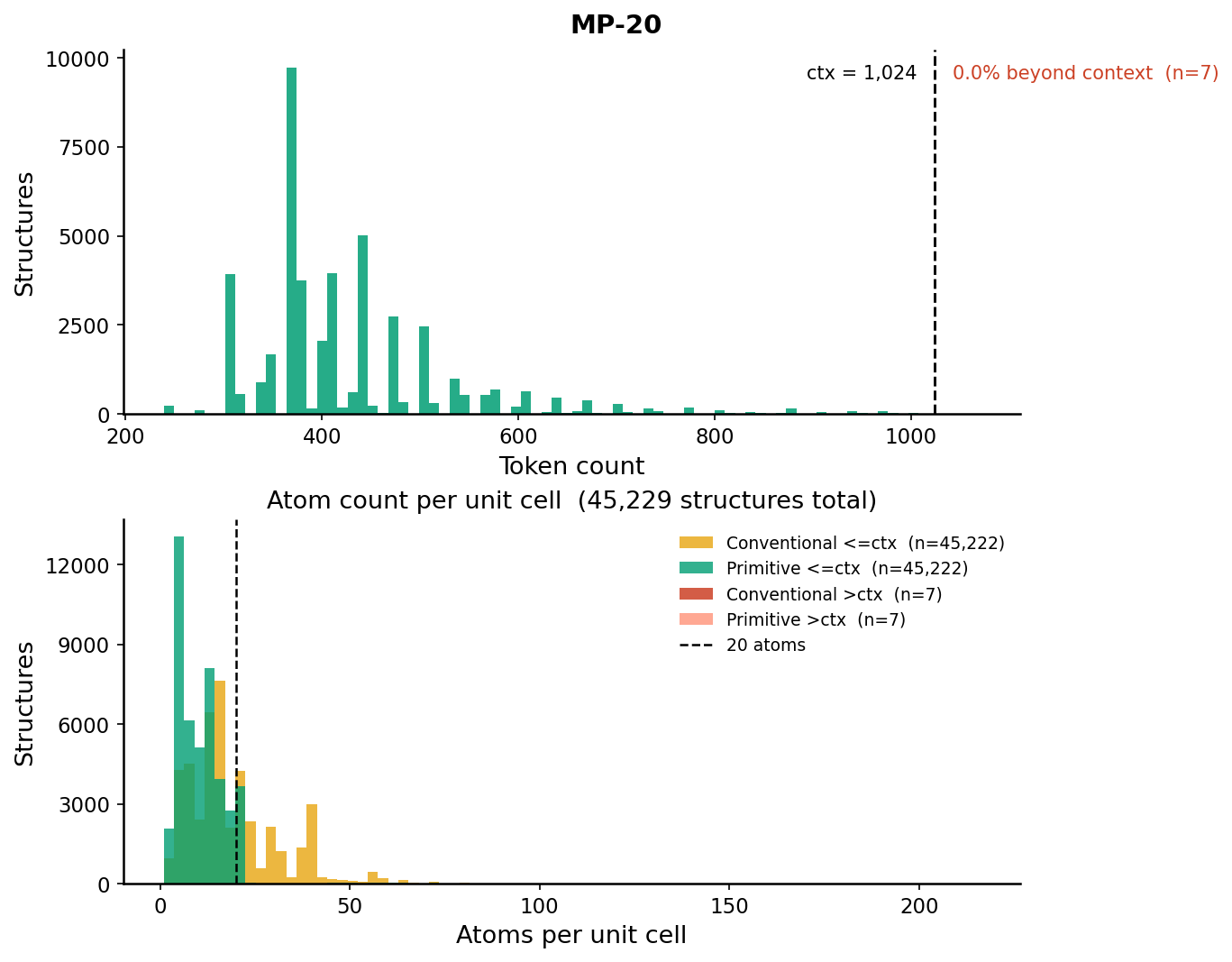}
    \includegraphics[width=0.44\textwidth]{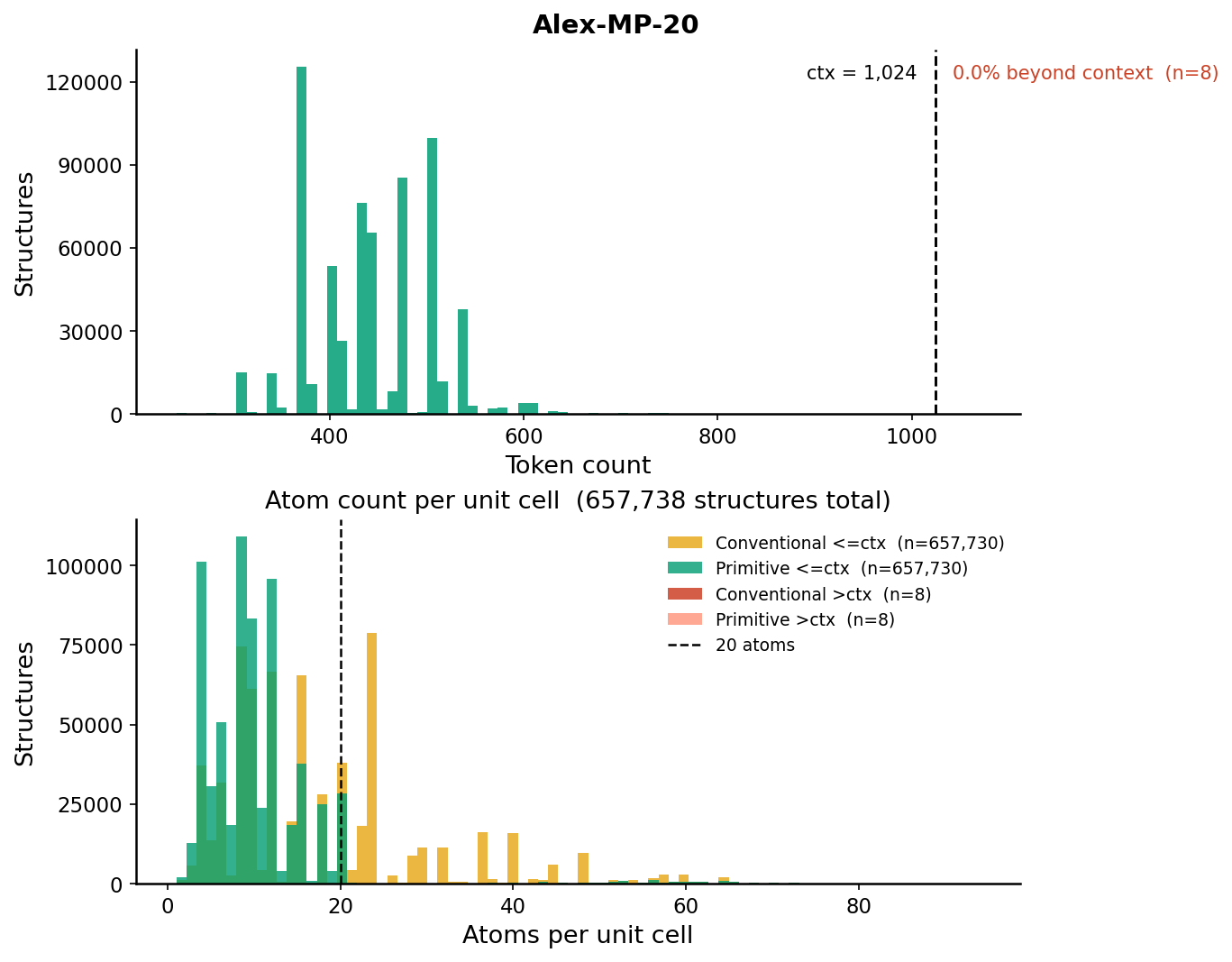}
    \includegraphics[width=0.44\textwidth]{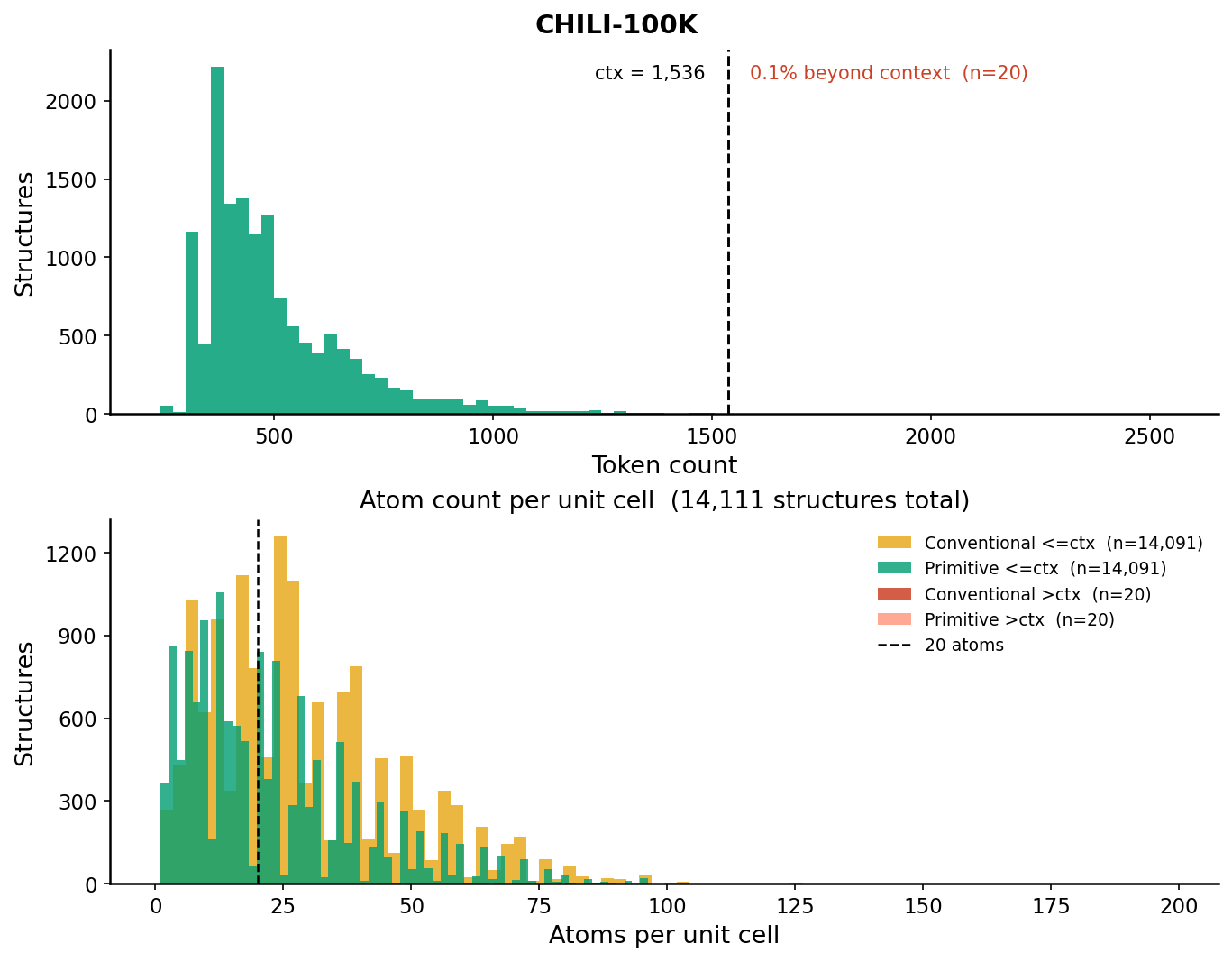}
    \includegraphics[width=0.44\textwidth]{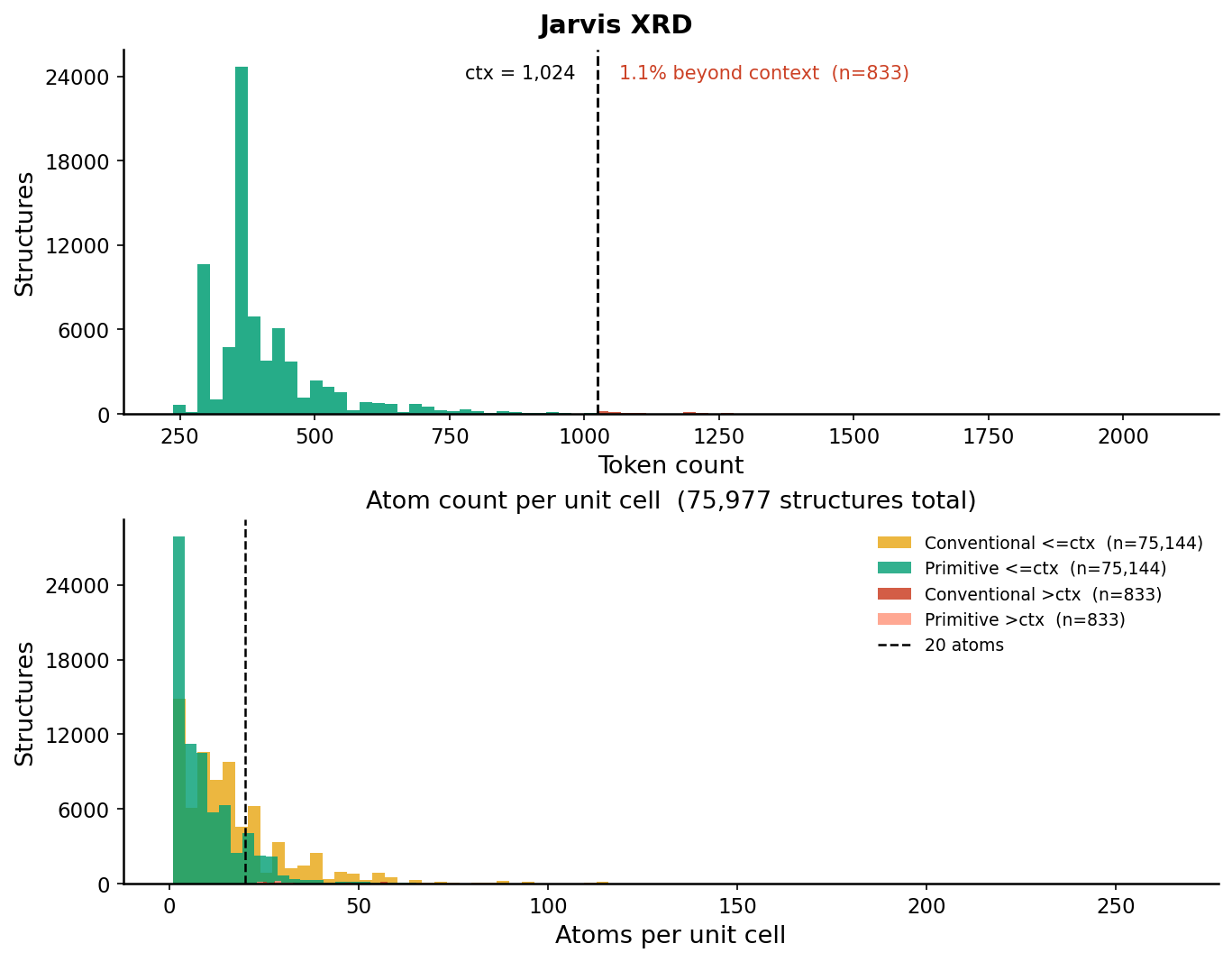}
    \includegraphics[width=0.44\textwidth]{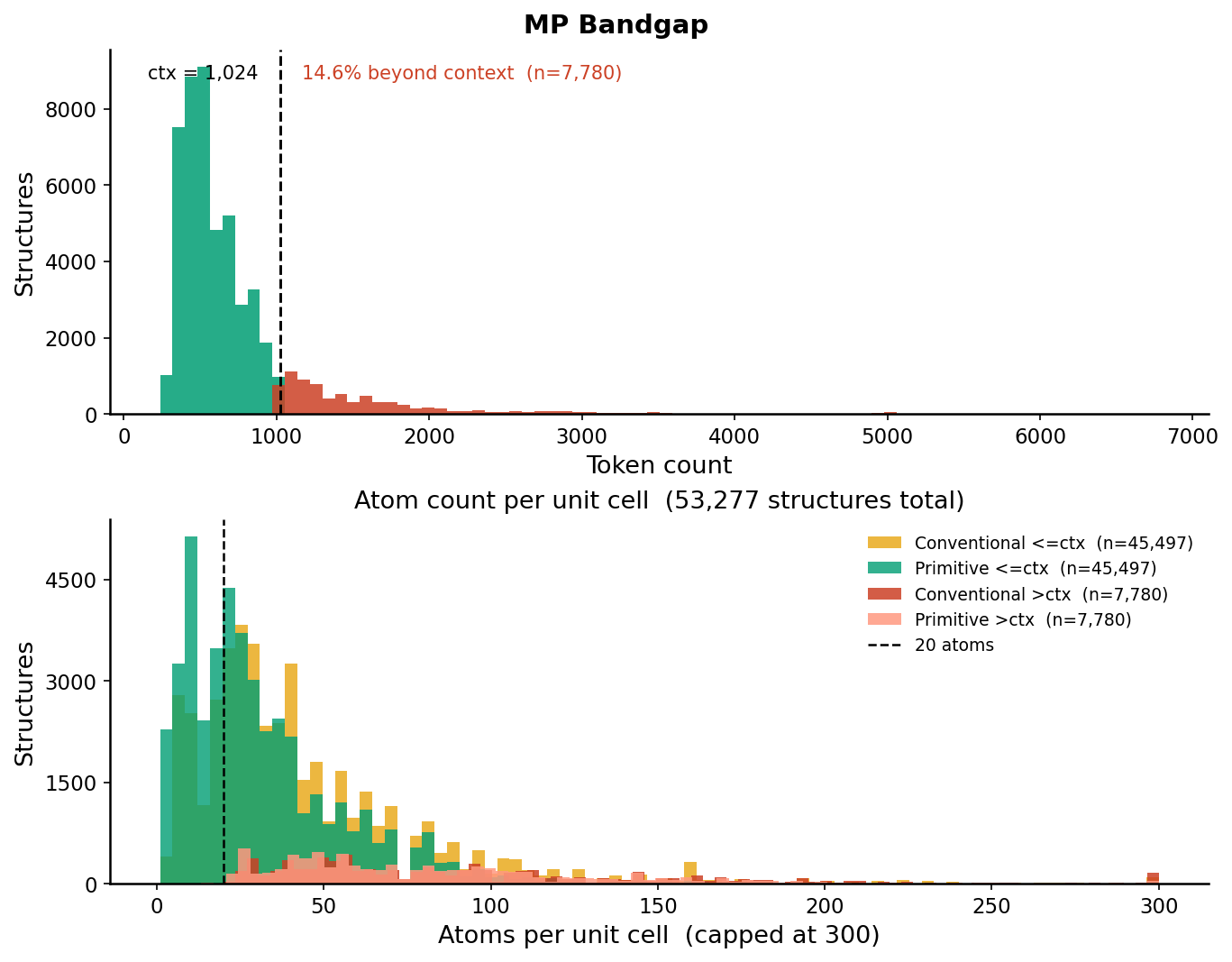}
    \includegraphics[width=0.44\textwidth]{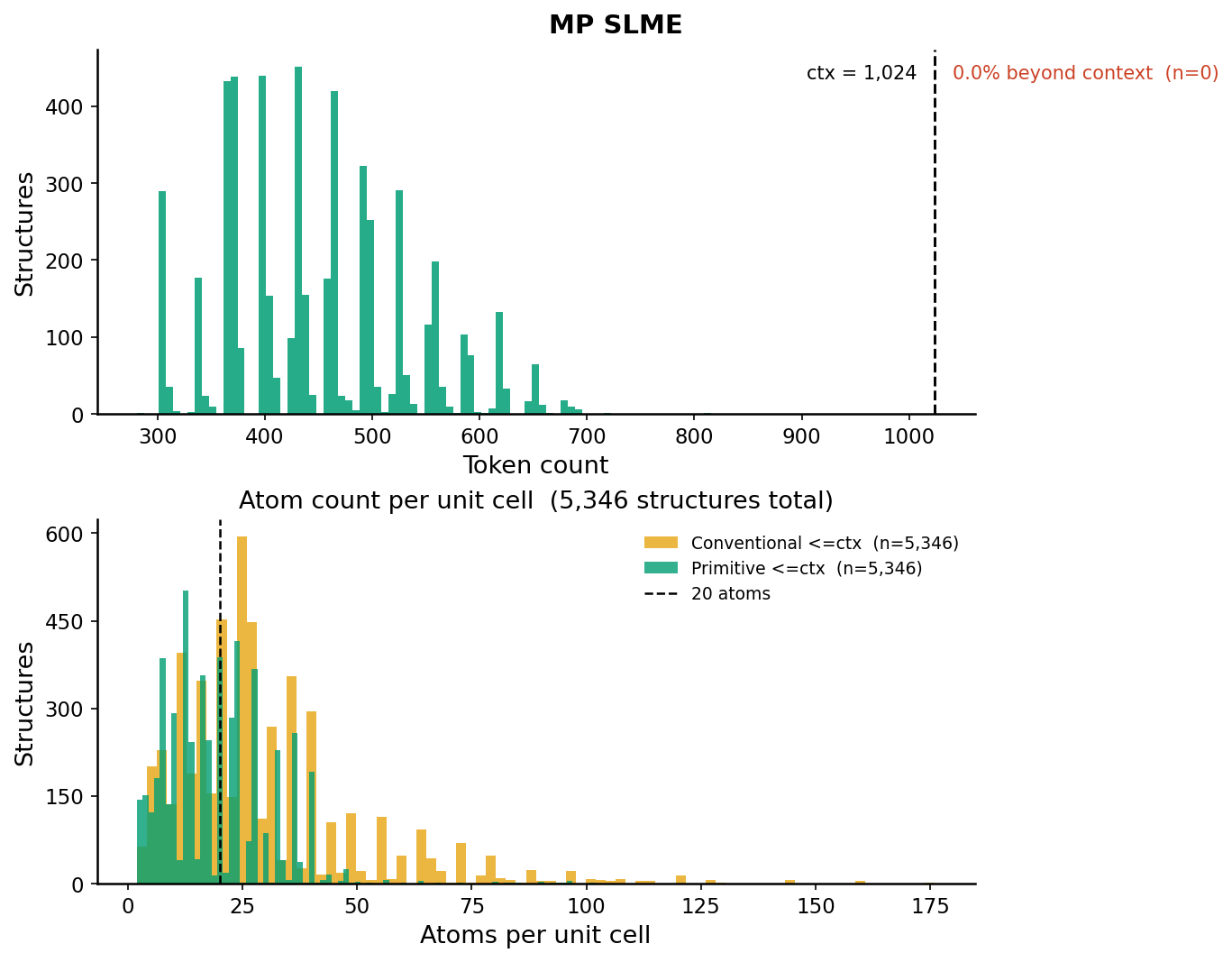}
    \caption{Distribution of token counts and atom counts for pre-processed Crystallographic Information Files (CIFs) across all fine-tuning datasets. In each panel, the top subplot shows the token-count distribution relative to the model context limit, which is 1024 tokens for standard datasets and 1536 tokens for the Chili-100K XRD dataset. The bottom subplot shows the atom-count distribution per CIF. The histograms compare conventional-cell representations used for CrystaLLM-\scalebox{1.3}{$\pi$} with primitive representations used for comparison to graph diffusion models such as MatterGen~\cite{zeniGenerativeModelInorganic2025}. Vertical lines mark the 20-atom threshold often used to restrict diffusion-model datasets, including MP-20~\cite{jainCommentaryMaterialsProject2013} and Alex-MP-20~\cite{zeniGenerativeModelInorganic2025}. Green and yellow denote structures within the context window, whereas red and orange denote structures excluded for exceeding it. Together, these distributions show that CIF tokenisation can represent many systems well beyond 20 atoms within the available context window because only the asymmetric unit is written explicitly.}
    \label{fig:token_distributions}
\end{figure*}

\clearpage

\subsection{Preprocessing and Batching}
\label{app:pre_processing}
The base CrystaLLM-\scalebox{1.3}{$\pi$} model trains to learn patterns on a large corpus of CIFs. Before this, the relevant data needs to be preprocessed into an optimised machine-readable format.
In essence, preprocessing steps include CIF deduplication, augmentation and tokenisation. They follow the same core logic as the original CrystaLLM work~\cite{antunesCrystalStructureGeneration2024}. 

\textbf{Deduplication} was performed by keeping the lowest volume per formula unit structure for identical cell composition and space group.
For \textbf{data augmentation}, symmetry information was embedded using pymatgen's symmetry finding with a tolerance of~\SI{0.1}{\angstrom} and floating point numbers were all rounded to 4 decimal places. Additionally, brackets were added to the CIF, encapsulating parts that vary between crystal structures as described in Section~\ref{sec:backbone_theory}, like ``data\_[Ge4]'' where text inside the brackets is considered \textit{variable} and outside the brackets is considered \textit{fixed}. For an example of a fully augmented CIF, see Appendix~\ref{app:augmented_cif}.

During \textbf{tokenisation}, CIFs were wrapped in \texttt{<bos>} and \texttt{<eos>} to favour learning of termination behaviour and clear example boundaries. Tokenisation was performed on CIF chunks such as digits, atom types or CIF tags in the same manner as CrystaLLM~\cite{antunesCrystalStructureGeneration2024}, producing a final token vocabulary of 377. A \textbf{condition vector} for each material was generated by concatenating associated properties of interest when property conditioning methods were activated.

For \textbf{batching}, sequences longer than the context length were truncated and shorter ones were stacked in a round-robin manner until the length was reached, detailed in Pseudo-code~\ref{algo:rr_batching}. The input sequences were allowed to contain multiple CIFs; however, when conditioning, only the first CIF's condition vector was applied. Training data was shuffled at each epoch during training, so CIF stacking was never the same twice during training. 

\begin{algorithm}[H]
\caption{Round-robin packing of CIF token sequences to context length $C$}
\begin{algorithmic}[1]
\Require List of $N$ token sequences $\{X_0, X_1, \dots, X_{N-1}\}$ (Each $X$ is a CIF in dataloader)
\Ensure Packed sequence $out$ with length exactly $C$ (context length)
\State
\Function{RoundRobin}{features, $C$}
  \For{$i \gets 0$ to $N-1$}
    \State $X \gets \text{features}[i].X$
    \State
    \If{$\text{len}(X) \geq C$} \Comment{CIF is longer than context, slice it}
      \State $out \gets X[0:C]$
      \State \Return $out$
    \EndIf
    \State
    \State $out \gets X$ \Comment{Start with current CIF}
    \State $current\_index \gets i$
    \State
    \While{$\text{len}(out) < C$} \Comment{Append more until context reached}
      \State $current\_index \gets (current\_index + 1) \bmod N$ \Comment{Next in batch}
      \State $next\_CIF \gets \text{features}[current\_index].X$
      \State $remaining\_space \gets C - \text{len}(out)$
      \State $tokens\_to\_take \gets \min(\text{len}(next\_CIF),\, remaining\_space)$
      \State Append $next\_CIF[0:tokens\_to\_take]$ to $out$
    \EndWhile
    \State
    \State $out \gets out[0:C]$ \Comment{trim to exact length if needed}
    \State \Return $out$
  \EndFor
\EndFunction
\end{algorithmic}
\label{algo:rr_batching}
\end{algorithm}

\clearpage
\subsection{Augmented CIF}
\label{app:augmented_cif}

\begin{figure*}[htbp!]
    \centering
    \includegraphics[width=0.5\textwidth]{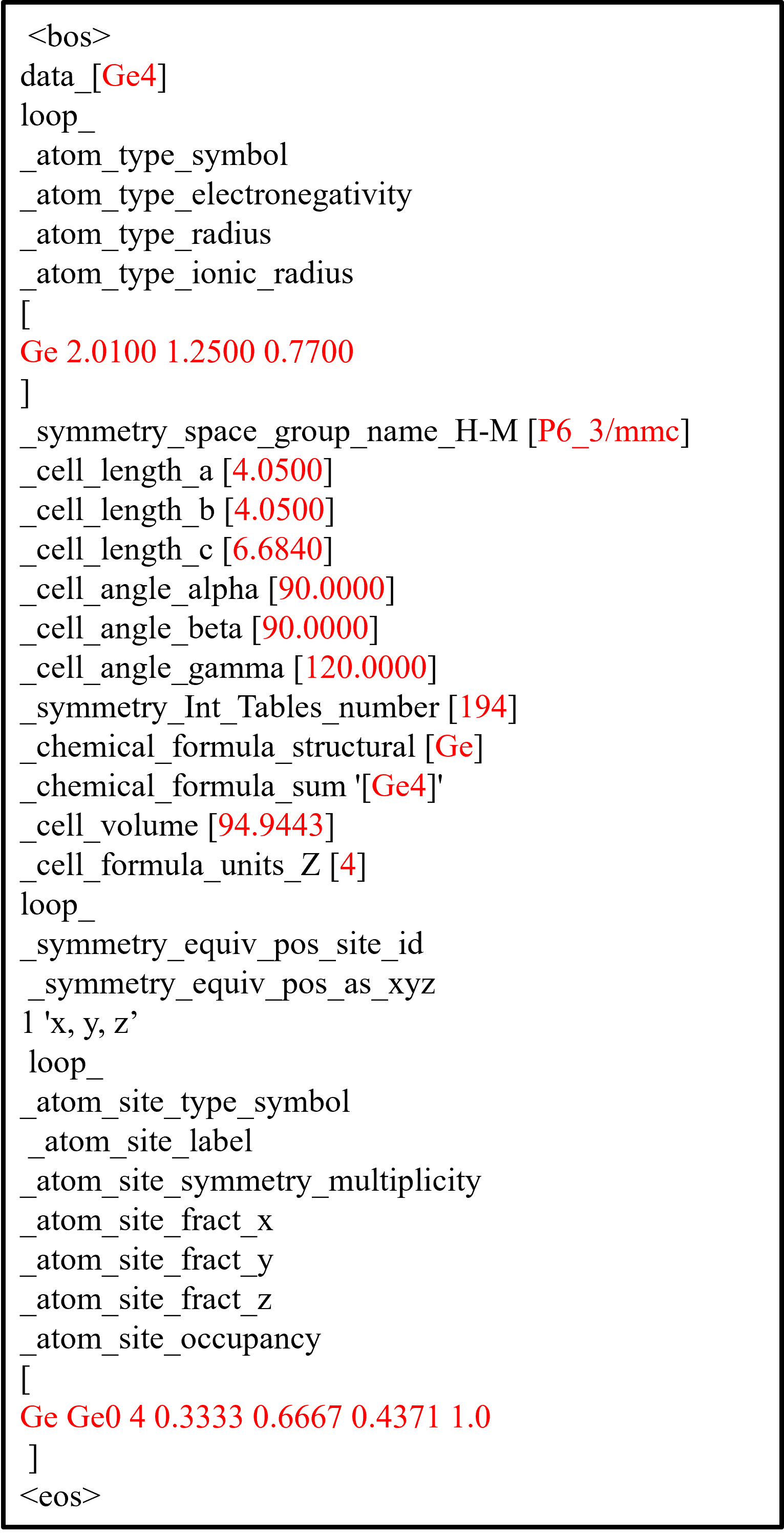}
    \caption{Example of an augmented CIF for \ce{Ge4}. Bracketed fields mark the special tokens introduced during augmentation. Variable tokens are shown in red and fixed tokens are shown in black. This representation physically separates learnable structural content from invariant CIF syntax.}
    \label{fig:augmented_cif}
\end{figure*}

\clearpage
\subsection{LeMat-Benchmark}
\label{app:lemat-bench}

\begin{table*}[htbp!]
    \centering
    \begin{tabular}{llcccccccccc}
        \toprule
        \textbf{Model} & \textbf{Method} & \textbf{Valid} & \textbf{Unique} & \textbf{Novel} & \textbf{Stable} & \textbf{Metastable} & \textbf{SUN} & \textbf{MSUN} & $\mathbf{E_{\text{hull}}}$ \textbf{(eV)} & $\mathbf{E_{\text{form}}}$ \textbf{(eV)} & \textbf{RMSD (\AA)} \\
        & & $\uparrow$ & $\uparrow$ & $\uparrow$ & $\uparrow$ & $\uparrow$ & $\uparrow$ & $\uparrow$ & $\downarrow$ & $\downarrow$ & $\downarrow$ \\
        
        \midrule
        \multicolumn{12}{c}{\textbf{MP-20 Non-Pre-Relaxed}} \\
        \midrule
        Crystal-GFN ~\cite{ai4science2026crystalgfnsamplingcrystalsdesirable} & GFN & 51.7\% & 51.7\% & 51.7\% & 0.0\% & 0.0\% & 0.0\% & 0.0\% & 2.0858 & \textbf{-1.3036} & 1.8665 \\
        LLaMat3 ~\cite{mishra2025foundationallargelanguagemodels} & AR & 15.4\% & 15.2\% & 10.5\% & 0.1\% & 2.1\% & 0.0\% & 0.2\% & 1.7067 & 0.7423 & 1.0057 \\
        ADiT ~\cite{joshi2025allatom} & Diff & 90.6\% & 87.8\% & 26.0\% & 0.4\% & 36.5\% & 0.0\% & 1.0\% & 0.3333 & -0.7339 & \textit{0.3794} \\
        LLaMat2 ~\cite{mishra2025foundationallargelanguagemodels} & AR & 84.4\% & 81.4\% & 30.0\% & 0.7\% & 34.7\% & 0.1\% & 2.1\% & 0.4395 & -0.4726 & 0.5363 \\
        SymmCD ~\cite{levy2025symmcd} & Diff & 73.4\% & 73.0\% & 47.0\% & 1.4\% & 18.6\% & 0.1\% & 2.4\% & 0.8761 & -0.0195 & 0.8720 \\
        CrystalFormer ~\cite{caoSpaceGroupInformed2024} & AR & 69.9\% & 69.4\% & 31.8\% & 1.4\% & 28.8\% & 0.0\% & 3.1\% & 0.7039 & -0.1689 & 0.6585 \\
        Zatom-1-WD ~\cite{morehead2026zatom1multimodalflowfoundation} & Flow & 67.2\% & 65.4\% & 24.2\% & 0.2\% & 22.8\% & 0.0\% & 0.4\% & 0.5189 & -0.3687 & 0.4773 \\
        \textbf{CrystaLLM-$\pi$ (ours)} & AR & 86.8\% & 84.9\% & 24.9\% & \textit{2.9\%} & \textbf{49.0\%} & \textbf{0.3\%} & 3.6\% & 0.3205 & -0.5903 & \textbf{0.3631} \\
        DiffCSP++ ~\cite{jiao2024space} & Diff & 95.3\% & \textbf{95.1\%} & 62.0\% & 1.0\% & 26.4\% & \textit{0.2\%} & 5.0\% & 0.4093 & -0.5189 & 0.6933 \\
        DiffCSP ~\cite{jiao2023crystal} & Diff & \textbf{95.7\%} & \textit{94.8\%} & \textit{66.2\%} & 2.3\% & 29.8\% & 0.1\% & 8.5\% & 0.2747 & -0.6367 & 0.5857 \\
        Chemeleon1 ~\cite{parkExplorationCrystalChemical2024} & Diff & \textit{95.4\%} & \textit{94.8\%} & 64.0\% & \textbf{3.6\%} & 34.4\% & \textbf{0.3\%} & \textit{11.3\%} & \textit{0.2055} & \textit{-0.7590} & 0.4907 \\
        Chemeleon2 ~\cite{park2025guidinggenerativemodelsuncover} & Diff + RL & 95.2\% & 88.1\% & \textbf{71.6\%} & 0.0\% & \textit{39.8\%} & 0.0\% & \textbf{21.2\%} & \textbf{0.1557} & -0.7167 & 0.4226 \\
        
        \midrule
        \multicolumn{12}{c}{\textbf{MP-20 Pre-Relaxed}} \\
        \midrule
        WyFormer ~\cite{kazeev2503wyckoff} & AR & 93.4\% & 93.0\% & \textit{66.4\%} & 0.5\% & 15.7\% & 0.1\% & 1.9\% & 0.4988 & -0.4306 & 0.8121 \\
        WyFormer-DFT ~\cite{kazeev2503wyckoff} & AR & 95.2\% & 95.0\% & \textit{66.4\%} & 3.7\% & 24.8\% & 0.4\% & 7.8\% & 0.2708 & -0.6660 & 0.4173 \\
        PLaID++ ~\cite{xu2025plaidpreferencealignedlanguage} & AR + RL & 96.0\% & 77.8\% & 24.2\% & 12.4\% & \textit{60.7\%} & 1.0\% & 7.6\% & 0.0854 & -0.4975 & 0.1286 \\
        Zatom-1-WD ~\cite{morehead2026zatom1multimodalflowfoundation} & Flow & 88.3\% & 86.0\% & 37.9\% & 4.7\% & 50.5\% & 0.3\% & 8.8\% & 0.1093 & -0.8202 & 0.3387 \\
        MatterGen ~\cite{zeniGenerativeModelInorganic2025} & Diff & 95.7\% & 95.1\% & \textbf{70.5\%} & 2.0\% & 33.4\% & 0.2\% & 15.0\% & 0.1834 & -0.7020 & 0.3878 \\
        MiAD ~\cite{okhotin2025miadmirageatomdiffusion} & Diff & 96.2\% & 94.3\% & 40.2\% & 6.4\% & \textbf{62.0\%} & 1.0\% & 16.6\% & \textit{0.0804} & -0.8436 & 0.2494 \\
        OMatG ~\cite{hollmer2025open} & Diff + Flow & 96.4\% & \textit{95.2\%} & 51.2\% & 11.6\% & 49.8\% & 1.0\% & 18.0\% & 0.0956 & \textbf{-0.9227} & \textit{0.0759} \\
        MCFlow ~\cite{seong2026multimodalcrystalflowanytoany} & Flow & \textbf{97.2\%} & \textbf{96.3\%} & 52.2\% & 11.9\% & 49.3\% & 0.7\% & \textit{18.9\%} & 0.0987 & -0.8513 & 0.1696 \\
        Crystalite ~\cite{veljkovic2026crystalitelightweighttransformerefficient} & Diff & \textbf{97.2\%} & \textit{95.8\%} & 53.2\% & \textit{12.7\%} & 51.6\% & \textit{1.5\%} & \textbf{22.6\%} & 0.0905 & -0.8916 & 0.1322 \\
        OMatG-FC ~\cite{hollmer2025open} & Diff + Flow & \textbf{97.2\%} & 92.8\% & 28.9\% & \textbf{18.4\%} & 57.9\% & \textbf{1.7\%} & 12.0\% & \textbf{0.0694} & \textit{-0.9149} & \textbf{0.0685} \\
        
        \midrule
        \multicolumn{12}{c}{\textbf{Alex-MP-20 Non-Pre-Relaxed}} \\
        \midrule
        \textbf{CrystaLLM-$\pi$ (ours)} & AR & 93.4\% & 93.2\% & 23.9\% & \textbf{4.8\%} & \textbf{71.3\%} & \textit{0.3\%} & 12.5\% & \textbf{0.0941} & -0.5811 & \textbf{0.1914} \\
        Chemeleon1 ~\cite{parkExplorationCrystalChemical2024} & Diff & \textit{94.0\%} & \textbf{93.9\%} & \textit{59.8\%} & \textit{4.0\%} & 54.4\% & \textbf{1.0\%} & \textit{28.0\%} & 0.1120 & \textbf{-0.6344} & 0.3089 \\
        Chemeleon2 ~\cite{park2025guidinggenerativemodelsuncover} & Diff + RL & \textbf{97.4\%} & \textit{93.5\%} & \textbf{93.7\%} & 0.3\% & \textit{67.4\%} & 0.2\% & \textbf{62.4\%} & \textit{0.0959} & \textit{-0.6086} & \textit{0.2678} \\
        
        \midrule
        \multicolumn{12}{c}{\textbf{Alex-MP-20 Pre-Relaxed}} \\
        \midrule
        None &  &  &  &  &  &  &  &  &  &  &  \\
        
        \bottomrule
    \end{tabular}
    \caption{Generative performance metrics for models evaluated on the MP-20 and Alex-MP-20 benchmarks within LeMat-Bench~\cite{betala2025lemat}. The table distinguishes non-pre-relaxed and pre-relaxed generation protocols, where the latter includes a structural relaxation step before candidate evaluation and the former evaluates out-of-the-box unrelaxed structures. Entries within each section are sorted by ascending cumulative Stable, Unique, Novel (SUN) and Meta-Stable, Unique, Novel (MSUN) scores (bottom row is best). Upward arrows ($\uparrow$) specify metrics where higher values indicate superior generative capacity (Valid, Unique, Novel, Stable, Metastable, SUN, MSUN). Downward arrows ($\downarrow$) indicate metrics where lower values represent greater stability and structural accuracy ($E_{\text{hull}}$, $E_{\text{form}}$, RMSD). The method column denotes the model family: autoregressive (AR), diffusion (Diff), generative flow network (GFN), flow matching (Flow) and reinforcement learning (RL). Bold and italics denote the best and second-best metrics within each subdivision.}
    \label{tab:combined_benchmarks_acronyms}
\end{table*}

\clearpage

\subsection{Output Logit Study}
\label{app:logit_study}

The custom digit-wise tokenisation scheme induces a hierarchical generation process for atomic coordinates. The model predicts leading integers and primary decimals with high certainty to establish coarse spatial boundaries, while subsequent trailing decimals display broader probability distributions. This enables stochastic exploration within defined structural limits.

\begin{figure}[htbp!]
    \centering
    \includegraphics[width=0.9\columnwidth]{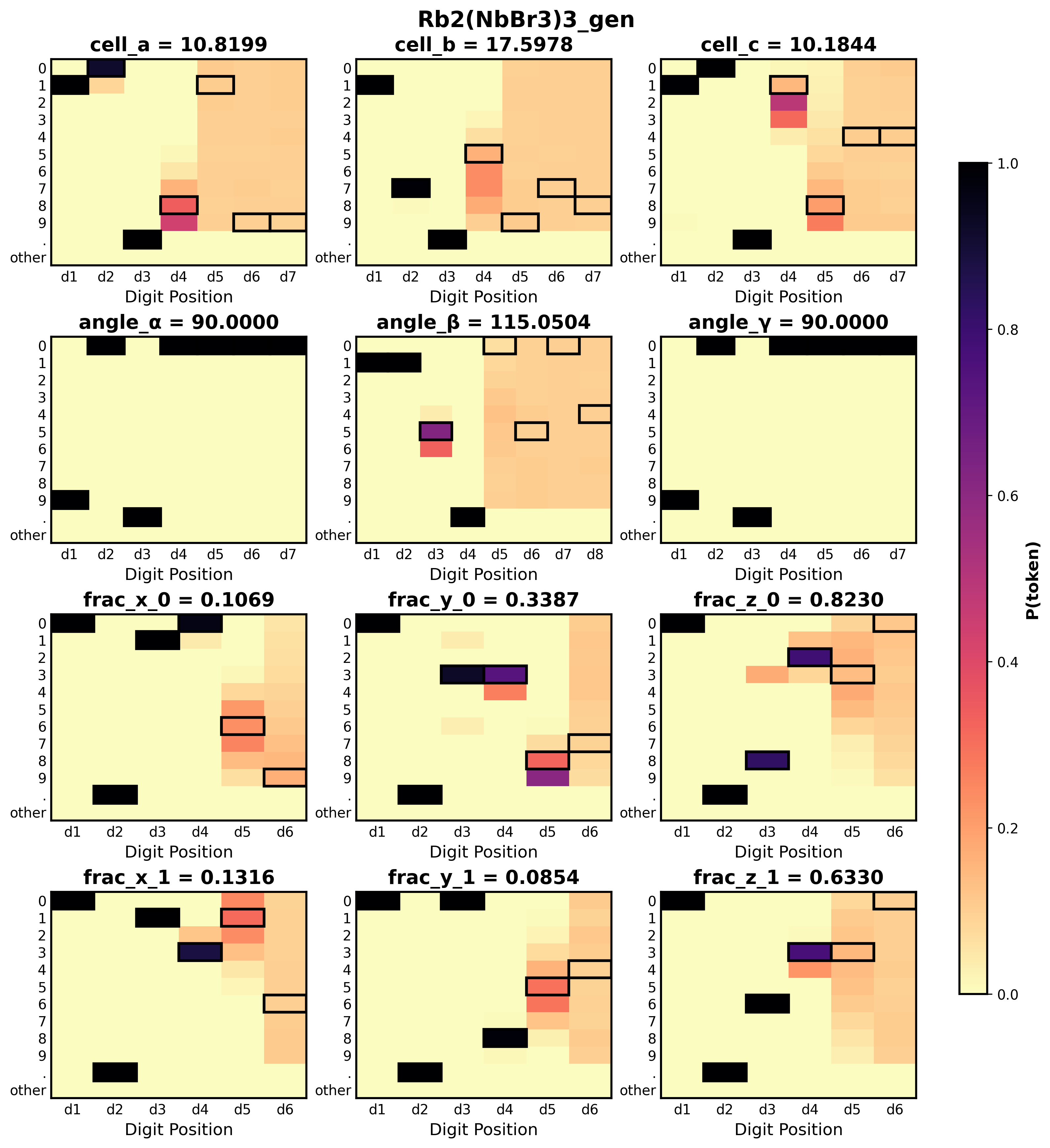}
    \caption{Autoregressive logit analysis for continuous crystallographic parameters in the generated \ce{Rb2(NbBr3)3} structure sequence. Each heatmap corresponds to one parameter shown above the subplot. Columns denote digit positions in the generated token sequence, while rows denote vocabulary bins consisting of digits $0$ to $9$, the decimal point ``.'' and an aggregated ``other'' category. The colour scale shows the predicted probability mass $P(\text{token})$, and black boxes mark the tokens sampled in the final \ce{Rb2(NbBr3)3} sequence. This material was absent from both the pre-training and fine-tuning datasets (Sections~\ref{sec:slme} and~\ref{app:slme_candidates}). The distributions show that the model concentrates probability within the digit space for continuous-valued fields, with early digits fixing coarse geometric constraints at high confidence and later decimal positions remaining broader to support fine structural variation.}
    \label{fig:logit_study_hierarchy}
\end{figure}

\clearpage

\subsection{Architectural details for Encoders}
\label{app:encoders}

\begin{figure*}[htbp!]
    \centering
    \includegraphics[width=0.7\textwidth]{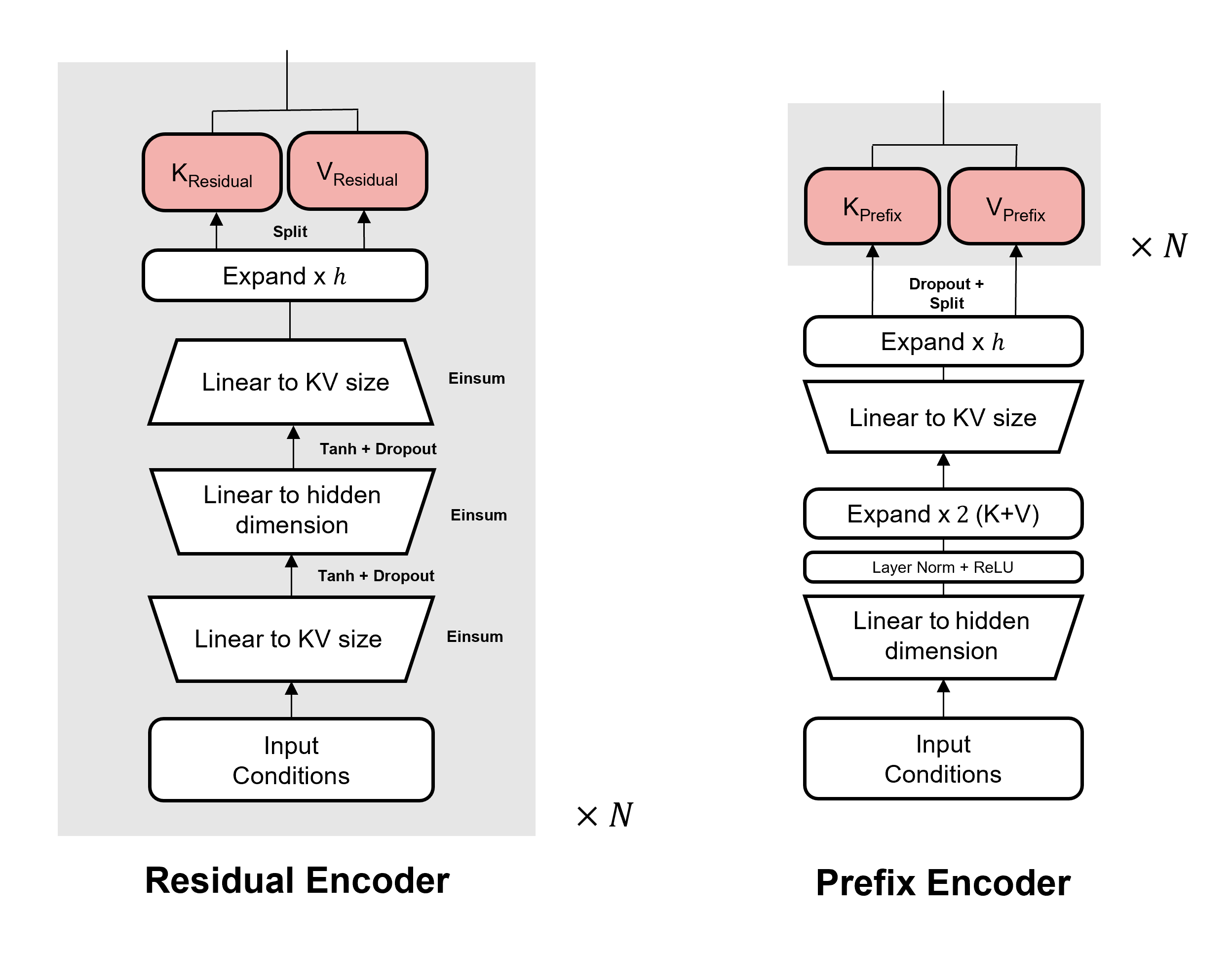}
    \caption{Residual and Prefix conditioning encoder architectures. The left panel shows the Residual encoder, and the right panel shows the Prefix encoder. Both map the input condition vector to learned key-value tensors that are replicated across the $N$ decoder blocks of the transformer. The Residual encoder uses einsum-based expansions so that individual input properties occupy distinct regions of the latent space and can be associated with separate key-value pairs. The Prefix encoder instead learns a shared set of key-value tensors that is not directly attributable to individual input properties. This difference mirrors the downstream attention design, where Residual attention preserves property-specific structure and Prefix attention injects a shared conditioning context.}
    \label{fig:Prefix_Residual_encoders}
\end{figure*}

\clearpage

\subsection{Architectural details for PrependGPT}
\label{app:prepend_gpt}

\begin{figure}[htbp!]
    \centering
    \includegraphics[width=0.95\textwidth]{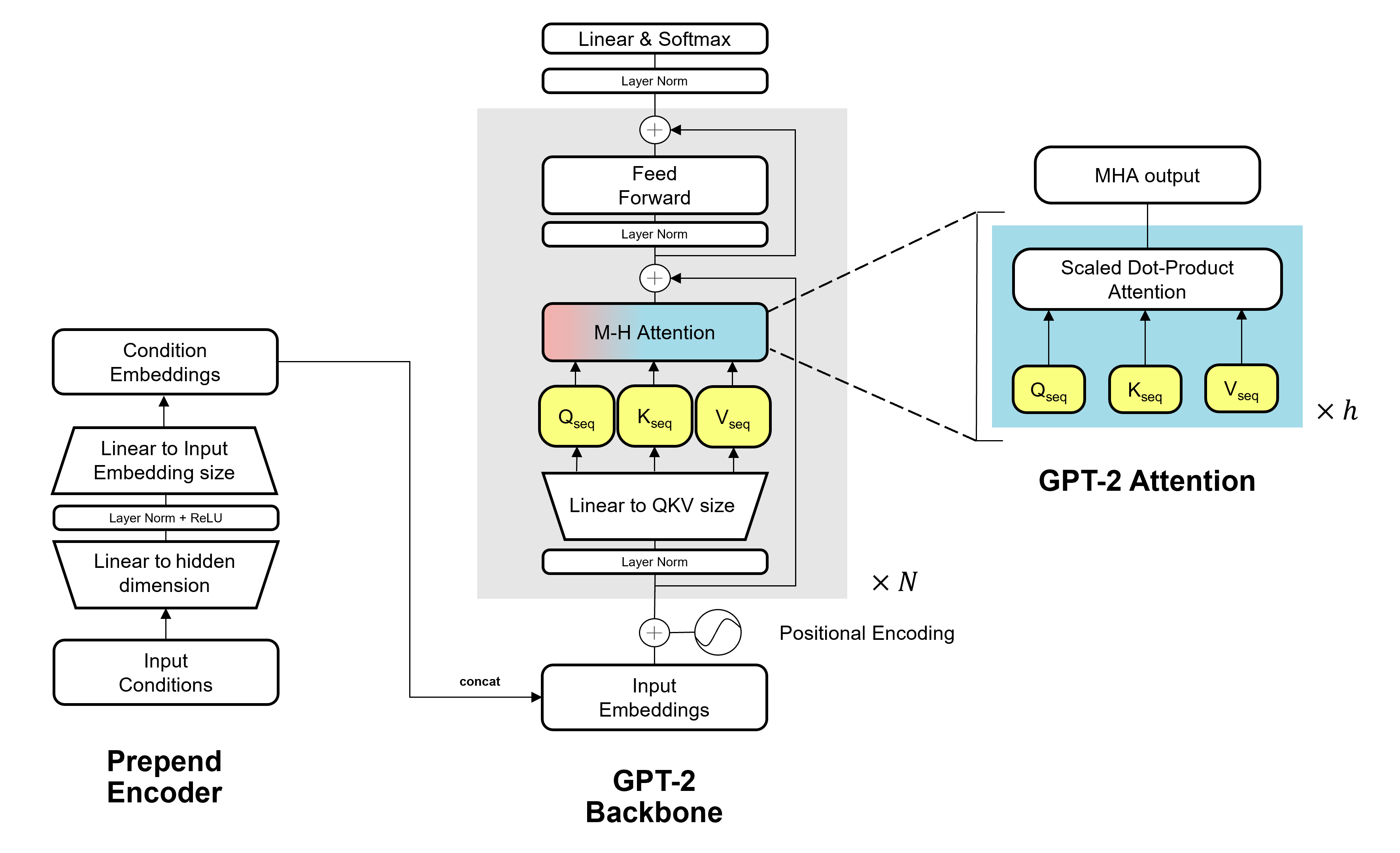}
    \caption{Prepend conditioning architecture. The Prepend encoder maps the input condition vector to the same embedding dimension as the input sequence embeddings. These condition embeddings are concatenated with the token embeddings before entering the GPT-2 backbone. The downstream attention mechanism then remains unchanged relative to the standard GPT-2 block. This makes Prepend conditioning a sequence-level intervention rather than an attention-level modification.}
    \label{fig:prependgpt_diagram}
\end{figure}

\clearpage

\subsection{Experimental protocols}
\label{app:experimental_protocols}
Structures for the LeMat-Bench evaluation were generated using models pre-trained on either the standard MP-20 or Alex-MP-20 datasets (Section~\ref{sec:base_model}). Following benchmark requirements, 2.5K CIFs were sampled. For the MP-20 dataset, sampling parameters were set to \texttt{T} = 1.0, \texttt{top-K} = 50 and \texttt{top-p} = 0.95. For the Alex-MP-20 dataset, parameters were adjusted to \texttt{T} = 1.25, \texttt{top-K} = 50 and \texttt{top-p} = 0.95. \textit{Ab initio} prompting was employed uniformly without specifying target cell compositions.

The MP~Bandgap dataset was used for model training in Section~\ref{sec:pre-training}. All four conditioning architectures were evaluated to assess information retention and baseline conditioning capabilities. Models were either trained from scratch or fine-tuned from the pre-trained backbone. Sampling across both initialisation regimes generated 1K CIFs per condition using \texttt{T} = 0.7, \texttt{top-K} = 30 and \texttt{top-p} = 0.95, with \textit{ab initio} prompting. To evaluate performance across regions of varying training data density, eight target band gap conditions were tested: \SIlist[list-units=single]{0.07;0.18;0.68;1.55;2.44;3.72;6.78;12.1}{\electronvolt} (inputted as normalised values). A secondary condition of $E_{\mathrm{hull}} = 0~\si{\electronvolt/atom}$ was applied uniformly across all trials.

\begin{figure}[htbp!]
    \centering
    \includegraphics[width=0.47\textwidth]{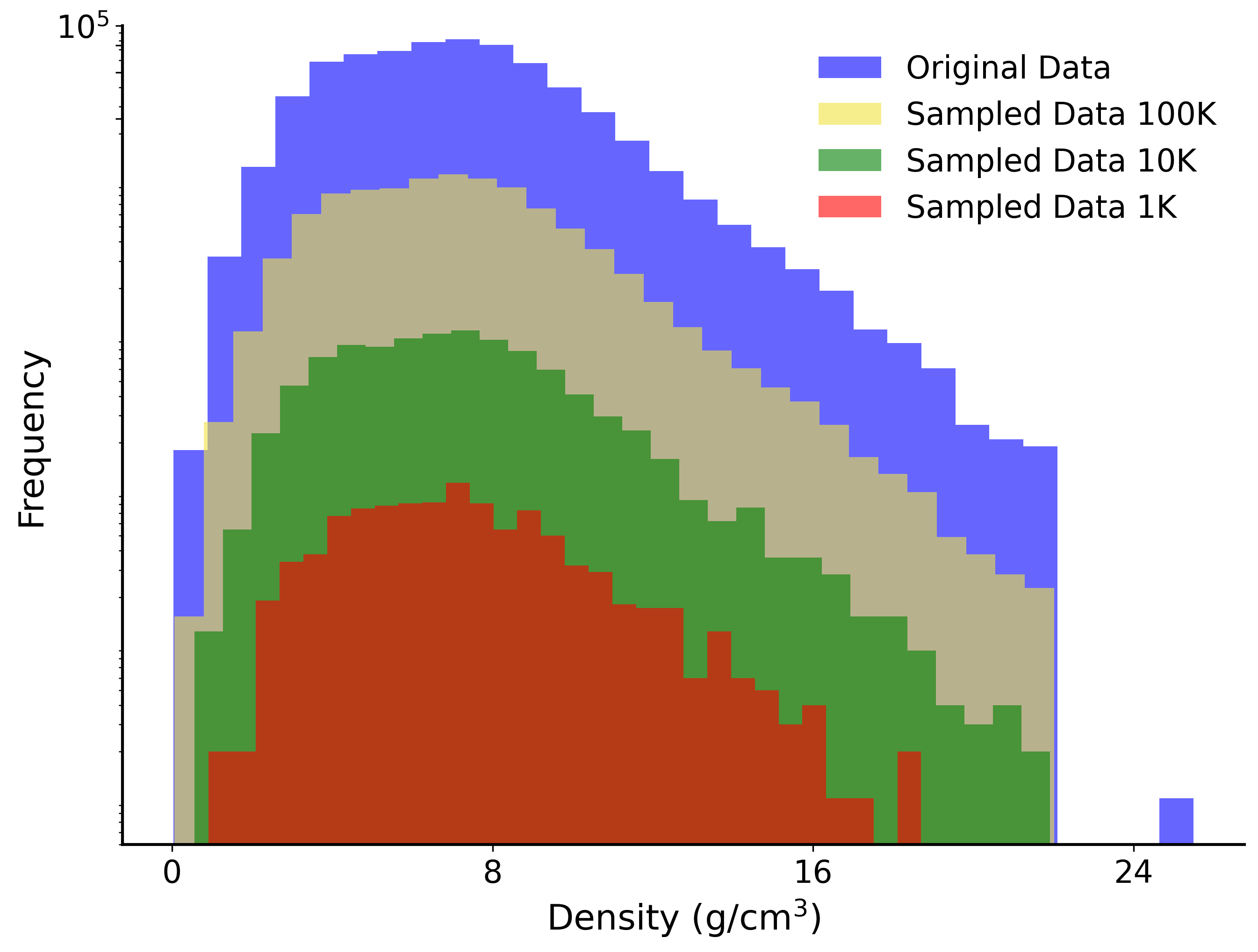}
    \caption{Density distributions of the original MatterGen density dataset and its randomly sampled subsets. The histogram shows the frequency of material densities (\si{g/cm^3}) on a logarithmic scale for the full dataset together with random subsets of 100K, 10K and 1K entries. Smaller subsets progressively lose coverage in the high-density tails, illustrating how limited random samples can underrepresent the full target space.}
    \label{fig:dataset_size_dist}
\end{figure}

For Section~\ref{sec:dataset_size_study}, the MatterGen Density dataset was randomly sampled to produce subsets containing 1K, 10K and 100K structures (Figure~\ref{fig:dataset_size_dist}). Pre-trained models were fine-tuned on each subset to evaluate conditional generation performance under constrained data regimes.

Sampling generated 1K CIFs per condition using \texttt{T} = 1.0, \texttt{top-K} = 15 and \texttt{top-p} = 0.95, with \textit{ab initio} prompting. Eight density conditions were evaluated: \SIlist[list-units=single]{1.27;2.55;6.37;10.20;12.75;15.30;19.12;22.94}{g/cm^3} (inputted as normalised values), with $E_{\mathrm{hull}} = 0~\si{\electronvolt/atom}$ applied as a secondary condition.

For the discovery task, the complete MP~SLME dataset was used to fine-tune the base model for SLME-conditioned CIF generation. Sampling produced 100K CIFs using \texttt{T} = 1.0, \texttt{top-K} = 50 and \texttt{top-p} = 0.95, with \textit{ab initio} prompting. The elevated \texttt{top-K} parameter expanded the accessible chemical space during atomic species selection, preventing collapse toward common elements and increasing the elemental diversity of the generated structures.

Materials recovery tasks used the XRD datasets. Given the highly constrained nature of these tasks, like incorporating multiple conditions, heterogeneous data, cell compositions and occasionally space group information, a ``1-perplexity'' sampling scheme was employed for crystal structure predictions (ablation detailed in Appendix~\ref{app:ablation_studies}). Sampling parameters were fixed at \texttt{T} = 0.75, \texttt{top-K} = 10 and \texttt{top-p} = 0.95. The temperature parameter was determined via a systematic sweep that optimised match rates and RMSD metrics on a sub-sampled test set for both ``20-consistent'' and ``1-perplexity'' scored CIFs (see Section~\ref{sec:sampling} for sampling schemes and Appendix~\ref{app:ablation_studies} for the ablation study). All models were prompted with target cell compositions alongside XRD spectra. For the polymorph recovery study (Section~\ref{sec:experimental_recovery}), we used \texttt{T} = 1.1, \texttt{top-K} = 15 and \texttt{top-p} = 0.95. Models were additionally evaluated with space group prompting and supercell prompting by doubling the input stoichiometry relative to the true cell composition.

\clearpage
\subsection{Pre-training Study Extras}
\label{app:pre-train}

\begin{table}[htbp!]
    \centering
    \renewcommand{\arraystretch}{1.3}
    \setlength{\tabcolsep}{3.0pt}
    \begin{tabular}{@{}llccc@{}}
        \toprule
        \textbf{Regime} & \textbf{Model} & $\mathbf{N_{\mathrm{Valid}}}$ & $\mathbf{R^2}$ & \begin{tabular}[c]{@{}c@{}}\textbf{MAE}\\ \textbf{($\pm$ SE)}\end{tabular} \\
        
        \midrule
        \multirow{5}{*}{Scratch}
        & MatterGen & 6585 & 0.56 & 1.58 (0.039) \\
        \cmidrule{2-5}
        & Prefix & 3788 & 0.60 & \textbf{1.03 (0.093)} \\
        & Residual & \textit{4036} & \textit{0.61} & 1.40 (0.099) \\
        & Prepend & \textbf{4162} & \textbf{0.65} & 1.30 (0.098) \\
        & Raw & 956 & 0.22 & \textit{1.27 (0.383)} \\
        
        \midrule
        \multirow{3}{*}{Pre-tnd}
        & Prefix & \textbf{5759} & \textbf{0.59} & \textit{1.48 (0.047)} \\
        & Residual & 5457 & 0.51 & 1.69 (0.056) \\
        & Prepend & \textit{5670} & \textit{0.58} & \textbf{1.46 (0.046)} \\
        \bottomrule
    \end{tabular}
    \caption{Band gap conditioning performance across Scratch and Pre-trained (Pre-tnd) learning regimes. The autoregressive architectures are compared with the graph diffusion baseline MatterGen~\cite{zeniGenerativeModelInorganic2025}. $N_{\mathrm{Valid}}$ is the total number of structurally valid generations. $R^2$ and MAE $\pm$ SE are computed between requested and predicted band gaps for $Q_{\mathrm{VSUN}}$ materials only. Bold and italics denote the best and second-best autoregressive results within each regime.}
    \label{tab:pre-train-regime-metrics}
\end{table}

\begin{figure*}[htbp!]
    \centering
    \includegraphics[width=0.8\textwidth]{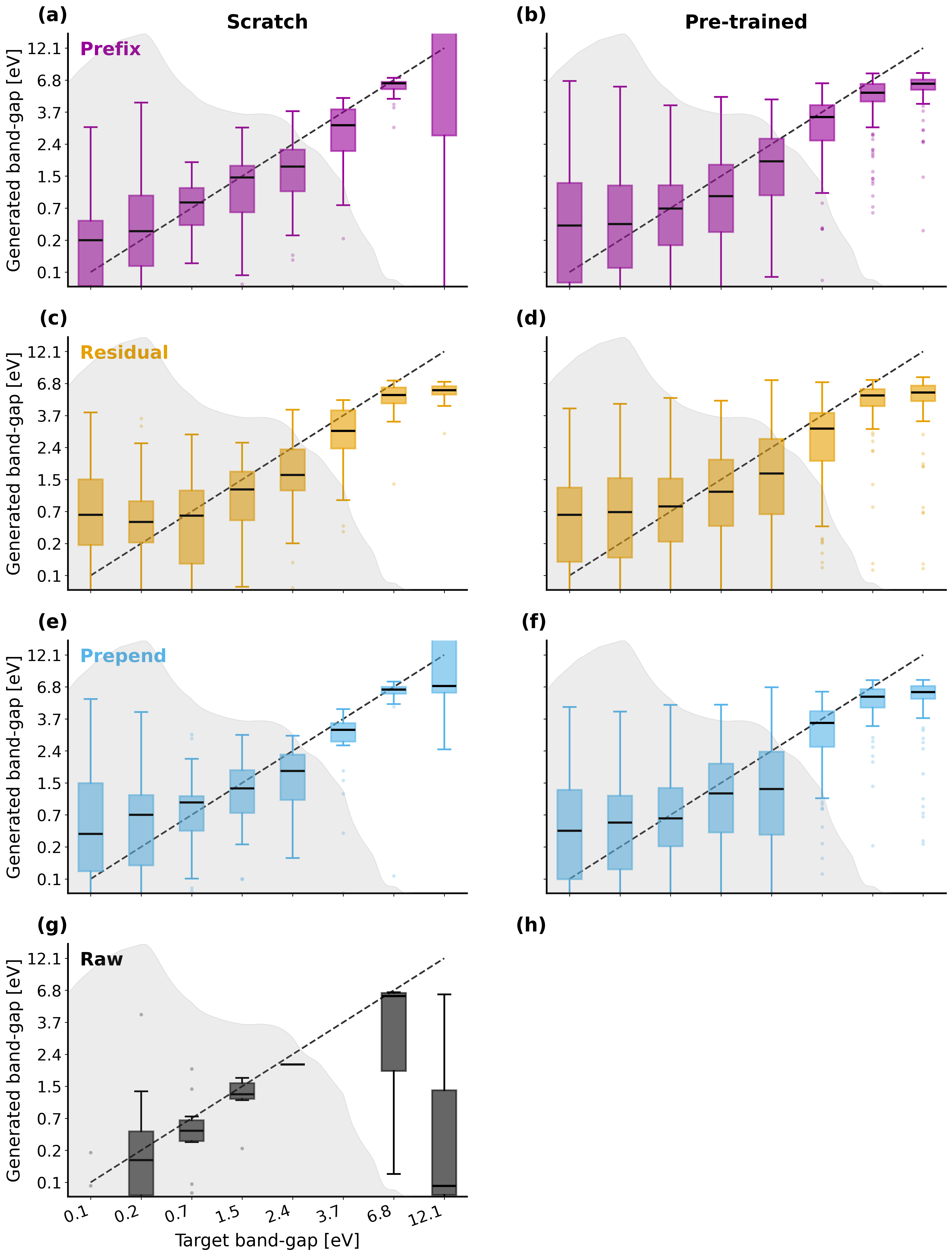}
    \caption{Parity distributions comparing ALIGNN-predicted band gaps of generated materials against requested targets across conditioning architectures and training regimes. Columns distinguish the from scratch and pre-trained learning regimes, while rows correspond to the Prefix, Residual, Prepend and Raw conditioning models. Box plots summarise the distributions of predicted band gaps for $Q_{\mathrm{VSUN}}$ structures at discrete, non-linearly spaced target values. The dashed diagonal line marks perfect calibration between requested and generated values. The grey background profile shows the band gap distribution of the training dataset. Panel (h) is empty because the Raw baseline failed to generate Valid structures after pre-training. Together with the quantitative results in Table~\ref{tab:pre-train-regime-metrics}, these distributions indicate that the pre-trained Prefix architecture provides the most accurate and consistent target tracking across the sampled property range.}
    \label{fig:pre-training_parity_plot}
\end{figure*}

\clearpage

\clearpage
\subsection{Dataset Size Study extras}
\label{app:dataset-size-study}

\begin{table}[htbp!]
    \centering
    \renewcommand{\arraystretch}{1.3}
    \setlength{\tabcolsep}{3.5pt}
    \begin{tabular}{llccccc}
    \toprule
    \textbf{Size} & \textbf{Model} & $\mathbf{N_{\mathrm{Valid}}}$ & $\mathbf{R^2}$ & \begin{tabular}[c]{@{}c@{}}\textbf{MAE}\\ \textbf{($\pm$ SE)}\end{tabular} & \begin{tabular}[c]{@{}c@{}}\textbf{Valid-Hit}\\ \textbf{(\%)}\end{tabular} & \begin{tabular}[c]{@{}c@{}}\textbf{Q-Hit}\\ \textbf{(\%)}\end{tabular} \\
    
    \midrule
        \multirow{3}{*}{1K}
        & Prefix & \textit{6681} & \textbf{-0.27} & \textbf{6.42 (0.096)} & \textbf{8.26} & \textbf{3.15} \\
        & Residual & \textbf{7170} & -0.69 & 7.68 (0.096) & \textit{6.34} & \textit{2.69} \\
        & Prepend & 5914 & \textit{-0.53} & \textit{7.35 (0.104)} & 5.26 & 2.45 \\
        
        \midrule
        \multirow{3}{*}{10K}
        & Prefix & \textbf{7618} & \textbf{-0.12} & \textbf{5.97 (0.081)} & \textbf{10.38} & \textbf{4.59} \\
        & Residual & 7526 & \textit{-0.61} & \textit{7.44 (0.091)} & 6.44 & \textit{3.05} \\
        & Prepend & \textit{7609} & -0.65 & 7.56 (0.090) & \textit{6.55} & 2.97 \\
        
        \midrule
        \multirow{3}{*}{100K}
        & Prefix & 7700 & \textbf{0.83} & \textbf{1.98 (0.039)} & \textbf{36.93} & \textit{12.40} \\
        & Residual & \textit{7734} & 0.77 & \textit{2.35 (0.046)} & \textit{34.61} & \textbf{13.05} \\
        & Prepend & \textbf{7735} & \textit{0.78} & 2.39 (0.045) & 33.45 & 12.28 \\
        
        \midrule
        \multirow{3}{*}{Full}
        & Prefix& 7679 & \textbf{0.97} & \textbf{0.72 (0.016)} & \textbf{56.70} & \textit{16.32} \\
        & Residual & \textbf{7745} & 0.95 & 1.00 (0.021) & 49.33 & 16.14 \\
        & Prepend & \textit{7686} & \textit{0.96} & \textit{0.84 (0.019)} & \textit{55.43} & \textbf{16.51} \\
    \bottomrule
    \end{tabular}
    \caption{Density-conditioning performance across fine-tuning dataset sizes and architectures. $N_{\mathrm{Valid}}$ is the total number of structurally valid generations. Valid-Hit is the percentage of those within $\pm$~\SI{1.0}{g/cm^3} of the requested target, and Q-Hit is the percentage of those generations that also satisfy the $Q_{\mathrm{VSUN}}$ criteria. $R^2$ and MAE $\pm$ SE are computed between requested and predicted densities for $Q_{\mathrm{VSUN}}$ materials only. Bold and italics denote the best and second-best results within each dataset size.}
    \label{tab:dataset-size-metrics}
\end{table}

\clearpage
\subsection{Comparison with graph diffusion}
\label{app:diffusion-comparison}

\begin{figure*}[htbp!]
    \centering
    \includegraphics[width=0.8\textwidth]{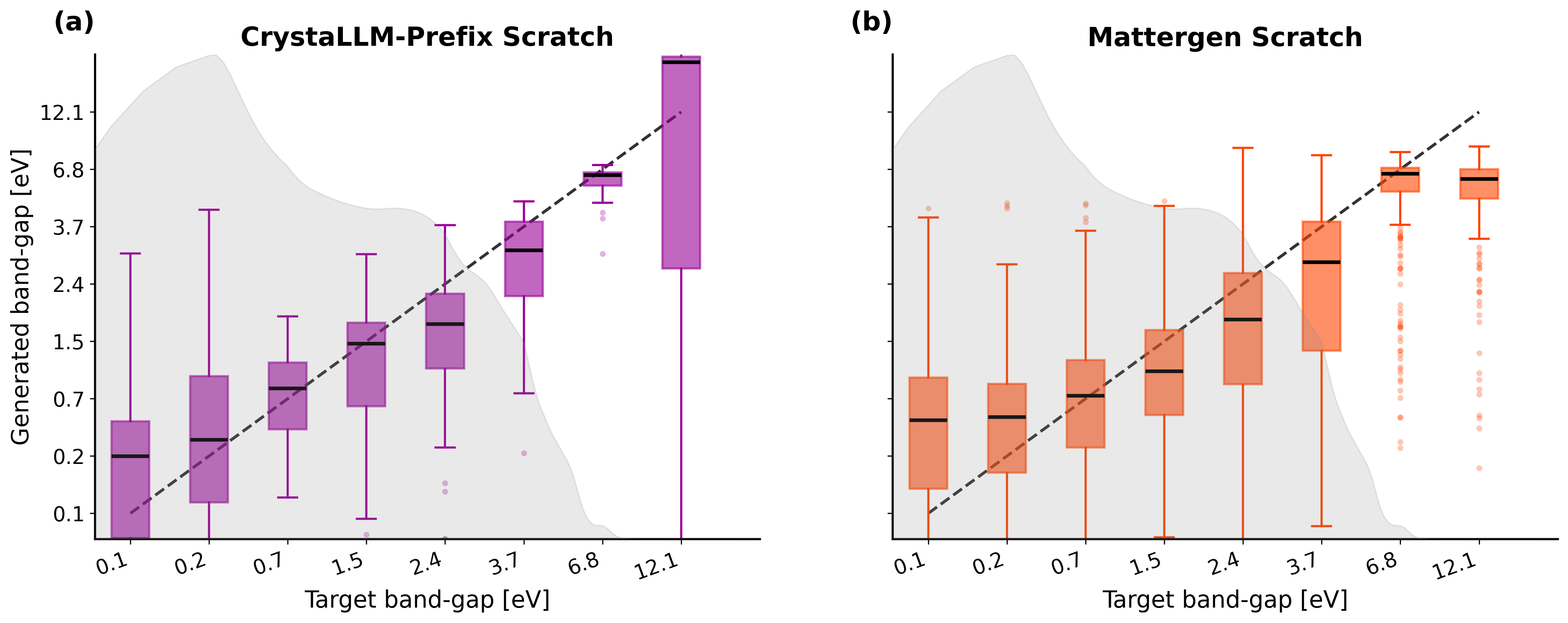}
    \caption{Parity distributions comparing ALIGNN-predicted band gaps against requested target values for materials generated by the CrystaLLM-\scalebox{1.3}{$\pi$} Prefix model and the MatterGen graph diffusion baseline. Panel \textbf{a} shows the from scratch CrystaLLM-\scalebox{1.3}{$\pi$} Prefix model and panel \textbf{b} shows the from scratch MatterGen baseline. Box plots summarise the distributions of predicted band gaps for $Q_{\mathrm{VSUN}}$ generations at discrete, non-linearly spaced target values. The dashed diagonal line marks ideal calibration between requested and generated values. The shaded grey background profile shows the band gap distribution of the training dataset. These distributions indicate that the CrystaLLM-\scalebox{1.3}{$\pi$} Prefix architecture maintains tighter calibration and narrower spread across most of the sampled property range than the graph diffusion baseline.
    }
    \label{fig:mattergen_parity}
\end{figure*}

\begin{figure*}[htbp!]
    \centering
    \includegraphics[width=0.8\textwidth]{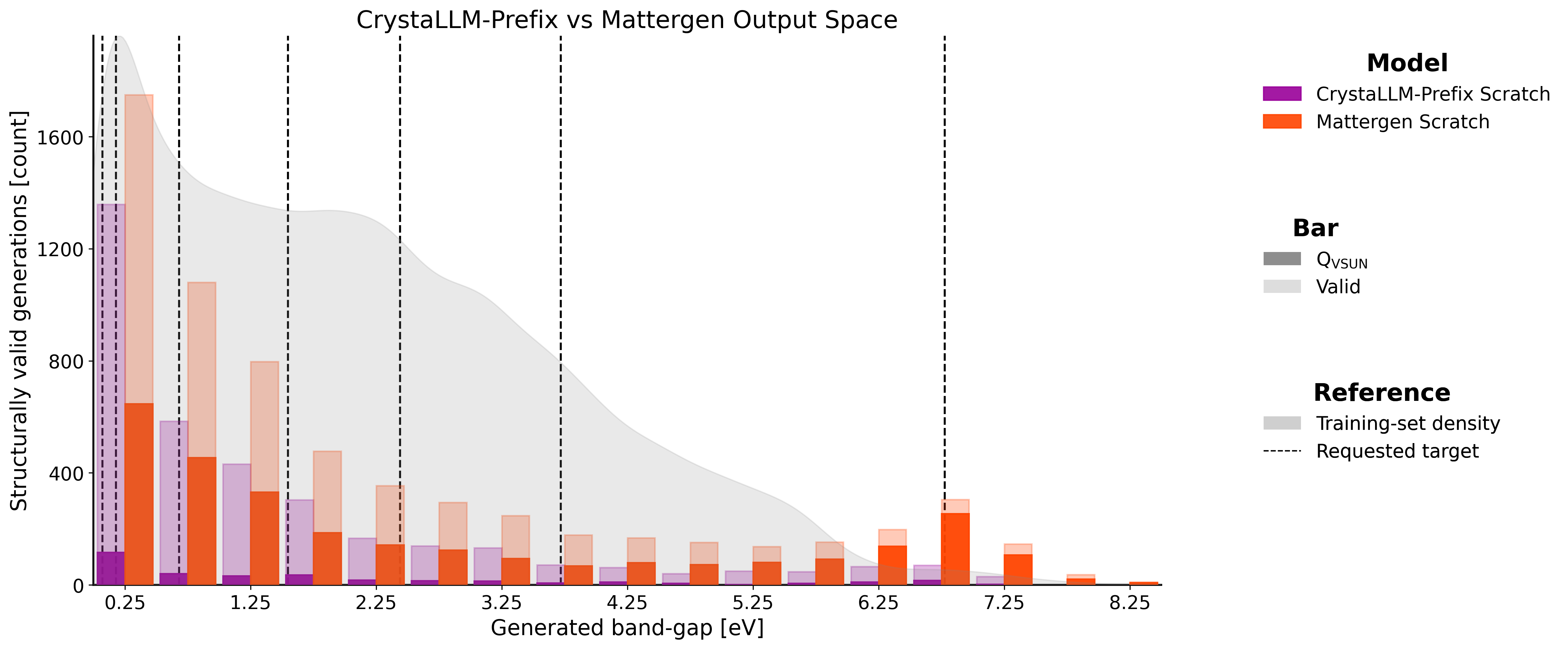}
    \caption{Output-space distribution of band gap-conditioned generations comparing the from scratch CrystaLLM-\scalebox{1.3}{$\pi$} Prefix architecture with the MatterGen graph diffusion baseline. The histogram reports counts across generated band gap bins, where translucent bars denote structurally valid generations and solid bars denote the subset satisfying the comprehensive $Q_{\mathrm{VSUN}}$ metric (Validity, Stability, Uniqueness and Structural Novelty; Methods~\ref{sec:eval_metrics}). Dashed vertical lines mark the requested target band gaps, and the shaded grey background shows the training-set band gap distribution. No method generated structures satisfying the \SI{12.1}{\electronvolt} target, which explains its absence in the output space. These distributions show that MatterGen produces more Valid and $Q_{\mathrm{VSUN}}$ structures across most of the sampled range, while both methods fail at the most extreme target.}
    \label{fig:mattergen_output_space}
\end{figure*}

\begin{figure*}[htbp!]
    \centering
    \includegraphics[width=0.8\textwidth]{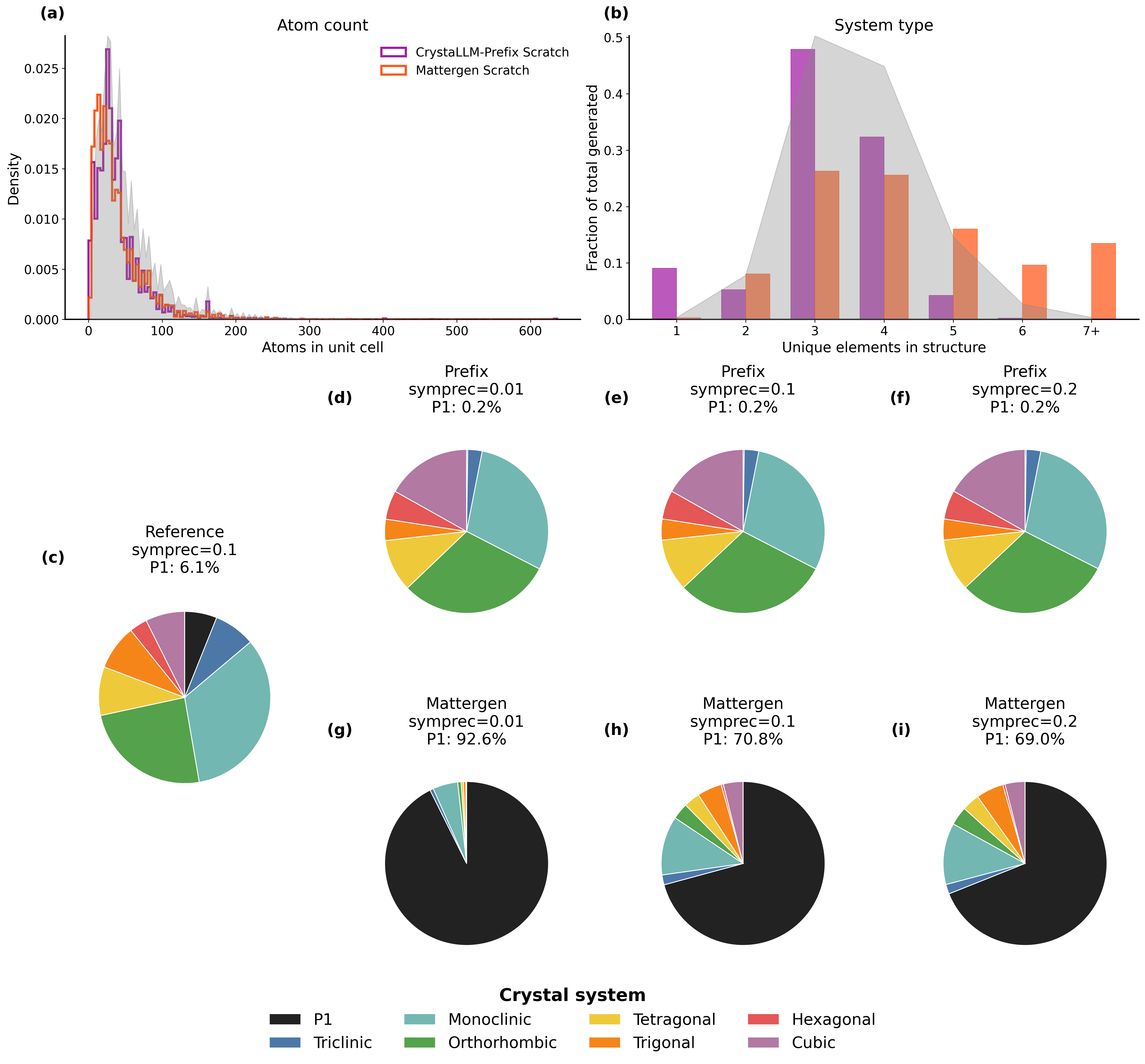}
    \caption{Distributional analysis of structural parameters for 8K materials generated by the MatterGen baseline and the from scratch CrystaLLM-\scalebox{1.3}{$\pi$} Prefix architecture across eight target band gaps. Panel \textbf{a} shows the probability density of the total atom count per unit cell. Panel \textbf{b} shows the fraction of generated structures as a function of compositional complexity, where the terminal 7+ bin collects all outliers, including MatterGen generations with up to 31 distinct elements. Panels \textbf{c} through \textbf{i} summarise symmetry outcomes as crystal-system pie charts: panel \textbf{c} reports the reference training-set distribution at $\mathrm{symprec}=0.1~\si{\angstrom}$, panels \textbf{d}, \textbf{e}, and \textbf{f} report CrystaLLM-\scalebox{1.3}{$\pi$} Prefix generations at $\mathrm{symprec}=$~\SIlist[list-units=single]{0.01;0.01;0.2}{\angstrom}, and panels \textbf{g}, \textbf{h}, and \textbf{i} report MatterGen generations at the same $\mathrm{symprec}$ values. In these pies, P1 is treated as its own category, separate from triclinic. The shaded grey background in panels \textbf{a} and \textbf{b} shows reference distributions extracted from the training dataset. Together, these comparisons show that CrystaLLM-\scalebox{1.3}{$\pi$} Prefix tracks the training-set distributions more closely in atom count, compositional complexity and crystal-system assignment, whereas MatterGen produces more high-complexity compositions and remains strongly biased toward P1 even as $\mathrm{symprec}$ increases.
    }
    \label{fig:mattergen_structural_dists}
\end{figure*}

Figures~\ref{fig:mattergen_parity},~\ref{fig:mattergen_structural_dists} and~\ref{fig:mattergen_output_space} show behavioral differences between MatterGen and CrystaLLM-\scalebox{1.3}{$\pi$}. Figure~\ref{fig:mattergen_output_space} highlights a marked difference in compositional complexity between the CrystaLLM-\scalebox{1.3}{$\pi$} and MatterGen~\cite{zeniGenerativeModelInorganic2025}. The former more often produces unary to senary compositions, whereas the graph diffusion baseline sometimes generates structures with substantially larger numbers of distinct elements, reaching up to 31 in a single unit cell in this benchmark. One likely contributor is that the autoregressive CIF representation encodes stoichiometry through discrete tokens, while the graph diffusion representation does not impose the same explicit compositional scaffold during generation. Because convex-hull analysis scales poorly for very large chemical systems, these were intractable on available hardware, so structures with more than six unique elements were excluded from the stability calculations.

CrystaLLM-\scalebox{1.3}{$\pi$} also yields substantially higher symmetry retention in the unrelaxed outputs considered here. Because the model generates discrete crystallographic fields hierarchically (Appendix~\ref{app:logit_study}), many outputs remain compatible with higher-order space-group assignments before post-processing. By contrast, the unrelaxed MatterGen outputs are strongly biased toward P1 in this analysis, consistent with small coordinate deviations obscuring symmetry before relaxation. At $\mathrm{symprec}=0.2~\si{\angstrom}$, approximately $69\%$ of MatterGen structures still fail to recover higher-order symmetry, compared with $0.2\%$ for CrystaLLM-\scalebox{1.3}{$\pi$}. This comparison helps explain why relaxation is particularly important when evaluating continuous-coordinate diffusion outputs.

Finally, Table~\ref{tab:compute_costs} shows a large computational gap between the two frameworks in this benchmark. CrystaLLM-\scalebox{1.3}{$\pi$} was run on a 24G L4 GPU with a peak VRAM demand of~\SI{8.99}{\giga\byte}, whereas the MatterGen baseline required~\SI{40.35}{\giga\byte} on 40G A100 GPUs. Training and inference were also substantially faster for the autoregressive model under the reported settings. These differences are consistent with the lower sampling complexity of the 1D autoregressive pipeline relative to graph-based diffusion. In this setting, the discrete token approach therefore offers a lighter route to conditional generation, although the trade-off with structural yield remains as discussed in Section~\ref{sec:pre-training}.

\vspace{5mm}

\begin{table}[htbp!]
    \centering
    \renewcommand{\arraystretch}{1.3}
    \setlength{\tabcolsep}{4.0pt}
    \begin{tabular}{@{}llccc@{}}
        \toprule
        \textbf{Model} & \textbf{GPU} & \textbf{\begin{tabular}[c]{@{}c@{}}Train Time\\ (HH:MM:SS)\end{tabular}} & \textbf{\begin{tabular}[c]{@{}c@{}}Inference Time\\ (1K Gens)\end{tabular}} & \textbf{\begin{tabular}[c]{@{}c@{}}Max VRAM\\ (GB)\end{tabular}} \\
        \midrule
        CrystaLLM-$\pi$ & L4 (24G) & 03:47:47 & 00:02:26 & 8.99 \\
        MatterGen & A100 (40G) & 325:30:06 & 15:06:08 & 40.35 \\
        \bottomrule
    \end{tabular}
    \caption{Computational resource requirements and processing times for CrystaLLM-\scalebox{1.3}{$\pi$} and MatterGen on the MP~Bandgap dataset. In the evaluation set, $95\%$ of structures contain between 1 and 104 atoms per unit cell. Reported metrics are the hardware used, peak video random access memory (VRAM), total training duration and inference time required to generate 1K structures. MatterGen~\cite{zeniGenerativeModelInorganic2025} was trained for 1100 epochs across an unconditional Phase 1 stage and a conditional Phase 2 stage with a classifier-free guidance adapter. CrystaLLM-\scalebox{1.3}{$\pi$} was trained for 24 epochs.}
    \label{tab:compute_costs}
\end{table}

\clearpage

\vspace{5mm}

\clearpage

\subsection{Losses}
\label{app:loss_landscapes}

\begin{figure}[htbp!]
    \centering
    \includegraphics[width=0.6\columnwidth]{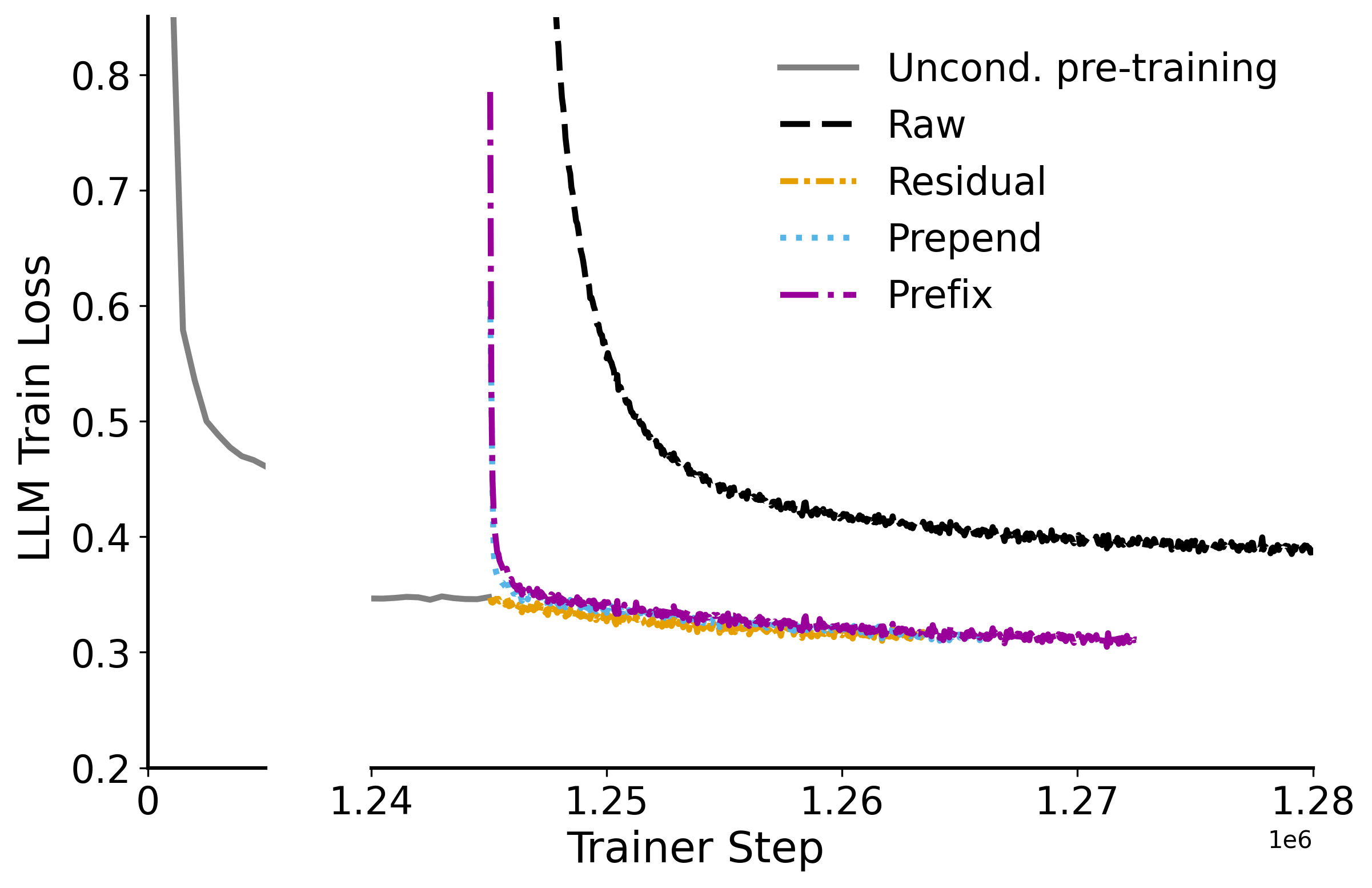}
    \caption{Evolution of the total training loss across the transition from unconditional pre-training to conditional fine-tuning on the MP~Bandgap dataset. The horizontal axis uses a broken scale to show the initial pre-training descent (left) together with the late-stage convergence and fine-tuning transition near $1.25 \times 10^6$ trainer steps (right). The solid grey curve traces the unconditional pre-training loss. The Raw, Residual, Prepend and Prefix conditioning architectures were then fine-tuned for up to 50K additional steps with matched hyperparameters and early stopping. These trajectories show that the Residual architecture undergoes the least disruption at the transition, whereas the Raw method destabilises strongly and does not recover the pre-training loss level.}
    \label{fig:fine-tuning_losses}
\end{figure}

\begin{figure}[htbp!]
    \centering
    \includegraphics[width=0.6\columnwidth]{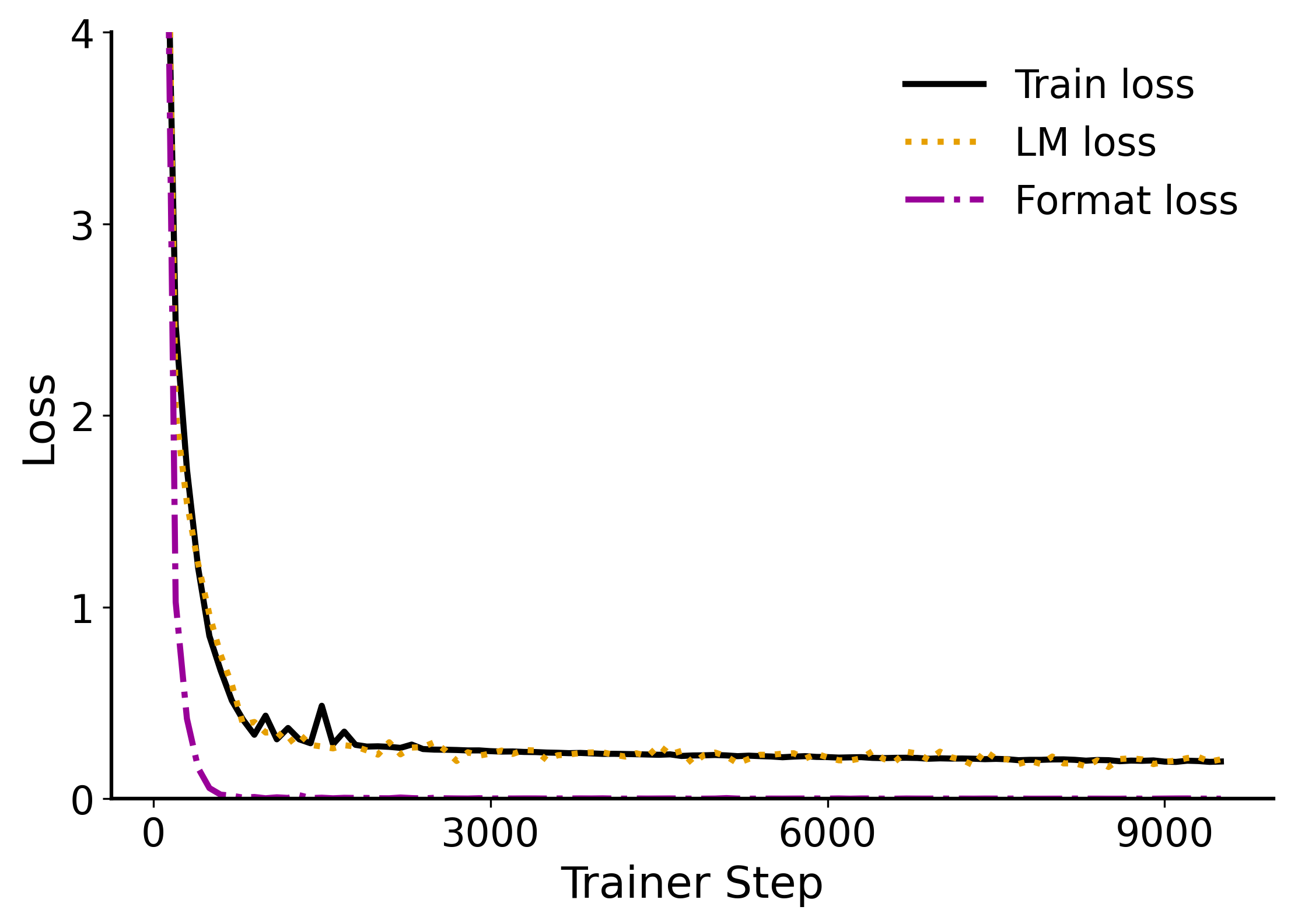}
    \caption{Evolution of individual training loss components for the generative model trained from scratch on the MP-20 XRD benchmark. The plot shows the total train loss, language-modelling (LM) loss and format loss evaluated with component weighting $\{n_1, n_2\} = \{1,1\}$. The format loss decays to near zero within the first 1K steps, after which the total loss is dominated by the LM term. This indicates that the model learns the invariant Crystallographic Information File (CIF) syntax early and then spends most of its training on the variable structural content.}
    \label{fig:format_losses}
\end{figure}

\clearpage
\subsection{Sampling Scheme Parameter Studies for Structure Recovery Tasks}
\label{app:ablation_studies}
To determine optimal hyperparameters and generation strategy for the crystal structure recovery tasks, a series of ablation studies were conducted on a 1K structure subset of the MP-20 test set.

The first study, shown in Table~\ref{tab:temp_ablation}, gives a sampling temperature of \texttt{T} = 0.75 as a good balance between recovery accuracy and generative diversity. While a lower temperature of \texttt{T} = 0.5 yields a marginally better RMSD, selecting \texttt{T} = 0.75 increases the 1-perplexity match rate by $1.4\%$ points, maximising the probability of identifying the correct structure in a single closed loop attempt. Higher temperatures offer a slight increase in the match rate for 20-consistent CIF generation attempts, but at the cost of reduced structural accuracy.

The second study, shown in Table~\ref{tab:ranking_ablation}, confirms the benefits of both XRD conditioning and perplexity-based ranking. Providing XRD information consistently improves both the 1-perplexity match rate ($+3.3\%$ points) and the RMSD ($-0.0058$). Then, ranking candidates by perplexity provides an improvement again for both metrics without any trade-off.

\begin{table*}[htbp!]
    \centering
    \begin{tabular}{lccccc}
        \toprule
        \textbf{Model} & \textbf{Temp} & \textbf{Ranking Scheme} & \textbf{Match Rate (\%)} & \textbf{RMSD} & \textbf{Avg. Perplexity} \\
        
        \midrule
        \multirow{12}{*}{\textbf{CrystaLLM-\scalebox{1.3}{$\pi$}}} & 0.5 & 20-consistent & 73.2 & \textbf{0.0379} & \textbf{1.151} \\
        & 0.75 & 20-consistent & 75.3 & 0.0393 & 1.164 \\
        & 1.0 & 20-consistent & 76.4 & 0.0467 & 1.172 \\
        & 1.25 & 20-consistent & 76.3 & 0.0461 & 1.176 \\
        & 1.5 & 20-consistent & \textbf{76.6} & 0.0532 & 1.181 \\
        & 1.7 & 20-consistent & 76.2 & 0.0580 & 1.181 \\
        
        \cmidrule{2-6}
        & 0.5 & 1-perplexity & 63.6 & \textbf{0.0408} & \textbf{1.136} \\
        & 0.75 & 1-perplexity & \textbf{65.0} & 0.0465 & 1.148 \\
        & 1.0 & 1-perplexity & 63.2 & 0.0504 & 1.154 \\
        & 1.25 & 1-perplexity & 61.0 & 0.0473 & 1.154 \\
        & 1.5 & 1-perplexity & 58.4 & 0.0511 & 1.158 \\
        & 1.7 & 1-perplexity & 56.6 & 0.0511 & 1.156 \\
        \bottomrule 
    \end{tabular}
    
    \caption{Ablation study for sampling-temperature optimisation. The model was trained from scratch on the MP-20 XRD dataset. ``$N$-consistent'' denotes generation until $N$ self-consistent CIFs are obtained, where self-consistent means passing basic chemical validity checks. ``1-perplexity'' denotes selecting the lowest-perplexity output among those $N$-consistent CIFs for matching, as described in Section~\ref{sec:sampling}.}
    \label{tab:temp_ablation}
\end{table*}

\begin{table*}[htbp!]
    \centering
    \begin{tabular}{lcccccc}
    \toprule
        \textbf{Model} & \textbf{Temp} & \textbf{Ranking Scheme} & \textbf{XRD Info} & \textbf{Match Rate (\%)} & \textbf{RMSD} & \textbf{Avg. Perplexity} \\
        
        \midrule
        \multirow{6}{*}{\textbf{CrystaLLM-\scalebox{1.3}{$\pi$}}} & 0.75 & 20-shot & No & 73.8 & 0.0427 & -- \\
        & 0.75 & 20-shot & Yes & 75.3 & 0.0418 & -- \\
        & 0.75 & 20-consistent & Yes & \textbf{75.3} & \textbf{0.0393} & 1.164 \\
        
        \cmidrule{2-7}
        & 0.75 & 1-shot & No & 59.5 & 0.0589 & -- \\
        & 0.75 & 1-shot & Yes & 62.8 & 0.0531 & -- \\
        & 0.75 & 1-perplexity & Yes & \textbf{65.0} & \textbf{0.0465} & 1.148 \\
        \bottomrule
    \end{tabular}
    
    \caption{Ablation study on the utility of XRD conditioning and perplexity-based ranking. The model was trained from scratch on the MP-20 XRD dataset. ``$N$-shot'' denotes matching against $N$ generated candidates. ``$N$-consistent'' denotes generation until $N$ self-consistent CIFs are obtained, where self-consistent means passing basic chemical validity checks. ``1-perplexity'' denotes selecting the lowest-perplexity output among those $N$-consistent CIFs for matching, as described in Section~\ref{sec:sampling}. All models were trained with XRD information, and the XRD ablation removes that information only at inference time.}
    \label{tab:ranking_ablation}
\end{table*}

\clearpage

\subsection{Data Leakage Analysis for XRD Structure Recovery}
\label{app:data_leakage}

Large-scale pre-training on the 4.35 million structures introduces the risk of data leakage when evaluating downstream tasks. To decouple the model's intrinsic structure-solving capabilities from potential memorisation of the pre-training data, the novelty of each material in the test set was defined against the union of the pre-training (LeMaterial) and fine-tuning (Jarvis-DFT train/validation) datasets.
\begin{enumerate}
    \item \textbf{Seen:} Materials where both the exact structural configuration and the reduced formula are present in the reference data using pymatgen's \texttt{StructureMatcher} with the VSUN metric parameters (See Section~\ref{sec:diversity_metrics})
    \item \textbf{Novel Polymorph (Novel Polym.):} Materials where the reduced formula exists in the reference data, but the specific structural polymorph is absent.
    \item \textbf{Novel Composition (Novel Comp.):} Materials where the reduced formula and structure is completely absent from all reference datasets.
\end{enumerate}

Under the ``20-consistent'' evaluation scheme, the match rate increases from $84.99\%$ for ``Seen'' materials to $87.85\%$ for ``Novel Compositions'', while the RMSD improves from \SIrange{0.0360}{0.0288}{\angstrom}. A plausible interpretation is that memorised structure priors can sometimes compete with the conditioning signal for ``Seen'' compositions, whereas this competition is weaker for novel compositions. However, the split is also affected by sample composition: although the average atom counts are similar ($16.5$ for ``Seen'' versus $15.9$ for ``Novel Compositions''), the ``Seen'' subset contains larger outliers, reaching $232$ atoms compared with $116$ for novel targets. The observed advantage for novel compositions should therefore be interpreted as dataset-specific rather than as a general rule that out-of-distribution recovery is easier. 

\begin{table}[htbp!]
    \centering
    \setlength{\tabcolsep}{4pt}
    \begin{tabular}{l l l c c c c c c c c c}
        \toprule
        \textbf{Model} & \textbf{Num} & \textbf{Subset} & \textbf{N in} & \textbf{Match} & \textbf{RMSD} & \textbf{MAE} & \textbf{MAE} & \textbf{MAE} & \textbf{MAE} & \textbf{Avg.} & \textbf{Max} \\
        & \textbf{gens} & & \textbf{bucket} & \textbf{Rate (\%)} & \textbf{(\AA)} & \textbf{a (\AA)} & \textbf{b (\AA)} & \textbf{c (\AA)} & \textbf{V (\AA$^3$)} & \textbf{Atoms} & \textbf{Atoms} \\
        \midrule
        
        CrystaLLM-\scalebox{1.3}{$\pi$} & 20 & Seen & 4251 & 84.99 & 0.0360 & 0.131 & 0.131 & 0.216 & 9.359 & 16.5 & 232 \\
        CrystaLLM-\scalebox{1.3}{$\pi$} & 20 & Novel Polym. & 2716 & 85.68 & 0.0380 & 0.158 & 0.137 & 0.216 & 11.960 & 15.6 & 160 \\
        CrystaLLM-\scalebox{1.3}{$\pi$} & 20 & Novel Comp. & 609 & 87.85 & 0.0288 & 0.142 & 0.150 & 0.213 & 23.125 & 15.9 & 116 \\
        
        \midrule
        CrystaLLM-\scalebox{1.3}{$\pi$} & 1 & Seen & 4251 & 66.67 & 0.0343 & 0.096 & 0.097 & 0.186 & 9.013 & 17.0 & 232 \\
        CrystaLLM-\scalebox{1.3}{$\pi$} & 1 & Novel Polym. & 2716 & 65.46 & 0.0373 & 0.104 & 0.106 & 0.188 & 7.100 & 16.1 & 160 \\
        CrystaLLM-\scalebox{1.3}{$\pi$} & 1 & Novel Comp. & 609 & 66.50 & 0.0260 & 0.062 & 0.067 & 0.140 & 7.474 & 17.3 & 116 \\
        \bottomrule
    \end{tabular}%
    \caption{Stratified performance metrics on the Jarvis-DFT benchmark for evaluating data leakage. The test set is split by strict novelty criteria relative to the combined pre-training and fine-tuning datasets. Structural fidelity is assessed by root mean square displacement (RMSD), mean absolute errors (MAE) of lattice parameters and cell volume, and the average and maximum atom counts of successfully recovered structures. The ``20'' and ``1'' ``Num gens'' columns correspond to the ``20-consistent'' and ``1-perplexity'' generation settings, respectively.}
    \label{tab:jarvis_stratified}
\end{table}

\clearpage

\subsection{Default Hyperparameters for Pre-training and Framework Validation}
\label{app:training_hyperparams}

The model backbone consists of 26 million parameters with an embedding size of 512, 8 decoder layers and 8 attention heads, following CrystaLLM's small model architecture~\cite{antunesCrystalStructureGeneration2024}. Training employed DeepSpeed Stage 2 optimisation across 2 NVIDIA L4 GPUs with early stopping based on validation loss plateau or increase to avoid overfitting. For the exact hyperparameter values used during training, see Table~\ref{tab:hp_settings} and configuration files are available in the \href{https://github.com/C-Bone-UCL/CrystaLLM-pi/tree/main/_config_files}{\texttt{project repository}}~\cite{configfiles}.

Unless hyperparameter searches were carried out, default training hyperparameters were used in the various studies of this work. For default pre-training on the structure-only dataset, the model used cosine learning rate scheduling with the AdamW~\cite{loshchilovDecoupledWeightDecay2019} optimiser and pre-training on the text-only MP-20 and Alex-MP-20 for the LeMat-Benchmark (Section~\ref{app:lemat-bench}) used the Muon~\cite{jordan2024muon} optimiser. Conditional fine-tuning continued from pre-trained checkpoints with reduced learning rates for backbone parameters, while newly initialised conditional encoders used higher learning rates. The conditional encoder hidden dimension was set to 1024 with reduced regularisation compared to pre-training.

Models trained from scratch in comparative studies used default fine-tuning hyperparameters with higher regularisation. Training duration varied by study design, with pre-training benefits experiments using 50K steps and dataset size studies scaling from 1.4K to 400K steps based on dataset size. Exact configuration files for all training and fine-tuning runs are available in the \href{https://github.com/C-Bone-UCL/CrystaLLM-pi}{\texttt{model repository}}~\cite{modelrepository}.

\begin{table*}[htbp!]
    \centering
    \begin{tabular}{lcc}
        \toprule
        \textbf{Hyperparameter} & \textbf{pre-training} & \textbf{fine-tuning} \\
        
        \midrule
        Optimiser & AdamW & AdamW \\
        Learning Rate & $10^{-3}$ & $5\times10^{-6}$ \\
        LR Schedule & Cos. to 0.1\%LR & Cos. to 1\% LR \\
        $\beta_1$ & 0.85 & 0.9 \\
        $\beta_2$ & 0.98 & 0.999 \\
        Gradient Clipping & 1.0 & 1.0 \\
        Batch Size & 32 & 32 \\
        Dropout & 0.1 & 0.1 \\
        Weight Decay & 0.1 & 0.01 \\
        Context Length & 1024 & 1024 \\
        Training Steps & 1.250M & Varies \\
        Warmup & 0 steps & 2\% of steps \\
        
        \midrule
        Cond. Hidden Dim. & -- & 1024 \\
        Cond. Dropout & -- & 0.01 \\
        Cond. LR & -- & $5\times10^{-4}$ \\
        Cond. Weight Decay & -- & 0.01 \\
        \bottomrule
    \end{tabular}
    
    \caption{Hyperparameter settings for the pre-training and conditional fine-tuning studies spanning 50K samples (pre-training benefits) and 1.4K--400K samples (dataset-size study). From-scratch models used a backbone LR of $10^{-3}$, a conditional LR of $10^{-3}$, weight decay of 0.1 and 15K training steps. ``Cond.'' denotes values specific to the conditioning encoder. LR denotes learning rate. ``Cos.'' denotes a cosine-annealed learning-rate schedule to a target fraction of the base learning rate.}
    \label{tab:hp_settings}
\end{table*}

\clearpage

\subsection{Discovery and Recovery hyperparameters}
\label{param_search}
Hyperparameter optimisation for specialised conditional generation tasks employed either Optuna with a MedianPruner~\cite{akibaOptunaNextgenerationHyperparameter2019} or a Weights \& Biases (WandB) Bayesian sweep across 50 trials. The optimisation objective targeted the minimisation of validation loss. Validation sets were generated by withholding $5\%$ of the data for SLME studies and $10\%$ for the Chili-100K and Jarvis dataset. Early stopping prevented overfitting during both the search and final training. Exact configuration files for all optimised training and fine-tuning runs are available in the \href{https://github.com/C-Bone-UCL/CrystaLLM-pi}{\texttt{model repository}}~\cite{modelrepository}.

\begin{table*}[htbp!]
    \centering
    \setlength{\tabcolsep}{5pt}
    \begin{tabular}{lccccc}
        \toprule
        \textbf{Dataset} & \textbf{Architecture} & \textbf{Initialization} & \textbf{Holdout} & \textbf{Trials} & \textbf{Optimized Parameters} \\
        
        \midrule
        MP~SLME & Prefix & Pre-trained & 5\% & 50 & Cond. Prefix Tokens: $\{1, 2, 3, 4, 5\}$ \\
        & & & & & Cond. Hidden Dim.: $\{512, 1024\}$ \\
        & & & & & Base LR: $\log U(5\times10^{-7}, 5\times10^{-5})$ \\
        & & & & & Cond. Dropout: $U(0.0, 0.2)$, Warmup steps: $U(50, 500)$ \\
        & & & & & Base WD: $U(0.01, 0.2)$, Cond. WD: $U(0.01, 0.3)$ \\
        
        \midrule
        MP-20 XRD & Residual & Scratch & Existing & 50 & Base LR: $\log U(1\times10^{-5}, 1\times10^{-2})$ \\
        & & & & & Cond. Dropout: $U(0.0, 0.2)$, Warmup: $U(0.01, 0.10)$ \\
        & & & & & Base WD: $U(0.01, 0.2)$, Cond. WD: $U(0.01, 0.3)$ \\
        & & & & & Backbone Dropout: $U(0.0, 0.2)$ \\
        
        \midrule
        Jarvis XRD & Residual & Pre-trained & 10\% & 50 & Base LR: $\log U(5\times10^{-7}, 5\times10^{-5})$ \\
        & & & & & Cond. LR: $\log U(1\times10^{-4}, 1\times10^{-3})$ \\
        & & & & & Warmup: $U(0.01, 0.10)$, Cond. Dropout: $U(0.0, 0.2)$ \\
        & & & & & Base WD: $U(0.01, 0.2)$, Cond. WD: $U(0.01, 0.3)$ \\
        
        \midrule
        Chili-100K XRD & Residual & Sequential & 10\% & 50 & Base LR: $\log U(5\times10^{-4}, 5\times10^{-3})$ \\
        & & & & & Cond. LR: $\log U(5\times10^{-3}, 5\times10^{-2})$ \\
        & & & & & Warmup Ratio: $U(0.01, 0.10)$, Cond. Dropout: $U(0.0, 0.2)$ \\
        & & & & & Base WD: $U(0.01, 0.1)$, Cond. WD: $U(0.01, 0.1)$ \\
        \bottomrule
    \end{tabular}
    
    \caption{Hyperparameter search spaces for specialised conditional generation across multiple materials datasets. Continuous distributions are denoted by $U$ (uniform) and $\log U$ (log-uniform), while discrete options are enclosed in braces. Validation loss optimisation used Optuna (MedianPruner, TPESampler) for all studies except Chili-100K, which used a Weights \& Biases Bayesian sampler, and all searches were limited to 50 trials. Total training steps were SLME (16K), MP-20 scratch (30K) and Jarvis (16K). The Chili-100K XRD models underwent sequential fine-tuning with an initial Phase 1 adaptation on MatterGen XRD (500K steps) followed by Phase 2 Chili-100K optimisation (2.5K steps). Abbreviations: LR (learning rate), WD (weight decay), Cond. (conditioning encoder parameters).}
    \label{tab:hp_optimisation}
\end{table*}

\clearpage

\subsection{Open Sourced Models}
\label{app:available_models}

\begin{table}[htbp!]
    \centering
    \label{tab:model-summary}
    \setlength{\tabcolsep}{4pt}
    \begin{tabular}{llclccc}
        \toprule
        \textbf{Model Name} & \textbf{Study} & \textbf{Availability} & \textbf{Model} & \textbf{Training} & \textbf{Method} & \textbf{Property Vector} \\
        & & & \textbf{Parameters} & \textbf{Dataset} & & \\
        
        \midrule
        CrystalLLM-\scalebox{1.3}{$\pi$} Base &~\ref{sec:base_model} & Open-Source & 25.9M & LeMaterial & Base & None \\
        & & & & & & \\
        CrystalLLM-\scalebox{1.3}{$\pi$} MP-20 &~\ref{sec:base_model} & Open-Source & 25.9M & MP-20 & Base & None \\
        & & & & & & \\
        CrystalLLM-\scalebox{1.3}{$\pi$} Alex-MP-20 &~\ref{sec:base_model} & Open-Source & 25.9M & Alex-MP-20 & Base & None \\
        & & & & & & \\
        CrystalLLM-\scalebox{1.3}{$\pi$} Band gap &~\ref{sec:pre-training} & Open-Source & 61.6M & MP~Bandgap & Prefix & [Band gap, $E_{\text{hull}}^{\text{PBE}}$] \\
        & & & & & & \\
        CrystalLLM-\scalebox{1.3}{$\pi$} Density &~\ref{sec:dataset_size_study} & Open-Source & 61.6M & MatterGen & Prefix & [Density, $E_{\text{hull}}^{\text{PBE}}$] \\
        & & & & Density & & \\
        CrystalLLM-\scalebox{1.3}{$\pi$} Chili-100K-XRD &~\ref{sec:experimental_recovery} & Open-Source & 47.7M & Chili-100K XRD & Residual & [$20 \times 2\theta$, $20 \times \text{intensity}$] \\
        & & & & & & \\
        CrystalLLM-\scalebox{1.3}{$\pi$} SLME &~\ref{sec:slme} & Open-Source & 38.7M & MP~SLME & Prefix & [SLME] \\
        & & & & & & \\
        \bottomrule
    \end{tabular}
    
    \caption{Overview of open-source CrystaLLM-\scalebox{1.3}{$\pi$} model variants. The table summarises the model releases associated with this work, including their training datasets, parameter counts and the conditioning architecture used. Although additional trained models are available upon request, this table lists the public releases for the main studies in the paper. The models can be found in the paper's \href{https://huggingface.co/c-bone/models}{\texttt{Hugging Face repository}}~\cite{HuggingFacerepository}.}
    \label{tab:available_models}
\end{table}

\clearpage

\subsection{SLME Candidate Structures}
\label{app:slme_candidates}

\begin{table*}[htbp!]
    \centering
    \footnotesize
    \setlength{\tabcolsep}{0pt}
    \renewcommand{\arraystretch}{1.1}
    \begin{tabular*}{\textwidth}{@{\extracolsep{\fill}}lcclcccl}
        \toprule
        \textbf{Reduced} & \textbf{Pred.} & \textbf{Sustainability} & \textbf{Novelty} & \textbf{MACE} & \textbf{DFT} & \textbf{DFT} & \textbf{Outcome / Status} \\
        \textbf{Formula} & \textbf{SLME (\%)} & \textbf{($HHI_{dist}$)} & \textbf{(Str. $|$ Comp.)} & \textbf{$E_{\text{hull}}$} & \textbf{$E_{\text{hull}}$} & \textbf{SLME (\%)} & \\
        
        \midrule
        \multicolumn{8}{c}{\textit{Condition: High SLME Optimisation}} \\
        
        \midrule
        \ce{Cs3(GeBr3)4} & 30.8 & 7224 & ft-pt $|$ ft-pt & 0.040 & -- & -- & Excluded: Unusual oxidation states \\
        \textbf{\ce{Rb2(NbCl3)3}} & \textbf{30.4} & \textbf{7203} & \textbf{ft $|$ ft} & \textbf{0.021} & \textbf{0.003} & \textbf{11.6} & \textbf{Validated: Stable} \\
        \textbf{\ce{Rb2(NbCl3)3}} & \textbf{29.9} & \textbf{7203} & \textbf{ft $|$ ft} & \textbf{0.015} & \textbf{0.003} & \textbf{11.3} & \textbf{Validated: Stable} \\
        \textbf{\ce{Rb2(NbBr3)3}} & \textbf{29.8} & \textbf{8747} & \textbf{ft-pt $|$ ft-pt} & \textbf{0.017} & \textbf{0.001} & \textbf{13.25} & \textbf{Validated: Stable} \\
        \ce{Cs2(NbCl3)3} & 28.9 & 7344 & ft $|$ ft & -0.011 & -- & -- & Excluded: Structural Redundancy \\
        \ce{Ba3GaAs3} & 28.5 & 4361 & ft-pt $|$ ft & 0.086 & -- & -- & Excluded: Structural Redundancy \\
        \ce{CsSnTe2} & 28.4 & 5724 & ft-pt $|$ ft & 0.072 & -0.329 & Metallic & \textit{Interested: but DFT showed metallic} \\
        \ce{TaTeI3} & 28.4 & 6166 & ft $|$ ft & 0.024 & -- & -- & Excluded: Unusual oxidation states \\
        \textbf{\ce{Cs2NaInAs2}} & \textbf{28.1} & \textbf{6298} & \textbf{ft $|$ ft} & \textbf{-0.070} & \textbf{-0.605} & \textbf{26.4} & \textbf{Validated: High Efficiency and Stable} \\
        \ce{TaI6} & 28.1 & 6512 & ft-pt $|$ ft & 0.016 & -- & -- & Excluded: Unusual oxidation states \\
        \ce{Ba4Bi2O} & 27.9 & 5502 & ft $|$ ft & 0.015 & -- & -- & Excluded: Low Novelty \\
        \ce{K3InAs2} & 27.9 & 5204 & ft $|$ ft & -0.013 & -- & -- & Excluded: Structural Redundancy \\
        \textbf{\ce{Cs2NaGaAs2}} & \textbf{27.9} & \textbf{6731} & \textbf{ft $|$ ft} & \textbf{-0.042} & \textbf{-0.589} & \textbf{24.4} & \textbf{Validated: High Efficiency and Stable} \\
        \ce{K2NaSnSb2} & 27.9 & 6198 & ft $|$ ft & 0.000 & 0.035 & Metallic & \textit{Interested: but DFT showed metallic}\\
        \ce{Cs2NaSnAs2} & 27.8 & 6114 & ft $|$ ft & 0.008 & -- & -- & Excluded: Structural Redundancy \\
        
        \midrule
        \multicolumn{8}{c}{\textit{Condition: Combined Sustainability (HHI$_{\mathrm{dist}}$) and SLME Optimisation}} \\
        
        \midrule
        \ce{KAg3S2} & 25.5 & 2207 & ft $|$ None & 0.045 & -- & -- & Excluded: Low Novelty \\
        \ce{Cu3GeS4} & 25.8 & 2440 & ft $|$ ft & -0.032 & -- & -- & Excluded: Low Novelty \\
        \ce{KCu3S2} & 25.2 & 2574 & ft-pt $|$ ft & 0.029 & -- & -- & Excluded: Chemical Heuristics (Cu-II) \\
        \ce{SnS} & 25.5 & 2644 & ft-pt $|$ None & 0.057 & -- & -- & Excluded: Known Material \\
        \ce{Ba(CuS)2} & 25.8 & 2656 & ft $|$ None & 0.057 & -- & -- & Excluded: Low Novelty \\
        \ce{K2Ni3S4} & 25.4 & 2706 & ft $|$ None & 0.001 & -- & -- & Excluded: Low Novelty \\
        \ce{K2(TiCl3)3} & 27.1 & 2748 & ft $|$ ft & 0.026 & -- & -- & Excluded: Unusual oxidation states \\
        \ce{K2(TiCl3)3} & 26.0 & 2748 & ft $|$ ft & 0.018 & -- & -- & Excluded: Unusual oxidation states \\
        \ce{K2(TiCl3)3} & 26.5 & 2748 & ft $|$ ft & 0.022 & -- & -- & Excluded: Unusual oxidation states \\
        \ce{Cu2SiSe4} & 26.0 & 2794 & ft-pt $|$ ft & 0.047 & -- & -- & Excluded: Chemical Heuristics (Cu-II) \\
        \ce{K2Cu3S4} & 26.4 & 2797 & ft-pt $|$ ft-pt & 0.064 & 0.183 & Metallic & \textit{Interested: but DFT showed metallic} \\
        \textbf{\ce{NaHfCuS3}} & \textbf{25.1} & \textbf{2886} & \textbf{ft $|$ ft} & \textbf{-0.076} & \textbf{-0.013} & \textbf{23.3} & \textbf{Validated: High Efficiency and Stable} \\
        \ce{Ba(CuSe)2} & 26.2 & 2974 & ft $|$ None & 0.045 & -- & -- & Excluded: Low Novelty \\
        \ce{KCu2Se3} & 26.9 & 3014 & ft-pt $|$ ft & 0.052 & -- & -- & Excluded: Chemical Heuristics (Cu-II) \\
        \ce{KCu2Se3} & 26.3 & 3014 & ft $|$ ft & 0.048 & -- & -- & Excluded: Chemical Heuristics (Cu-II) \\
        \bottomrule
    \end{tabular*}
    
    \caption{Screening summary of the top 30 generated photovoltaic candidates sorted by their conditioning metric. \textbf{Bold} entries denote candidates validated by DFT as stable or high-efficiency, while entries with outcome status in \textit{italics} denote materials that were investigated further with DFT but did not remain satisfactory. The top panel lists candidates conditioned purely on optimal SLME ($>30\%$), while the bottom panel targets a multi-objective condition of high SLME and low supply-chain risk (Methods~\ref{sec:eval_metrics}). Novelty is reported relative to the fine-tuning dataset only (ft) or to both the fine-tuning and pre-training datasets (ft-pt). ``Unusual oxidation states'' exclusions indicate cases where charge balance would require chemically unusual oxidation states. Cu(II)-containing compounds are excluded because d$^9$ copper is difficult to treat accurately with DFT. $E_{\text{hull}}$ values are reported in \si{\electronvolt/atom}. Following Yu and Zunger, materials with SLME above $20\%$ are classified as high efficiency~\cite{yuIdentificationPotentialPhotovoltaic2012}.}
    \label{tab:slme_candidates_full}
\end{table*}

\clearpage

\bibliography{apssamp}% Produces the bibliography via BibTeX.

\end{document}